\documentstyle[12pt]{article}
\begin{document}

\bibliographystyle{unsrt}

\baselineskip=24pt
\newcounter{figurenumber}
\newenvironment{FIGURE}
{\stepcounter{figurenumber}\noindent{\bf Figure \thefigurenumber\/}:}{}
\def\gnt{\gamma^{\theta}_{\nu}}
\def\sumij{\sum_{<ij>\sigma}}\def\sumijl{\sum_{<ij>\sigma\sigma^{\prime}}}
\def\sigmi{\sum_{i\sigma}}
\def\si{\bf S_{i}}
\def\sj{\bf S_{j}}
\def\sip{S_{i}^{+}}
\def\sib{S_{i}^{-}}
\def\sjp{S_{j}^{+}}
\def\sjm{S_{j}^{-}}
\def\siz{S_{i}^{z}}
\def\sjz{S_{j}^{z}}
\def\ciud{c_{i\uparrow}^{\dagger}}
\def\cidd{c_{i\downarrow}^{\dagger}}
\def\ciu{c_{i\uparrow}}
\def\cid{c_{i\downarrow}}
\def\gabpm{G^{AB}_{+- \beta\beta^{\prime}}({\bf q}, \omega)}
\def\gabmp{G^{AB}_{-+ \beta\beta^{\prime}}({\bf q}, \omega)}
\def\gbbpm{G^{BB}_{+- \beta\beta^{\prime}}({\bf q}, \omega)}
\def\gbbmp{G^{BB}_{-+ \beta\beta^{\prime}}({\bf q}, \omega)}
\def\paqp{\Pi^{A}_{+ {\bf q}}}
\def\paqm{\Pi^{A}_{- {\bf q}}}
\def\pbqp{\Pi^{B}_{+ {\bf q}}}
\def\pbqm{\Pi^{A}_{- {\bf q}}}
\def\lpm{L^{B}_{+-}}
\def\rpp{\delta_{+ \beta^{\prime}} \delta_{\beta -}}
\def\rmm{\delta_{- \beta^{\prime}} \delta_{\beta +}}
\def\opb{\omega + 2R_{B}}
\def\opa{\omega + 2R_{A}}
\def\omb{\omega - 2R_{B}}
\def\oma{\omega - 2R_{A}}
\def\k{(\bf k)}
\def\s{\sum}
\def\tc{\tilde{\chi}}
\def\bpb{BaPb$_{1-x}$Bi$_{x}$O$_{3}$}
\def\bkb{Ba$_{1-x}$K$_{x}$BiO$_{3}$}
\def\bab3{BaBiO$_{3}$}
\def\bap3{BaPbO$_{3}$}
\def\dos{density of states}
\def\br{barium}
\def\biss{bismuthates}
\def\bis{bismuthate}
\def\syms{systems}
\def\sys{system}
\def\stc{$T_{c}$}
\def\hb{Hubbard}
\def\negu{negative-$U\/$}
\def\exhb{extended-Hubbard}
\def\secor{semiconductor}
\def\secin{semiconducting}
\def\sucor{superconductor}
\def\sucin{superconducting}
\def\cdw{CDW}

\onecolumn
\begin{flushright}{\today}
\end{flushright}
\renewcommand{\thefootnote}{\fnsymbol{footnote}}
\centerline {\large \bf The Exotic Barium Bismuthates}
\centerline{A. Taraphder\footnotemark[1]}
\centerline{Laboratoire d'etudes des Proprietes Electroniques
des Solides,}
\centerline{Centre National de la Recherche Scientifique, B.P.
166, 38042 Grenoble, France,}
\centerline{and}
\centerline{Rahul Pandit\footnotemark[2], H. R.
Krishnamurthy\footnotemark[2], and T.V. Ramakrishnan\footnotemark[2]} 
\setcounter{footnote}{2}
\addtocounter{footnote}{-1}
\footnotetext[1]{Present address: Mehta Research Institute of 
Mathematics and Mathematical Physics, 10 Kasturba Gandhi Marg,
Allahabad-211002, India.} 
\footnotetext[2]{also at Jawaharlal Nehru Centre for Advanced
Scientific Research, Bangalore 560012, India.}

\centerline{Department of Physics, Indian Institute of Science,
Bangalore - 560012, India}

\centerline{\bf Abstract}

We review the remarkable properties, including
superconductivity, charge-density-wave ordering, and
metal-insulator transitions, of lead- and potassium-doped barium
bismuthate.  We discuss some of the early theoretical studies of
these systems.  Our recent theoretical work, on the
negative-$U\/$, extended-Hubbard model for these systems, is
also described.  Both the large- and intermediate-$U\/$ regimes of
this model are examined, using mean-field and random-phase
approximations, particularly with a view to fitting various
experimental properties of these bismuthates.  On the basis of
our studies, we point out possibilities for exotic physics in
these systems.  We also emphasize the different consequences of
electronic and phonon-mediated mechanisms for the negative
$U.\/$ We show that, for an electronic mechanism, the \secin
\,\,phases of these bismuthates must be unique, with their transport
properties {\it dominated by charge $\pm 2e$ Cooperon bound
states}. This can explain the observed difference between the
optical and transport gaps. We propose other experimental tests
for this novel mechanism of charge transport and comment on the
effects of disorder.


\vfill
\pagebreak

The materials \bpb, \, and \bkb, \, derived from \br\,
bismuthate,\, \bab3,\, have many remarkable properties
 \cite{AWS1,SU1,LFM1,LFM4,RJC3,RJC1,RJC2,SP1,AT1,AT2}.  These include
structural and metal-semiconductor transitions as $x\/$ and the
temperature $T\/$ are varied; diamagnetic, semiconducting (CDW
ordered?) phases even in the presence of a partially filled
Bi\,6s-O\,2p$_{\sigma}\/$ antibonding band, with optical and
transport gaps ($2\/$ eV and $0.24\/$ eV, respectively, at $x =
0\/$) differing by nearly an order of magnitude; and \sucin \,\,
\stc 's (13K and 34K, respectively, for Pb and K doping) that
are $3-5\/$ times higher than for other three-dimensional oxide
systems with similar, {\it low} densities of states at the Fermi
level. However, partly because of all the attention devoted to
the high-$T_{c}\/$ cuprates, these bismuthates have not been
studied extensively, either experimentally or theoretically. 

We review here the experimental properties of lead- and
potassium-doped barium bismuthate, give a summary of some early
theoretical studies of these systems, and then provide an
overview of our recent theoretical work \cite{AT1,AT2},
which uses the negative-$U\/$, extended-Hubbard model for these
systems.  The principal goal of our review is to consolidate 
what is known (experimentally and theoretically) about these
remarkable materials.  We have organised the review 
as follows: Section I is devoted to the phenomenology of lead-
and potassium-doped barium bismuthate.  Section II is devoted to
a survey of the models that have been put forth for these
systems and to some of the early theoretical work on them.
Sections III and IV are devoted to our work on these systems based on the
negative-$U,\/$ extended-Hubbard model: Both the large-$U\/$
(Sec.III) and intermediate-$U\/$ (Sec.IV) regimes of this
model are examined, using mean-field and random-phase
approximations, particularly with a view to fitting various
experimental properties of these bismuthates.  Section IV ends
with some concluding remarks in which we point out, on the basis
of our studies, possibilities for exotic physics in these
systems, the different consequences of electronic and
phonon-mediated mechanisms for the negative $U,\/$ and new
experimental tests for them.

\vspace{1.5cm}
\noindent{\large \bf I. Phenomenology of Barium Bismuthates}
\vspace{5mm}

This Section contains a detailed account of the normal and
superconducting properties of the barium bismuthates. We begin
with an introduction (Sec.I.1), summarize the remarkable
experimental properties of these systems (Sec.I.2), and then
discuss various experimental observations and their
interpretations (Sec.I.3). In Sec.I.4 we give a short 
account of the electronic structure and configuration of these
systems as obtained from band-structure calculations.
\vspace{7.5mm}

\noindent{\large \bf I.1  Introduction}
\vspace{5mm}

Compounds containing elements of Group VA (e.g., Sb, Bi, etc.)
have drawn considerable attention \cite{MBR} from chemists for a
long time. According to their electronic configurations
($ns^{2}p^{3}\/$ in their outer shell $n\/$; e.g., for Bi $n\/$
is $6\/$ and the electronic configuration is
[Xe]4$f^{14}5d^{10}6s^{2}6p^{3}\/$, where [Xe] corresponds to
filled shells of xenon) they should form compounds with
oxidation states $3^+\/$ or $5^+\/$, as these would yield
closed, outer-shell configurations for the corresponding
cations.  Indeed, these elements generally form compounds with
valences $3^+\/$ or $5^+\/$, {\em skipping} the intermediate
valence $4^+\/$.  Thus chemists were intrigued when they were
able to prepare compounds in which these cations have a nominal
valence $4^+\/$. However, the actual valences of these cations
(in these compounds) turned out to be controversial; e.g., the
valence of Sb in the halides of the type
A$^{1+}_{2}\/$SbX$^{1-}_{6}\/$, (X represents a halogen and A a
monovalent cation like NH$_{4}\/$): Originally 
these compounds were regarded as instances of tetravalent Sb.
Finally, careful Sb-M\"ossbauer studies  \cite{MBR} confirmed that
the actual compounds are like
A$^{1+}_{4}\/$Sb$^{3+}\/$Sb$^{5+}\/$X$^{1-}_{12}\/$, i.e., the
Sb$^{4+}\/$ ions charge disproportionate into Sb$^{3+}\/$ and
Sb$^{5+}\/$ ions.

A similar controversy has existed regarding the valence of Bi in
BaBiO$_{3}$, which was thought to be an example of a compound
with tetravalent Bi.  Here also there is some experimental
evidence (see later) that Bi charge disproportionates into
Bi$^{3+}\/$ and Bi$^{5+}\/$ configurations, thus skipping the
intermediate valence $4^+\/$.  (Since Bi is M\"ossbauer
inactive, a direct verification, as in the case of Sb Halides,
is not possible.) The renewed and vigorous interest in this
system was motivated by the discovery \cite{AWS1} that these
systems show superconductivity when doped with lead, with a
maximum transition temperature $T_{c}\/$ of 13K.  For a long
time thereafter, this was the highest superconducting $T_{c}\/$
observed amongst compounds containing no transition metal. More
recently, it has been reported
 \cite{LFM1,LFM4,RJC3,NJ,DGH2,DGH3} that, on doping with potassium
(which replaces Ba), this system shows an even higher
superconducting transition temperature ($T_{c}\simeq 30 K\/$ at
$40\%\/$ K doping).  

There are, in fact, about fifteen elements (not restricted to
Group VA) in the periodic table that show this extraordinary
property of {\em valence skipping}  \cite{MBR,CMV}. For example, 
Ga, In and Tl show valences of $1^+\/$ and $3^+\/$, Ge, Sn and
Pb of $2^+\/$ and $4^+\/$ and Bi and Sb of $3^+\/$ and $5^+\/$.
V and Nb form compounds with valence states $3^+\/$ and $5^+\/$,
Ti with $2^+\/$ and $4^+\/$. In all these cases the intermediate
valence states are either unstable or metastable.  The formal
valence state of these cations determines, to a large extent,
the shape and orientation of the other orbitals (principally of
the anion) in the compound they form; hence any theoretical
calculations and predictions of the chemical nature and physical
properties of these compounds would necessarily require this
information. In addition, the microscopic physics underlying
valence skipping is extremely important for the construction of
suitable theoretical models.

\vspace{7.5mm}

\noindent{\large \bf I.2  Summary of Important  Experimental Observations}
\vspace{5mm}

The compounds BaPb$_{1-x}$Bi$_{x}$O$_{3}$ and
Ba$_{1-x}$K$_{x}$BiO$_{3}$ and those in which Pb is replaced by
Sb  \cite{AWS1}, Ba by Sr, and K by Rb
 \cite{LFM1,LFM4,RJC3,NJ,DGH2,DGH3} have many interesting
properties  \cite{SU1,RJC1,RJC2,koyama,minami,AWS2,RS,BB3}.

\begin{description}
\item[(1)] BaBiO$_{3}$ is an insulator even though it has a
half-filled  Bi\,6s-O2p$_{\sigma}\/$ band. 
\item[(2)] The alloys exhibit a metal-semiconductor transition
as the concentration of the dopant, say Pb or K, is varied
 \cite{AWS1,SU1,LFM1,LFM4}. 
\item[(3)] However, at all dopant concentrations these systems are
diamagnetic  \cite{SU1,BB3,YJU,BB4,BB5},
so their semiconducting phases are clearly not Mott insulators. 
\item[(4)] In their semiconducting phases these alloys exhibit
two energy gaps differing by nearly an order of magnitude
($2\/$ eV and $0.24\/$ eV in BaBiO$_3$); the larger gap shows up
clearly in measurements of the optical conductivity and the
smaller one in the temperature dependence of the resistivity
 \cite{SU1,AT1,KM,blanton}. The transport gap seems to be
inaccessible to optical-reflectance, photoconductivity, or
photoacoustic spectroscopy. 
\item[(5)] The semiconducting phases of the lead-doped alloys
show many properties (e.g., two Bi-O distances) that indicate
the presence of charge-density-wave (CDW) ordering
 \cite{SU1,RJC1,RJC2,SP1,AWS2,DEC,GUK,SS2}.  
\item[(6)] These alloys undergo various structural transitions 
as functions of the concentration $x\/$ and the temperature
$T\/$; the structures exhibited by these alloys - ranging 
from monoclinic to cubic - can be obtained by slight distortions
of an underlying cubic perovskite structure \cite{SP1,DEC,SP2}. 
\item[(7)]   Both the lead- and potassium-doped systems are type-II
superconductors with $T_{c}$'s that are $3-5\/$ times higher than
those for other three-dimensional oxide systems with similar
densities of states at the Fermi level (calculated from experimentally
obtained values of $\gamma\/$, the Sommerfeld parameter
 \cite{SU1,RJC1,RJC2,BB1,BB2,BB6}). 

\end{description}

Thus these systems are quite unusual and a theoretical
understanding of their properties is a challenging task, further
complicated by the limited amount of experimental data available
and by the lack of agreement among experimentalists on many of
the observations and their interpretations.

Recently \cite{licheron} a family of Bi-based layered
superconducting compounds have been synthesized which are
isostructural with the tetragonal cuprate superconductor
La$_2$CuO$_4$. These compounds,
Ba$_{2-x}$K$_x$Pb$_{1-y}$Bi$_y$O$_{4-y}$, show an insulator to metal
transition as a function of potassium doping. The metallic
phase shows superconductivity with a highest transition
temperature of about 14K. We will not discuss these compounds in
this review.  

\vspace{7.5mm}
\noindent{\large \bf I.3  Experimental Overview}
\vspace{5mm}

This Section contains an overview of the experimental data
available for these bismuthates. We have tried, as far as
possible, to collect results which are backed by more than one
group and which have been obtained by using single-crystal samples. 
\vspace{5mm}

\noindent{\bf I.3.1 Phase Diagram}
\vspace{3mm}

The structures of  barium bismuthates can all
be obtained from slight distortions of an underlying cubic
perovskite structure shown in Fig.I.1(a). In this structure, Bi
(or Pb)\, atoms occupy the corners of a cubic unit cell, O\,
atoms the face centres, and Ba (or K\,) atoms the body centres.
There are octahedral clusters of O\, atoms around each Bi (or Pb
)\, atom. Slight rotations or breathing-mode distortions of
these octahedra lead to a variety of noncubic structures. For
example, the distortions that yield the\, pure 
BaBiO$_{3}\/$ structure are shown in Fig.I.1b.  Here there are
two distinct nearest-neighbour Bi-O\, distances -- 2.28 $\AA\/$
and 2.12 $\AA\/$: oxygen octahedra around one of the sublattices
of the underlying cubic structure contract and those around the
other sublattice expand.  This can be interpreted in terms of a
{\em freezing} of the breathing mode of the oxygen octahedra
leading to a CDW: Tilts of the oxygen octahedra further lower
the symmetry of BaBiO$_3\/$.  It is not surprising, therefore,
that both\, BaPb$_{1-x}$Bi$_{x}$O$_{3}\/$ and\,
Ba$_{1-x}$K$_{x}$BiO$_{3}\/$ show rich phase diagrams with many
structural transitions. Their phase diagrams in the temperature
$T\/$ and the concentration $x\/$ plane are shown schematically
in Figs. I.2a and I.2b, respectively \cite{SP1,AWS2,SP2}.

In addition to these transitions both these bismuthates show
metallic, superconducting, and semiconducting phases
(Figs. I.2a and I.2b), the transitions between which show up
most clearly in transport properties. The orders of these
transitions are not very clear: The metal-superconductor
transition is most probably second order. Some workers have
suggested that there is a first-order transition at finite
temperature from the metallic phase to the semiconducting
phase \cite{AWS2} (hence the hatched, two-phase region in the
phase diagram of Fig.I.2a). From the available experimental data it
is not clear whether the superconducting and semiconducting
phases coexist along any phase boundary and, if so, what the
order of the transition along this boundary is \cite{SU1,AT1}.
However, the maximum value of the superconducting $T_c\/$ seems
to occur at nearly the same value of $x\/$ as the
metal-semiconductor transition.  

 Electron diffraction study of the structural transitions 
in BaPb$_{1-x}$Bi$_x$O$_3$ has been performed on powder samples 
by Koyama and Ishimara \cite{koyama} and by Minami \cite{minami} 
recently. The phase diagram as a function of doping $x$ that 
Koyama and Isimara get differs from the one reported by
Cox and Sleight \cite{DEC} earlier. These authors do not get any orthorhombic
phase in their electron diffrcation studies. They find evidence for  
tetragonal and monoclinic phases at low temperature and a
high temperature ($\simeq$500K) cubic phase (the transition
occuring at $x=0.4$). They have also investigated the
existence of any incommensurate phase and concluded in the
negative, in agreement with Cox and Sleight.

In early studies of the lead-doped compound \cite{AWS1} it was
suggested that superconductivity could only occur in the
tetragonal phase; however, subsequent work seems to suggest that
it can also be obtained in specimens with orthorhombic
symmetry \cite{SU1}. In the potassium-doped compound,
superconductivity has been observed only in the cubic
phase \cite{SP2}. In both the potassium- and lead-doped systems,
the superconducting phase appears right after the
semiconductor-metal transition, with increasing lead or
potassium concentration.  

\vspace{5mm}

\noindent{\bf I.3.2 Normal-State Properties}
\vspace{3mm}

Much more experimental data are available for
BaPb$_{1-x}$Bi$_{x}$O$_{3}\/$
 \cite{SU1,RJC1,RJC2,SS2,BB1,BB2,SK1,SK2,KK2,KK1,TTh,TT,JEG,MO,TI,ST1}
than for Ba$_{1-x}$K$_{x}$BiO$_{3}\/$.  Some properties of the
metallic phase ($ 0 < x < 0.35\/$ ) of\,
BaPb$_{1-x}$Bi$_{x}$O$_{3}\/$ are given in Table I.1:\,
Resistivities are quite high, nearly independent of
temperature, and increase from $ 210 \mu \Omega cm\/$ ( at $ x =
0\/$) to $540 \mu \Omega cm\/$ ( at $ x = 0.24 \/$).  Carrier
concentrations, obtained from measurements of the Hall
coefficient  \cite{TTh,TT} range from $ 1.42 \times 10^{20}
cm^{-3}\/$ ( \,at $ x=0\/$) to $2 \times 10^{21} cm^{-3}\/$ ( at
$ x= 0.2 \/$ ); the effective mass at\, $ x = 0.2 \/$ is $ 0.3
m_{e}\/$ (obtained from plasma-edge data  \cite{KK1,ST1}). The
specific heat $C\/$ depends linearly on $T\/$ at low temperature, but
the co\-efficient of proportionality is rather low \cite{BB5,TT}
($ \simeq 0.6\/$ mJ/mole K$^{2}\/$ at $x=0 \/$ ). However, plots
of $ C/T \/$ vs $ T^{2} \/$ show some deviation from linearity
(requiring fits of the form  \cite{TT} $C=\gamma T +\beta
T^{3}+\alpha T^{5} \/$,\, with $ \gamma = 0.6\/$ mJ/mole 
K$^{2}\/$, \,$ \beta = 0.42\/$ mJ/mole K$^{4} \/$ and $\alpha =
1.1\/$ mJ/mole K$^{6} \/$). In the range $ 0 < x < 0.2 \/$
optical-reflectivity data can be fit to a Drude form as expected
in a conven\-tional metal. Plasma frequencies, as obtained from
optical experiments \cite{ST1}, increase with increasing $ x \/$
(i.e.,\, with increasing Bi concentration) in the above range;
this is consistent with the increase in carrier concen\-tration
with increasing $ x \/$ (Table I.1), as obtained from
Hall-effect measure\-ments.  LAPW,\, band-structure calculations
 \cite{LFM1,LFM2,LFM3} are fairly successful in accounting for
the observed properties of the metallic phase in the range $ 0 <
x < 0.2\/$ (Sec.I.4); and the most recent band-structure
calculations \cite{liech} have also obtained reasonable results
for the insulating phase. 

As the metal-semiconductor transition is approached by 
increasing $x\/$, the metallic phase starts showing unusual
transport properties  \cite{SU1,KK1,TT,ST3} (see Sec.I.3.6): 
the temperature coefficient of resistivity
becomes negative around $x = 0.3\/$ (Fig.I.3a) even though,
at low-enough temperatures, the system becomes superconducting
for $x < 0.35\/$, and only beyond $x = 0.35\/$ is true
semiconducting behaviour (i.e., an activated behaviour over a
large temperature range with the low-temperature resistivity
showing no sign of a superconducting transition) observed down
to the lowest temperatures measured (Fig.I.3a and I.3b).
Reflectivity data show a gradual departure from Drude fits
 \cite{ST3} in the metallic phase from $x = 0.25\/$ onwards as
the metal-semiconductor boundary is approached (Fig.I.4).

Various transport and optical properties in the semiconducting
phase of BaPb$_{1-x}$Bi$_{x}$O$_{3}\/$ have been measured, but
there is a dearth of such systematic data for the\,
Ba$_{1-x}$K$_{x}$BiO$_{3}\/$ system, though some measurements
have started coming out \cite{blanton,sato}. For
BaPb$_{1-x}$Bi$_{x}$O$_{3}\/$ the semiconducting phase extends
over a wide range: $0.35< x< 1\/$. The temperature dependence of
the resistivity in this phase yields a transport-activation
energy $E_{A}\/$ which goes from $ 0$ at $x=0.35\/$, to $ 0.2\/$
eV at\, $x = 0.6\/$, and finally to $0.24\/$ eV at $x=1\/$
(Fig.I.5).  By combining the resistivity and Hall-coefficient
data, the temperature dependence of the carrier concentration
in\, BaBiO$_{3}\/$ (i.e.,\,$ x = 1\/$) is found to be $
n(T)=1.1\times 10^{22}(cm^{-3}) exp(-0.24 eV/k_{B}T)\/$. The
pre-exponential factor is remarkably large \cite{SU1} and nearly
equal to the number of unit cells per $cm^{3}\/$, suggesting
that {\it the transport mechanism in this system is most probably 
intrinsic} (and not extrinsic, i.e., not because of dopants,
impurities, or defects).

Recently some measurements of the resistivity of 
single crystals of Ba$_{1-x}$Ki$_{x}$BiO$_{3}\/$ have been
carried out in the semiconducting phase \cite{hell1,fill1} for 
different values of $x\/$.  Hellman, et al. \cite{hell1} 
fit their resistivity data for single-crystal and thin-film
samples of Ba$_{1-x}$Ki$_{x}$BiO$_{3}\/$ to a variable-range
hopping law $\rho=\rho_{0}exp[(T_{0}/T)^{1/4}]\/$, with a very
high $T_{0}$ ($ \simeq 3-6\times 10^{8}$K). According to them
the states are very strongly localized and there is no
indication of any nonlinear transport in these systems, so
incommensurate CDW motion is possibly ruled out.  Resistivity
data \cite{dabr1} for superconducting
Ba$_{1-x}$Ki$_{x}$BiO$_{3}\/$ ceramics are also becoming
available. These measurements seem to indicate \cite{dabr1} that
there is a variable-range-hopping regime with activation energy
$T_{0} \simeq 10^{4}-10^{6}$ in the normal state. It is now
being argued by several groups \cite{hell1,fill1} that, even
in single-crystal samples of Ba$_{1-x}$Ki$_{x}$BiO$_{3}\/$, 
two or more phases with different chemical compositions are
present. (If so, then presumably one phase is metallic and the
other semiconducting.) 

Sato et al., \cite{sato} have carried out a series of
transport measurements on Ba$_{1-x}$Ki$_{x}$BiO$_{3}\/$
thin films at $x=0.45$ and found no evidence of
contributions from impurities or 
inhomogeneous distribution of potassium concentration in their
annealed samples. The resistivity data for the best samples have
a positive (but small) slope. Metallic temperature dependence is
observed down from T$_C$ upto 300K and the resisitivity at 300K is
lower than 500$\mu\Omega$cm. On the other hand negative
temperature coeffiecient of resistivity was observed in samples
that were grown at higher annealing temperature and these
samples have resistivities much higher than 700$\mu\Omega$cm. 

The Hall coefficient measurement \cite{sato} on
Ba$_{1-x}$Ki$_{x}$BiO$_{3}\/$ thin films indicates that the
{\em charge carriers are electrons}. This is in contradiction to the
expectation that the carriers should be holes as
monovalent potassium replaces divalent barium in
Ba$_{1-x}$Ki$_{x}$BiO$_{3}\/$. The value of Hall coeffiecient
R$_H$,  $(-) 3\times 10^4$ cm$^3$/C is about two times smaller than that
expected from band structure calculations. The temperature
dependence of R$_H$ is quite small. These workers have also done
Raman measurements and found that the height of the strong peak
at 570 cm$^{-1}$ that one observes at $x=0$, is reduced and it
broadens with potassium doping. This peak 
disappears close to the metal-insulator transition. The
reduction of the peak as well as its broadening with doping is
considered by these workers to be an indication that the remnant
CDW like ordering that one finds in the metallic phase of the lead doped
system \cite{ST3,ST2} (close to the metal-insulator transition)
is absent in the potassium doped system.  

Blanton, et al. \cite{blanton} have measured the reflectivity
spectrum of polycrystalline samples of
Ba$_{1-x}$Ki$_{x}$BiO$_{3}\/$ as a function of doping $x$ from
$x=0$\, to\, $x=0.4$ in the frequency range from $\simeq 250 $
\, to \, $25000$ cm$^{-1}$. Their data on optical conductivity
(obtained from the reflectivity data) indicate that the CDW
conductivity peak at 16000 cm$^{-1}$ at $x=0$ shifts downwards
in energy and broadens substancially with doping. The spectrum at $x=0.4$
shows a broad contribution to the conductivity that is centred
around 4000 cm$^{-1}$. Blanton, et al. claim that this broad
peak, has evolved from excitations across the single-particle
(CDW) gap of BaBiO$_3$ and is an indication that some remnant
CDW order persists even in the metallic phase close to the
transition as in the case of lead-doped barium bismuth
oxide \cite{ST3,ST2}.  Bozovic, et al. \cite{bozovic} have measured
the far-infrared reflectance spectra in thin film samples of the
superconducting Ba$_{1-x}$Ki$_{x}$BiO$_{3}\/$ at $x=0.4$. They
point out some overall similarity between the reflectance
spectra of the cuprate superconductors and the bismuthates at
optimal doping (corresponding to the highest T$_c$). All these
spectra are nearly linear in frequency $R(\omega)\simeq
A-B\omega$ (where $A,\,\,B$ are constants, $A \approx 1,
\,\, B/\hbar \approx 1$ eV$^{-1}$) below the plasma edge.



\vspace{5mm}

\noindent{\bf I.3.3 Charge Disproportionation and CDW ordering}
\vspace{3mm}

The actual valence state of Bi in the barium bismuthates is a
matter of some dispute (Sec.I.1). Evidence in favour or against
valence skipping by Bi in these systems is indirect. In
particular, there has been some controversy in the experimental
literature concerning the presence of a CDW in the
semiconducting phase of these compounds
 \cite{SU1,RJC1,RJC2,GUK,SS2}. The presence of the two valence
states, Bi$^{3+}$ and Bi$^{5+}$, is intimately connected to this
question.  Many consider that the controversy has been settled
more or less in favour of a CDW in the lead-doped system, but
the results for the potassium-doped system are not clear yet. It
is important to note that, to support a CDW in these systems,
the charge contrast need not necessarily be $2e\/$ 
 \cite{RJC1,RJC2}; indeed it is almost certainly less than
$2e\/$, as suggested by recent, high-energy-spectroscopy
experiments \cite{GUK}. The CDW controversy in these systems is
complicated further by the methods used for sample preparation
 \cite{RJC1,RJC2}: oxygen vacancies appear quite frequently in 
samples prepared in low-oxygen atmospheres, leading to
inhomogeneities that make it difficult to identify long-range
CDW ordering experimentally. Even in the lead-doped system, it
is still not clear how the CDW order disappears with increasing
inhomogeneity (or whether there is short-range CDW order away
from perfect stoichiometry). Optical-reflectivity spectra 
show that the CDW is quite robust (see below)
and persists throughout the semiconducting regime (upto a lead
concentration of $65~\%\/$).

Neutron-diffraction (powder) studies  \cite{SP1,SP2} of
BaK$_{1-x}$Bi$_{x}$O$_{3}\/$ show that the {\em frozen\/},
breathing-mode distortion of the oxygen octahedra leads to
long-range CDW order in some of the semiconducting phases of\,
BaK$_{1-x}$Bi$_{x}$O$_{3},\/$ but not in the semiconducting
phase abutting the superconducting phase; so the cause of the
semiconducting nature of this phase remains a
mystery \cite{SP1,AT2,SP2}.

\vspace{5mm}

\noindent{\bf I.3.4  Optical Properties}
\vspace{3mm}

A number of optical-reflectivity measurements
 \cite{SU1,SS2,ST1,ST3} have been performed on the lead-doped
system, both in the metallic and semiconducting phases, covering
a wide range of frequencies, from far-infrared to visible.  The
metallic regime shows a characteristic Drude behaviour with a
plasma edge, which shifts continuously towards high frequencies
with increasing $x\/$.  This edge is seen in the entire
composition range without an abrupt change anywhere. Such a
continuous variation of the reflectivity edge with $x\/$ is
indicative of the homogeneity of the system upto optical length
scales.  In the semiconducting phase the reflectivity spectra
are dominated by low-frequency optical phonons. No discontinuous
change is observed anywhere in the spectra in the
metal-semiconductor transition regime.

In optical-conductivity spectra, obtained from reflectivity
data, deviations from Drude behaviour occur as the transition is
approached from the metallic side. This transition is marked by
the transfer of the spectral weight into a high-frequency, {\em
quasi}\, peak  \cite{SU1,SS2,SU2,ST2}, which grows with
increasing $x\/$ into a peak containing most of the 
spectral weight in the semiconducting phase. The position of
this peak shifts towards higher frequencies with increasing
$x\/$ as shown in Fig.I.6a. (Fig.I.5 shows this peak position as
a function of $x\/$). It has been suggested  \cite{SU1} that this
peak (Fig.I.6a,b) has a profile which is like that of a CDW\,
system with a gap along 
the entire Fermi surface  (as in K$_{0.3}$MoO$_{3}$)\, and not
like the square-root absorption edge in conventional
semiconductors.  However, more recent work shows a broader peak
 \cite{blanton,hellm}. The {\it magnitude of the gap $E_{G}\/$, measured
by the position of this peak in the optical-conductivity
spectrum, is an order of magnitude higher than $E_{A}\/$, the
transport activation energy} \cite{SU1,AT1,AT2}, e.g., $E_{G} =
2.0\/$ eV and $E_{A} = 0.24\/$ eV at $x = 1\/$, whereas in a
conventional semiconductor $E_{A}=E_{G}/2$. The optical gap
decreases with doping (Figs. I.5 and I.6) and, at the
semiconductor-metal transition, it is quite small but {\em
nonzero}. (The transport activation energy vanishes at the
transition.) On the metallic side the plasmons show up strongly
in the reflectivity spectra and mask the details of the optical
gap. Throughout the semiconducting phase the optical gap is
larger than twice the transport gap.  The transport activation
energy or gap $E_{A}\/$ does not show up clearly in any optical,
photoconductivity, or photoacoustic studies.

\begin{sloppypar} 

Raman experiments  \cite{SS2,SS1} on BaBiO$_{3}\/$ show resonant
enhancement and the appearence of higher harmonics of the $569 cm^{-1}\/$
mode phonon (the breathing mode described above), indicating that the
optical gap $E_{G}\/$  is related to the strong, electron-phonon
interaction in the system  \cite{SS2,BB1,BB2,CKL,DGH1,JOS,jin,MS}.  The
resonant phonon energy for the potassium-doped system shows very similar
behaviour and, as in the lead-doped case, the resonance decreases with
increasing $x\/$. So far no one has reported any observation of two
distinct peaks in the Raman intensity which could be identified with the
two valence states of Bi directly.  If there were a strong charge
disproportionation  in BaBiO$_{3}$, one would have observed a split peak
with equal intensities.  It is possible that the laser power used in these
experiments creates oxygen vacancies and destroys the charge
disproportionation locally. We have not come across any discussion of
these issues in the experimental literature.

\end{sloppypar} 
\vspace{5mm}
\noindent{\bf I.3.5\, Magnetic Properties}
\vspace{3mm}

	Both the lead- and  potassium-doped systems are diamagnetic in
their normal phases over the entire temperature range of
measurement and for all values of doping. The diamagnetic
contribution from the core, which 
constitutes almost the entire susceptibility in both the systems, when
subtracted from the total susceptibility, leaves a tiny paramagnetic part
which, in the lead-doped system, shows a somewhat unusual behaviour: it
increases \cite{SU1} as the system becomes more insulating, over
the entire range of $x\/$ (Fig.I.7a).  For the potassium-doped
system the paramagnetic contribution remains nearly independent of
$x\/$ \cite{SU1,RJC1,RJC2,BB3,YJU,BB4,BB5}. The paramagnetic
contribution from the valence electrons is very small and lies
between $\approx$ 0.8 to 2 $\times 10^{-5}$ emu/mol in barium
potassium bismuth oxide. The temperature dependence of magnetic
susceptibility is shown in Fig.I.7b.

\vspace{5mm}

\noindent{\bf I.3.6 Metal-Semiconductor Transition}
\vspace{3mm}

The insulating nature of BaBiO$_{3}$ and the metal-semiconductor
transition that occurs on doping with Pb or K cannot be understood on the
basis of band-structure calculations  \cite{LFM1,LFM2,liech}: they yield a
half-filled conduction band for BaBiO$_{3}$. The simplest explanation for
the insulating nature of BaBiO$_{3}$ invokes a CDW instability which opens
a gap in the conduction band (see below).  Both Pb and K doping should
inhibit CDW ordering for two reasons: first, such doping moves the Fermi
level down from its position at half filling (K doping removes electrons
from the system; for Pb doping this follows from band-structure
calculations); and it also introduces disorder.

We now summarize the experimental findings close to the
metal-semiconductor transition in the \,BaPb$_{1-x}$Bi$_{x}$O$_{3}$
system. There is a strong precursor effect  \cite{YE} of the
metal-semiconductor transition as the Bi concentration is increased beyond
$x=0.2$. The resistivity data are shown in Figs. I.3a and I.3b.
The temperature coefficient of resistivity becomes negative for 
$x\stackrel{<}{\sim}0.2\/$.  This would normally indicate that
the system is already insulating. However, at low temperatures, the
resistivity drops to zero  upto  $x=0.35$ (Fig.I.3a), indicating  a
superconducting transition. Although the resistivity (for $0.25
\stackrel{<}{\sim} x \stackrel{<}{\sim}0.35$)  decreases with
increasing temperature, it cannot be fitted to an activated  behaviour
over a wide temperature range. Such a fit is possible only beyond
$x=0.35$ (Fig.I.3b) where, down to the lowest temperatures measured,
no superconducting transition is observed. Hence the criteria used by
experimentalists  \cite{SU1} for the metal-semiconductor
transition in these systems are the absence of superconductivity at low
temperatures and an activated behaviour of the resistivity over wide
temperature ranges. There is then really no clear separation between the
insulator and metal in the transition region \cite{la} and we might have a
transition {\em from a (Anderson-localized {\rm ?}) semiconductor to a
superconductor}. 

Reflectivity spectra (Fig.I.4) show a marked departure from Drude
behaviour from $x=0.2\/$ onwards. The temperature dependence of the
resistivity is unusual: From $x=0.2$ the resistivity becomes temperature
independent and around $x=0.25$ its temperature coefficient changes sign.
The system is, however,  metallic at that concentration (the resistivity
does not show an activated behaviour). The thermopower varies linearly
with temperature for $x < 0.2$; for $x > 0.2$ it no longer follows this
metallic behaviour.

Reflectivity spectra do not show any abrupt change across the
metal-semiconductor transition in the high-frequency region. In the
semiconducting phase there are several optic phonons in the low-frequency
region. As the metal-semiconductor transition is approached, the
low-frequency conductivity is reduced and the spectral weight is
transferred to the high-frequency quasi peak as discussed
earlier. The quasi peak has a tail extending into the
low-frequency regime indicating the 
possibility of states in the semiconducting energy gap. This tail
disappears completely beyond $x=0.7$.

\vspace{5mm}
\noindent{\bf I.3.7 Superconducting Properties}
\vspace{3mm}

The superconducting phase occurs
 \cite{AWS1,SU1,LFM4,RJC3,RJC1,RJC2} in the concentration range
$0.05< x <0.30 $ for\, BaPb$_{1-x}$Bi$_{x}$O$_{3}$ and $0.37 < x
< 0.60$ for Ba$_{1-x}$K$_{x}$BiO$_{3}$. The highest $T_{c}$ \,
is 13K at $ x=0.25$ for the former and 34 K at $x=0.4$ for the
latter .  The densities of states at the Fermi level for both
these bismuthate systems are quite low \cite{DGH3,BB5,TT}. Given
such meagre densities of states at the Fermi levels and the
three-dimensional structures of these bismuthates, the
superconducting transition temperatures are very high indeed.
The specific-heat anomaly associated with the superconducting
transition is very small \cite{BB1,BB2,TT}.
Tunneling \cite{BB5,JMV,ZS} and infrared reflectivity
measurements \cite{JMV,ZS} yield 
$2\Delta_{s}/k_{B}T_{c} \simeq 3.5$, the BCS\, value, where
$\Delta_{s}$ is the superconducting energy gap (measured from
the chemical potential).  For both these bismuthates there is a
substantial isotope effect with\, $T_{c} \propto M^{-\alpha}
\/$, and\,  $ \alpha = 0.2-0.3$  \cite{BB5} (A
higher value of nearly 0.4 has been reported by Hinks, et
al. \cite{DGH2,DGH3}, for the potassium-doped system.).
The superconducting properties of these systems are summarised in
Table I.2.  Note especially the coherence lengths ( $\xi_{GL}$ )
which are nearly $70-80 \AA\/$, i.e., about $30-40\/$ lattice
spacings \cite{CMV,BB5}.

Sato et al. \cite{sato} have measured the superconducting
properties of Ba$_{1-x}$Ki$_{x}$BiO$_{3}\/$ (at $x=0.45$) thin
film annealed samples. Infrared reflectivity spectrum in the
superconducting state at 5K (normalized by its value in the normal
state at 30K) shows a clear peak at 60 cm$^{-1}$ and a broad dip
at 135 cm$^{-1}$. The same features were earlier obtained by
Schlesinger et al. \cite{ZS} in high density cintered samples with
a transition width of 4K. The superconducting energy gap that
Sato et al. obtain is $3.7\pm 0.3$ meV, in good agreement with
Schlesinger et al. With the superconducting transition
temperature of their sample being 22.5 K, the value of
$2\Delta/k_{B}T_{c}$ is about 3.8. Similar values have also been
reported in the optical measurement \cite{ZS}, point contact
tunnelling measurement \cite{huang} on Ba$_{1-x}$Ki$_{x}$BiO$_{3}\/$
sintered samples at $x=0.4$ and tunnel junctions fabricated on
thin films grown by molecular-beam-epitaxy \cite{hellman} and laser
deposition method \cite{moon}. The conductance spectra for
several tunnel junctions have been measured \cite{sato,sharifi} from
10 to 100 $mV$ bias voltage. The conductance spectrum in the
region greater than 25 $mV$ has no observable phonon
structure \cite{sato,sharifi}. This result seems to suggest that
there is no strong electron-phonon coupling as has been assumed
by many workers \cite{jin}. 

The upper critical field in the barium potassium bismuthates
has been measured in thin films of polycrystalline samples
\cite{sato} and in sintered 
samples \cite{welp,kwok}. Unlike in the case of sintered samples,
the thin films show a reasonably sharp resistive transition in a
magnetic field. For the best thin film samples \cite{sato}
H$_{c2}$ shows almost linear temperature dependence down to 4.2K
and seems to saturate aound 1.7K.

The upper
critical field (H$_{c2}$) measurements in polycrystalline
samples of Ba$_{1-x}$K$_{x}$BiO$_{3}$ by Crabtree \cite{CT}, and
Filippini et al., \cite{CT}  show
that there is a significant upward curvature at fields beyond 10 Tesla.
There is also noticeable flattening of the H$_{c2}$ curve close
to the superconducting transition temperature. 
The value of the temperature
coeffiecient of H$_{c2}$ at the transition temperature
($dH_{c2}/dT $ at $T_c$) for Ba$_{1-x}$K$_{x}$BiO$_{3}$
presented in Table I.2 ($\simeq -0.5\,\, T/K$) is for  
polycrystalline samples (from Batlogg, et al., \cite{BB5}).
Measurements on thin films \cite{sato} indicate a much
larger value of nearly -1.0 T/K (at $x=0.45$).

Hysteretic Josephson junction
behavior has been observed in 
the $I-V$ characteristics of multidomain films of
Ba$_{1-x}$Ki$_{x}$BiO$_{3}\/$ at temperatures upto
8K by Martin et al., \cite{martin}. They fit this behaviour 
to a model of Josephson junctions in series coupled by a shunt
capacitor.

\noindent{\bf I.3.8  The Effect of Oxygen-Deficiency on the Properties of
BaPb$_{1-x}$Bi$_{x}$O$_{3}$ }
\vspace{5mm}

The effect of oxygen deficiency on the structural and transport
properties of the lead-doped system has been studied by a few
groups \cite{YE,MSu,CC1,CC2,mosley}.  Oxygen vacancies are easily
created by sputtering 
in a thin film of a single-crystalline BaPb$_{1-x}$BiO$_{3}$
sample. The principal effect of an increase in oxygen vacancies
is to increase the resistivity in the metallic regime and
decrease the superconducting $T_{c}$.  Furthermore the
resistivity of an oxygen-deficient metallic sample of
BaPb$_{1-x}$Bi$_{x}$O$_{3}$ (at $x=0.3$) {\em goes down } with
increasing temperature and the sharpness of this fall in
resistivity increases with increasing oxygen deficiency
(Fig.I.8a). Thus, at a fixed temperature, as the oxygen
deficiency increases, so does the resistivity.  However, the
``carrier density'', as inferred naively from the Hall
coefficient (for such a strongly disordered system), {\em
increases with oxygen deficiency} (Fig.I.8b).  The increase in
carrier density with oxygen deficiency is thus not reflected in
the resistivity results. Hence this suggests,  \cite{YE} that
there is a strong carrier localization effect because of the
defects created by the missing oxygen atoms, which more than
offsets the increase in ``carrier density''.

It is surprising that the carrier density should {\em increase} with
oxygen deficiency. For, as more and more oxygen is removed from
the system,  so are electrons, and
one would expect a {\em decrease} in the carrier (electron) density  (the
Hall coefficient is negative). At very low temperatures (Fig.I.8b) it
does seem that the more oxygen deficent  samples have lower  carrier
densities; however, this trend is reversed at high temperatures.  Enomoto,
et al. \cite{YE} argue that, at high temperatures, the
carriers that were localized by the oxygen deficiency are delocalized (the
activation energy for localization that fits their assumption is quite
small $\simeq 18$meV) and give rise to the enhanced carrier density; but
the resistivity data, (Fig.I.8b) do not reflect this enhanced carrier
density!  This clearly calls into question the naive interpretation of the
Hall-effect data.

Another interesting observation  \cite{YE} is that annealing
a sample of the lead-doped system with $x= 0.35\/$ at $873\/$ K and
$1400\/$ atm. (in an atmosphere of $80 \% \/$ Ar$_{2}$ and $20 \%
\/$ O$_{2}$) for $8\/$ hours changes it from a semiconductor to
a superconductor. This high-pressure annealing decreases the
resistivity markedly and, at $6\/$K, there is a sharp
superconducting transition with a width less than 1K.  However,
if the sample is annealed at a pressure of $1\/$ atm. (in an
oxygen atmosphere) and at 600 K, it shows no superconductivity.
This indicates that a metal-semiconductor transition can be
induced in the lead-doped system by increasing the number of O
vacancies. The sharpness of the superconducting transition is
also greatly enhanced as the O deficiency decreases.
A positron - annihilation study \cite{mosley} on single-crystal
samples of Ba$_{1-x}$K$_x$BiO$_3$ in the range  15K
to 300K shows that the sharpness of the superconducting
transition and the value of the transition temperature depends
strongly on oxygen vacancies, increasing as one approaches 
stoichiometry.  There is no signature of positron localization
and the positron lifetime is determined by the oxygen
stoichiometry. 

A naive interpretation of these results  \cite{YE} is most
probably fallacious.  Firstly, there should be strong {\em
Anderson-localization effects} in the electronic band, for,
apart from the O vacancies, the concentration of Bi is about $30
\%\/$.  Secondly, the
experimental system (at $x=0.3\/$) is quite close to its
metal-semiconductor transition and, as we saw earlier, this
region is quite unusual. In particular, it is not clear whether
the samples used in these experiments were metallic or
semiconducting. Thus a more careful analysis of these results
along with more controlled measurements are clearly necessary to
understand the unusual observations reported here.

\vspace{7.5mm}
\noindent{\large \bf I.4  The Electronic Structure of Barium Bismuthates}
\vspace{5mm}

The electronic structure of the barium bismuthates has been
studied well using detailed band-structure calculations
 \cite{LFM1,LFM2,liech} 
as well as various experimental methods  \cite{SS2}.  The
electronic configuration of a Bi atom is
[Xe]4$f^{14}5d^{10}6s^{2}6p^{3}$ and that of a Pb atom is
[Xe]4$f^{14} 5d^{10}6s^{2}6p^{2}$.  Hence in the $4^+\/$ valence
state (which is the nominal valence state of Bi and Pb in these
bismuthates) Bi$^{4+}$ has an open-shell configuration 6s$^{1}$,
whereas Pb$^{4+}$ has a closed outer-shell configuration.

In both the lead- and potassium-doped systems, each Bi atom is
surrounded by six O atoms which form an octahedron (Fig.I.1a).
There are ten orbitals in the ``valence band'' of the
bismuthates  \cite{LFM2,LFM3} given the BiO$_{3}\/$ unit cell,
each O atom contributes three 2p orbitals, and each Bi atom
contributes one 6s orbital (Fig.I.10).  The Bi-O bond lengths
are the same for all the six O atoms in the cubic phase.  We
first examine the electronic structure of the cubic phase and
later discuss the effects of the noncubic distortions.

\vspace{5mm}
\noindent{\bf I.4.1 The Electronic Structure of
BaPb$_{1-x}$Bi$_{x}$O$_{3}\/$} 
\vspace{3mm}

The electronic band structure of the bismuthates (Figs.I.11a -
I.11c) has been computed by using the LAPW method
 \cite{LFM2,LFM3}.  The
flat, low-lying bands are derived from Ba 5p states and the
uppermost complex of unoccupied states are derived from Ba 5d
and Bi 6p states. Note that the bands derived from Barium are
admixed only marginally with the valence-band manifold in the
middle.

Of the three 2p orbitals at each O atom, one forms a $\sigma\/$
bond (Fig.I.10) with the Bi/Pb 6s orbital (spherically symmetric) and the
rest form $\pi\/$ bonds.  The $\pi-\pi^{*}\/$ manifold falls within the
$\sigma-\sigma^{*}\/$ manifold (Figs.I.11 and I.12). The Fermi
level crosses 
the uppermost  $\sigma\/$ antibonding subband (Fig.I.11), leaving it
exactly half filled for BaBiO$_{3}\/$ (Fig.I.11c).  As Bi is
replaced by Pb, the filling of this $\sigma^{*}\/$ subband
changes (Fig.I.11a-b). In addition, LAPW 
band-structure calculations show that Pb doping also introduces a
noticeable chemical shift in the position of the 6s states.  In all cases
the Ba bands are well away from $E_{F}\/$. The principal features of these
bands can be understood in terms of a tight-binding model
 \cite{LFM2,LFM3}. Figure I.13a shows such a fit with a three-parameter,
tight-binding calculation.  The three parameters used are the orbital
energies \mbox{$\epsilon_{6s} = -4.1\/$ eV,} $\epsilon_{2p}=-1.9\/$ eV and
the nearest-neighbour, Bi-O hopping parameter (along the $\sigma\/$ bond)
$t_{sp \sigma}=2.2\/$ eV.  As there is no overlap in this model between
Bi\,6s and O\,2p$_{\perp}$, the entire $\pi-\pi^{*}\/$ manifold now
shrinks to a set of perfectly flat and degenerate  nonbonding bands.  A
five-parameter fit improves upon this, but the three-parameter fit clearly
shows the tight-binding, $\sigma$ character of the important bands. We
can, therefore, use this set of parameters to study the low-energy physics
of these systems.  This three-parameter, tight-binding model also predicts
an electron-like, $\Gamma$-centered, simple-cubic Fermi surface
for BaBiO$_{3}$.  The electron surface nests perfectly with
an identical hole surface centered at the zone corner (R) (Fig.I.13b).

Such a nested Fermi surface is susceptible to an electronic
instability that, in the bismuthates, leads to a Peierls distortion, in
which the O atoms surrounding every alternate Bi site are displaced
towards them. This is the breathing-mode displacement of the oxygen
octahedra, which doubles the unit cell and thereby opens up  a gap along
the entire, noninteracting Fermi surface, splitting the $\sigma^{*}$
subband into a filled and an unfilled band.  The effect of this Peierls
distortion is shown, in a folded-zone scheme in Fig.I.14 for only the
conduction-band states near $E_{F}\/$, using the three-parameter, 
tight-binding model.  This gap is estimated to be nearly 1eV in a cubic
BaBiO$_{3}$ structure, with alternate Bi-O bond lengths
differing by $0.16 \AA\/$  \cite{LFM2}. This lowering of the
energies of the states near $E_{F}\/$ comes entirely from the
Bi-O bond alteration and hence stabilizes such a distortion.
From the point of view of band-structure calculations, the role
of doping is merely to inhibit the frozen-in commensurate
distortion (since the Fermi level moves away from nesting),
thereby producing a metal in which the conduction electrons are
strongly coupled to the dynamic oxygen, breathing-mode phonons.

Band-structure calculations work fairly well in the metallic
regime of BaPb$_{1-x}$Bi$_{x}$O$_{3}$ \mbox{($0<x<0.2\/$)}. The
band crossing the Fermi level is derived from Bi/Pb\,6s-O\,2p
levels.  Since Bi atoms supply one extra electron per atom to
the conduction band, an increase in the Bi concentration pushes
up the Fermi level and increases the number of carriers in the
conduction band.  The plasma edge shifts towards higher
frequencies in this range exactly as predicted by the
band-structure calculations.  The observed low density of states
(DOS) at $E_{F}\/$ is also brought out by calculations as the
s-p character of the conduction band implies a low DOS at
$E_{F}\/$ (see below).  The Fermi surface for $x < 0.2$ is
nearly spherical as predicted by LAPW calculations.  However,
the band-structure results deviate from experimental
observations from $ x > 0.2$ onwards and cannot produce the
semiconducting phase for $x > 0.35$.  

\vspace{5mm}

\noindent{\bf I.4.2 The Electronic Structure of
Ba$_{1-x}$K$_{x}$BiO$_{3}$} 
\vspace{3mm}

Band-structure (LAPW) results  \cite{LFM1} are available for this
system at various values of $x\/$.  Figure I.15a shows these for
BaBiO$_{3}$ using the observed lattice parameters of
Ba$_{0.6}$K$_{0.4}$BiO$_{3}$.  The results are in close
agreement with the ones obtained for 
the lead-doped system.  The Fermi level crosses the uppermost of the
ten-band, Bi\,6s-O\,2p complex.  The flat, low-lying bands are Ba 5p
states, whereas the uppermost complex of unoccupied states is derived from
Ba 5d and Bi 6p states.The manifold of Bi\,6s-O\,2p bands features a pair
of broad ($\simeq$ 16eV) $\sigma$ and $\sigma^{*}$ subbands extending from
-12 eV to +4 eV at the zone corner.  The $\sigma^{*}$ band is half filled;
so, as a K$^{1+}$ ion replaces a Ba$^{2+}$ ion and removes an electron
from the system, the Fermi  level starts coming down the $\sigma^{*}$
band, and the occupancy of this band is reduced to $1-x\/$.  Here also a
three-parameter, tight-binding fit (similar to the Pb-doped case) works
quite well.

Recently Liechtenstein, et al. \cite{liech}, have
carried out what is perhaps the best band-structure study of
\bkb. They use the LDA and the full linear muffin-tin-orbital
method.  For $x = 0\/$ they find a lattice instability with a
combination of both tilt ($u_t\/$) and breathing ($u_b\/$)
distortions. This yields the experimentally
observed monoclinic structure.  For $x = 0.5\/$ they correctly
find the cubic phase, with $u_t = u_b = 0\/$, to be stable.

An interesting point that emerges from this 
band-structure study is that, when the details are put in,
\bab3 may have a rather small indirect band gap (Fig.I.15b). The 
actual value of this gap, gleaned from Fig.I.15b, is roughly
$0.025\/$ Ry or $0.34\/$ eV; though this may not be very reliable.
Hence one must consider 
the possibility that the small transport gap ($0.24\/$ eV)
reflects this aspect of the band structure.  However, optical
experiments must be able to see this as a phonon-assisted
threshold, starting around $0.68\/$ eV, in order to be 
consistent with the transport gap. All the data 
that we have seen have thresholds starting at $1 eV$. (For recent
spectroscopic data see Ref. \cite{hellm}.) If the
latter is taken as the indirect gap, the discrepancy with the 
transport gap is still too large to be understood in 
conventional terms.  (For example, if one were to attribute
it to polaronic or bipolaronic effects, as some groups have
done  \cite{SU1,hellm,rua}, the resulting Franck-Condon factors
would be too large, 
leading to self-trapped rather than mobile carriers as required
by the observed mobilities.) Thus one clearly needs an
electronic negative-$U\/$ mechanism
of the sort we have discussed elsewhere \cite{AT1,AT2} and
which we will discuss below. (Of course, for a proper 
comparison with experimental data and a better estimate of model
parameters it seems crucial to include the details of the band
structure.) 

\vspace{5mm}

\noindent{\bf I.4.3 Comparison Between the Electronic Structures of
Bismuthates and High-$T_{c}$ Cuprates}
\vspace{3mm}

It is interesting to compare the electronic structures of the
high-T$_{c}$ cuprates and bismuthates. There are some
similarities
 \cite{AT1,LFM3} which become apparent when viewed against the
electronic structures of conventional `low-$T_{c}\/$' oxide
superconductors, such as TiO.  There are also very significant
differences between the two classes of high-$T_{c}\/$ oxide
superconductors. Hence a comparison between their electronic
structures and the resulting types of ordering (spin or charge)
helps us in understanding the microscopic interactions that lead
to the observed properties of these systems.

Both these high-$T_{c}\/$ superconductors are characterized by
strongly hybridized bands (Cu\,3d-O\,2p for the cuprates or
Bi/Pb 6s-O\,2p for the bismuthates), in which the Fermi level
falls within the uppermost $\sigma\/$ antibonding subband
(Fig.I.12a-b), making it nearly half filled. In the cuprates
this band is formed by the $\sigma\/$ bonding between
Cu\,3d($x^{2}-y^{2}\/$) and the 2p orbitals of the four
surrounding oxygen atoms, and in the bismuthates, by the
$\sigma\/$ bonding between the spherically symmetric and
spatially extended Bi/Pb\,6s orbitals and the 2p orbitals of the
six surrounding oxygen atoms.  Note that, in low-$T_{c}$ oxides
(e.g., TiO, Li$_{1+x}$Ti$_{2-x}$O$_{4}\/$), the metal d states
lie well above the O-2p manifold and $E_{F}\/$ falls within the
$\pi-\pi^{*}$ manifold (Fig.I.12b).  This has two effects:
First, as the metal-3d orbitals are well above the O-2p orbitals
(e.g., in TiO this separation is nearly $5\/$ eV),the
hybridization between the metal d and oxygen p orbitals is quite
weak, so the states near $E_{F}\/$ for these low-$T_{c}$
materials are {\em predominantly of d-character} (which is
consistent with the observation that oxygen stoichiometry has
little effect on their electronic properties). Second, the
$\pi-\pi^{*}\/$ mainfold being {\em triply degenerate } (formed
by the hybridization between the triply degenerate t$_{2g}$
orbitals, viz., d$_{xy}$, d$_{yz}$ and d$_{zx}$, of the
transition metal and the O-2p orbitals of the neighbouring
oxygen atoms), the DOS at $E_{F}\/$, and hence the Sommerfeld
constant $\gamma\/$, is {\em much larger} for them than for 
the high-$T_{c}\/$ oxide superconductors, where the $\sigma^{*}$
band is derived from the overlap of O-2p orbitals with
nondegenerate Cu3d$-{x^2-y^2}$ or Bi\,6s orbitals. The
well-known $T_{c}$ vs.  $\gamma$ curve for various
superconductors (Fig.I.16) illustrates this feature: The
cuprates and bismuthates have a very low DOS at $E_{F}$ compared
to their superconducting $T_{c}\/$'s.

We have already discussed that a tight-binding fit for the
important bands works quite well for the bismuthates. In the
cuprates a similar, tight-binding fit also works well. For
La$_{2}$CuO$_{4}\/$ $t_{dp\sigma}=1.8\/$ eV and $\epsilon_{d} =
\epsilon_{p}=-3.2\/eV$; only two parametrs are
required here as the O-2p and Cu-\,3d levels are nearly
degenerate in the tight-binding fit.

Like the bismuthates, the cuprates also have a nested Fermi
surface ({\em two dimensional} as opposed to the
three-dimensional one in the bismuthates) at zero doping. A
system with a nested Fermi surface is susceptible to various
instabilities like spin-density wave (SDW), CDW, etc. The
nominal valence of Bi in the bismuthates is $4^{+}$ and that of Cu
in the cuprates is $2^{+}$, which means that Bi and Cu ions have
just one electron in their outer shells in these compounds. Thus
one should expect these oxides to show fully developed moments
in their insulating states; however, these two systems show very
different magnetic behaviour: The cuprates show a (somewhat
reduced) moment whereas the bismuthates are perfectly
diamagnetic and there is no indication of a moment (e.g., in ESR
measurements). In addition, the insulating phase of the cuprates
is antiferromagnetically ordered, but in the bismuthates it
exhibits CDW ordering.

These differences suggest that the low-energy degrees of freedom
in the cuprates are the spin degees of freedom, whereas those in
the bismuthates are the charge degrees of freedom. Thus the
dominant interaction in the cuprates is the onsite Coulomb
repulsion ({\em repulsive} $U_{dd}$) between two electrons (or
holes) at a Cu site and not the long-range part of the Coulomb
interaction, which favours CDW ordering. (An effective
(antiferromagnetic) spin model can be obtained from a strongly
correlated electron system via the Anderson superexchange
mechanism, if the correlation is repulsive.) However, in the
bismuthates, CDW formation and the consequent quenching of the
spin degrees of freedom suggests that the repulsive, onsite part
of the Coulomb interaction is suppressed relative to the
repulsive, long range Coulomb interaction.  The suppression can
arise from a strong electron-phonon coupling between the Bi
atoms and the breathing mode of the surrounding O atoms
 \cite{TMR,EJ1} or from an electronic mechanism  \cite{CMV}.
There are reasonably convincing arguments (see below) for the
onsite interaction in the bismuthates being effectively {\em
attractive} and dominant relative to the nearest-neighbour,
Coulomb repulsion.  Thus, at the simplest level, the cuprates
should be modelled by a Hubbard model with strong, onsite
repulsion whereas the bismuthates should be modelled by using an
extended Hubbard model with {\em onsite attraction} and {\em
nearest-neighbour repulsion}.
\vspace{5mm}

The experimental overview presented above suggests that the
bismuthates are qualitatively different from both low-$T_{c}\/$,
oxide superconductors and high-$T_{c}\/$, cuprate
superconductors. Their superconduting and unusual normal-state
properties make these bismuthates interesting, and an attempt to
understand them a challenging task. So far there has been no
fully successful theoretical treatment of these systems for all
ranges of doping.  However, there have been quite a few
attempts, with varying degrees of success, to understand some of
the experiments mentioned above as we discuss below.

\vspace{1.5cm}

\noindent{\large \bf II. Model Hamiltonians}
\vspace{5mm}

In this Section we review the various models and theoretical
treatments that have been put forth to understand some of the
experimental properties of the barium bismuthates, so as to put
our work in perspective.  These theories typically concentrate
on some subset of the available experimental observations.
Furthermore, most of these theories assume that the important,
low-energy physics of these bismuthates can be understood on the
basis of a negative-$U\/$, extended Hubbard model with an
attractive, onsite and a repulsive, nearest-neighbour, Coulomb
interaction. An important issue, which has much bearing on a
theoretical understanding of these systems, is the origin of the
attractive, onsite interaction.  Two points of view exist in the
literature:\newline
\noindent (1) that the attraction arises because of a strong
electron-phonon coupling \cite{TMR,EJ1}; and \newline 
\noindent (2) that the attractive interaction is electronic in
origin   \cite{CMV}.\newline
We discuss these two points of view below and the
theories that use them. 

\sloppypar	In Section II.1 we motivate and describe the one-band,
negative-$U,\/$ extended-Hubbard model for the bismuthates, and discuss
the origin of the negative $U.\/$ We also describe the (well-known)
mapping of this model onto a pseudospin model in the strong-$U$
limit.  In Sec.II.2 we discuss some earlier theoretical
treatments of these models and analyze them in the context of experimental
results.

\vspace{1.5cm}

\noindent{\large \bf II.1  The One-Band, Negative-$U\/$, Extended Hubbard
Model For The Bismuthates}
\vspace{5mm}

\sloppypar
	 At the BaPbO$_{3}\/$ end of the $T-x\/$ phase diagram of
BaPb$_{1-x}$Bi$_{x}$O$_{3}\/$ (i.e., in the region $x < 0.2\/$) most of
the properties of these bismuthates are conventionally regarded 
as being consistent with the simple band picture of Mattheiss and Hamman
(Sec.I.4).
A wide conduction band is formed because of the hybridization of Pb (or
Bi) 6s orbitals  with O-2p$_{\sigma}\/$ bonding orbitals.  An approximate,
tight-binding characterization of this band may be made by choosing $
E_{6s} ( Bi/Pb)= -4.1\/$ eV, $E_{2p} = -1.9\/$ eV, and $t_{sp\sigma} =
2.2\/$ eV.  The band is partially filled as Pb$^{4+}\/$ contributes no
conduction electrons to it (whereas O$^{2-}\/$ contributes six p electrons
per oxygen to the entire s-p band and two p electrons per oxygen to the
bonding band).  The superconducting transition temperature  $T_{c}\/$ is
low; however, it increases with $x\/$ in a manner that is consistent with
the increase (with $x\/$)  of both $N(0),\/$ the density of states at the
Fermi level, and $\lambda_{ep},\/$ the electron-phonon coupling constant
 \cite{SS1}.

	For the  regime $x > 0.2\/$, it is clear that such a simple band
picture is inadequate because electron-electron and electron-phonon
interaction effects are very important. It has been suggested
 \cite{TMR,EJ1} that some of the essential features of these
interaction effects can be incorporated in an {\em effective,
semi-phenomenological, multiband\/}, negative-$U,\/$ extended-Hubbard
model where only the O-2p$_{\sigma}\/$ and the Bi/Pb 6s orbitals are kept
explicitly, namely,

$$ {\cal H}_{mb} = -t_{sp\sigma} \sumij (c^{\dagger}_{i\sigma} p_{j\sigma}
+ H.C.) - U_{1}/2 \sigmi \hat{n}_{i\sigma}\hat{n}_{i-\sigma}+\epsilon_{p}
\sigmi p_{i\sigma}^{\dagger}p_{i\sigma} $$ $$ + V_{1}\sum_{<ij>\sigma
\sigma^{\prime}} (c_{i\sigma}^{\dagger}
c_{i\sigma})(p_{j\sigma^{\prime}}^{\dagger}p_{j\sigma^{\prime}}).
\eqno(II.1)$$ 

Here $c_{i\sigma}^{\dag}\/$ creates  Bi 6s  electrons at the $i\/$ sites
of a cubic reference lattice (at this level of modelling, noncubic
distortions are regarded as higher-order effects which can be incorporated
at a later stage), $p_{j}\/$ is the destruction operator for oxygen
electrons (in the $\sigma\/$ orbital), $t_{sp\sigma}\/$ is the hopping
matrix element for the transfer of an oxygen electron from the
$p_{\sigma}$ orbital to a nearest-neighbour bismuth (or lead) atom,
$U_{1}\/$ and $V_{1}\/$ account for onsite and nearest-neighbour Coulomb
interactions, respectively, and $\epsilon_{p}\/$ is the chemical potential
for oxygen electrons (with respect to the bismuth 6s level).

It is easy to show  \cite{AT2,CMV} that, in the limit $t_{sp\sigma} \ll
\epsilon_{p}\/$, this reduces to a negative-$U\/$, extended-Hubbard model
involving the Bi atoms only (i.e., involving electrons in the Wannier
orbitals, centered on Bi atoms, of the antibonding Bi 6s-O 2p$_{\sigma}\/$
band only), given by the Hamiltonian

$$ {\cal H}=-t \sumij (c^{\dag}_{i\sigma}c_{j\sigma}+h.c.)-U/2\sigmi
\hat{n}_{i\sigma}\hat{n}_{i-\sigma}+V \sum_{<ij>\sigma\sigma^{\prime}}
\hat{n}_{i\sigma}\hat{n}_{j\sigma^{\prime}}-\mu \sum_{i}
\hat{n}_{i},\eqno(II.2)$$

\noindent where $<ij>$ are nearest-neighbour pairs of sites, 
$\hat{n}_{i}=\sum_{\sigma}c^{\dag}_{i\sigma}c_{i\sigma}\/$, $\mu\/$ is the
chemical potential, $\hat{t}=|t_{ps}|^{2}/\epsilon_{p}\/$, $V = V_{1} z
|t_{sp\sigma}/\epsilon_{p}|^{2}$ and $U=U_{1}\/$.  However, the validity
of this reduction for realistic parameters and the relationship between
$t,V,\/$  and $U\/$ and $t_{sp\sigma},V_{1},$ and $U_{1}\/$ clearly needs
further study, for there are limits in which it must fail: in particular,
it must fail in the limit when most of the Bi atoms are replaced
by Pb atoms, so that BaPbO$_{3}\/$ turns out to be metallic as it is.
Nevertheless, this {\em effective, one-band} model, which has been
proposed by a number of authors \cite{AT2,CMV,TMR,EJ1,RM,MJR,EJ2,AT4,BKC,SR1},
can indeed serve as a starting point, at least in the semiconducting
regime, as it contains, the effects of the electron-electron and electron
phonon interactions (at the simplest level as discussed below).

In BaBiO$_{3}\/$,  $1/N \sigmi <\hat{n}_{i\sigma}>\/$ is constrained
to be $1\/$ ($N\/$ is the number of lattice sites) corresponding to a
nominal Bi valence of $4^{+}$.  In the
Ba$_{1-x}$K$_{x}$BiO$_{3}\/$ system, two important 
modifications are necessary:\\ (1) Since \mbox{K$^{1+}\/$} replaces
\mbox{Ba$^{2+}\/$}, one electron is removed from the system  and \\
\mbox{$1/N \sigmi < \hat{n}_{i\sigma}> = (1-x)\/$}, corresponding to a
nominal valence of $(4+x)^{+}$. (2) There is a random Coulomb potential
because of \mbox{K$^{1+}\/$}\ sites, which act as effective 1$^{-}$
impurities.

In order to model \cite{EJ1} the\, BaPb$_{1-x}$Bi$_{x}$O$_{3}$
system,  we must introduce quenched, random, site variables 
$q_{j} = 0\/$  (or 1), if site $j\/$ is
occupied by a Pb (or Bi) atom; i.e., $\overline{q}_{j} = x\/$, where the
bar denotes an average over the quenched randomness.  $U\/$ and $V\/$ in
Eq.(II.2) are assumed to act only on Bi sites, which can be modelled by
replacing $ \hat{n}_{i}$  in Eq.(II.2) by\, $ \hat{n}_{i}q_{i}$.  In
addition, we must introduce a random,  site-energy term,  $ \epsilon_{0}
\sigmi \hat{n}_{i\sigma}(1-q_{i}),$  with   $\epsilon_{0} = 2.5$eV,
corresponding to the difference between the local orbital energies of Pb
and Bi. Note only Bi sites contribute electrons to the antibonding band,
so  $ 1/N \sigmi <\hat{n}_{i\sigma}> = x\/$   (compare the case of doping
above); however, Pb sites do not give rise to a Coulomb potential.

The justification for the negative-$U\/$ terms in Eqs. (II.1)
and (II.2) is that it is the simplest phenomenological way of
modelling the {\em valence-skipping\/} phenomenon, i.e., the
stability of the Bi$^{3+}\/$ (6s$^{2}\/$) and Bi$^{5+}\/$
(6s$^{0}\/$) states relative to the Bi$^{4+}\/$ (6s$^{1}\/$)
state, that many believe is characteristic of the chemistry of
these bismuthates. As regards its origin, however, there are two
points of view: the first phonon mediated and the other
electronic.

The approach in which the negative $U\/$ arises from an
electron-phonon mechanism is similar to Anderson's \cite{PWA}
suggestion (in a different context) that an attractive,
electron-electron interaction can be generated if one couples
lattice vibrations to electronic degrees of freedom strongly.
Anderson had shown that the quenching of the magnetic moment in
the A-15 compounds and some amorphous semiconductors can be
explained if there is such an attractive interaction at each
site.  This idea was utilized by several authors to understand
the barium bismuthates, as outlined below.

Rice and Sneddon  \cite{TMR} and later Jurczeck and Rice
 \cite{EJ1}
used this idea and obtained the negative-$U\/$ model by coupling the
excess charge (Bi$^{4+}$ being regarded as the neutral configuration) on a
conventional positive-$U_{0}\/$ Bi site to the breathing-mode, optic
phonon of the surrounding oxygen octahedron.  Consider the BaBiO$_{3}$ end
of the composition range: The energy required to displace each O atom is

$$  H_{ij}=\frac{1}{2}C x^{2} - g x (q_{i}-q_{j}),$$

\noindent  where $q_{i} = (1-n_{i})\/$ is the excess charge on a Bi site, 
$(q_{i}-q_{j})$ is the charge difference between two
neighbouring Bi sites, $x\/$ is the (breathing mode)
displacement away from the equilibrium position of the oxygen
atom, $g$ is the electron-phonon coupling constant, and $C\/$ is
the spring constant of the oxygen breathing mode. The
minimization of $H_{ij}$ (with respect to $x\/$) yields $x\/$ in
terms $q_{i}$ and $q_{j}$ and thus eliminates the oxygen degrees
of freedom. The complete Hamiltonian that now describes the
conduction electrons in the Bi\,6s band, including the
long-range, Coulomb interactions between Bi sites is:

$$ H= \sum_{<ij>\sigma}t_{ij} c_{i\sigma}^{\dag}c_{j\sigma} - g^{2}/2C
\sum_ {<ij>}(q_{i}-q_{j})^{2} +\sum_{i\sigma} \frac{U_{0}}{2}
\hat{n}_{i\sigma}\hat{n}_{i-\sigma}+\sum_{i\neq j} (e^2/\epsilon r_{ij})
q_{i}q_{j}.\eqno(II.3)$$

The effective, onsite interaction $U_{eff}=U_{0}-zg^{2}/C $ is thus
strongly suppressed compared to $U_{0}$, and can even be
negative if $g\/$
is large enough ($z\/$, the coordination number, is 6 in the
three-dimensional 
cubic lattice considered here).  Furthermore, 
$ V_{nn} = (e^{2}/\epsilon r_{nn} + g^{2}/C) $ is the largest of
the repulsive energies, where 
$r_{nn}$ is the nearest-neighbour distance (lattice constant).

If we ignore the longer-range, repulsive interactions present in
the Hamiltonian (II.3), we get an effective, one-band,
negative-$U\/$, extended-Hubbard model involving Bi atoms only
(Eq. (II.2)).  We can incorporate the effects of the long-range part
of the repulsive interactions roughly by replacing\,
$e^{2}/\epsilon r_{nn}$ by\, $e^{2}\alpha/\epsilon r_{nn},\/$
where\, $\alpha\/$ is the Madelung constant.  In any case $V\/$
in Eq. (II.2) is treated as a phenomenological constant, which
must be obtained by fitting experimental data.

It has been suggested recently  \cite{CMV} that there is an
electronic mechanism by which  this attractive, onsite interaction can
arise in the barium bismuthates.  This mechanism is connected with the
physics of {\em valence  skipping, electronic polarizability and
screening} processes. There are nearly $15\/$ elements in the periodic
table in Groups III and $V\/$ which have a propensity for skipping a valence
state  (Ga, In and Tl favour valence states $1^{+}\/$ and $3^{+}\/$ and not
$2^{+}\/$; Bi and Sb favour $3^{+}\/$ and $5^{+}\/$ and skip
$4^{+}\/$ (Sec. I)).  The intermediate valence states are either
unstable or metastable. This was 
well known in chemistry and attributed to the favoured valence states
corresponding to closed-shell electronic configurations: The effective,
intra-atomic, repulsion energy $U_{n}$ for the $n^{th}\/$ charge-state of
an ion, $U_{n}  \equiv  E_{n+1} + E_{n-1}-2E_{n},\/$ where $E_{n}\/$ is
the total energy of the atom in the $n^{th}\/$ charge state, increases
along a period for most elements as $n\/$  increases, since the sizes of
the orbitals decrease with increasing $n\/$.  However, for the valence
skippers, it is clear  from Table II.1 (compiled from measured or
calculated ionization energies  \cite{CMV}) that $U_{n}\/$ has a
minimum at the valence that is skipped.  This reflects the larger
correlation energy of the neighbouring, closed-shell configurations. All
the $U\/$ values, including those at the skipped valence state, are still
positive, as is necessary for an isolated ion. Varma  \cite{CMV} argues
that this situation changes drastically when these ions are in a
crystal. The value of $U\/$ that should then be used is the effective,
onsite interaction after the screening effect of the environment has been
accounted for.  Consider BaBiO$_{3}\/$:  The formal valence of Bi is
$4^{+},\/$ so the configurations Bi$^{5+}$ and Bi$^{3+}$ are strongly
charged. Also, each Bi atom is surrounded by the highly covalent and
easily distortable O octahedra, thus nonlinear screening of the atomic
charges is important.  A negative value of $U\/$ (for Bi in the
valence state $4^{+}$ ) results if the
reductions in the energies of $5^{+}\/$ and $3^{+}\/$ configurations are
sufficiently large compared to twice the reduction in the energy of the
$4^{+}\/$ configuration.  Since the 6s$^{0}$ (Bi$^{5+}$) configuration is
most strongly charged compared to the $4+\/$ and $3+\/$ configurations, it
is screened most strongly by charge transfer from the O octahedra to the
6p states of Bi.  Density-functional calculations of Hamman (quoted in
Ref. \cite{CMV}) give $E(6s^{0}6p^{1}) - E(6s^{0}) = -43.3\/$ eV.
As shown in Table II.1, the electrostatic energy cost for the
transfer of an electron is less than that obtained by linear
interpolation between $U_{5+}\/$ and $U_{3+}\/$ by nearly $15\/$
eV.  Hence it is argued \cite{CMV} that nonlinear screening can
bring the effective $U_{4+}\/$ down to a negative value.

Thus both electron-phonon and electron-electron interactions can
lead to phenomenological, effective Hamiltonians of the
negative-$U\/$ type.  However, these two mechanisms are very
different when considered in detail.  In particular, if the
electron-phonon mechanism is involved, the effective, onsite,
attraction $U\/$ is negative only on energy scales smaller than
the frequencies of the breathing-mode phonons. If, on the other
hand, an electronic mechanism is involved, $U\/$ can be negative
over typical electronic energy scales (of order eV). This
difference has detectable experimental consequences (Sec.IV).

Models (II.1)-(II.3) have strong correlation effects and many
parameters, and are difficult to treat: Hence many workers  \cite{CMV,RM}
have tried to see whether the large-$U\/$,  limit
($U\gg zt$)  of model (II.2), which is much simpler (see below),  is
adequate for describing the properties of the Ba$_{1-x}$K$_{x}$BiO$_{3}$
system.  In this limit we can project out the singly occupied (Bi$^{4+}$)
states and keep only the empty ($|0>$ i.e.,\, Bi$^{5+}$) and
doubly occupied ($|2>$ i.e.,\,Bi$^{3+}$) states at each site.
The effective Hamiltonian in this reduced Hilbert space is
easily derived via second-order, degenerate, perturbation theory
(Appendix A) or canonical transformations, and can be written in
terms of the pseudospin operators $\bf{S}_{i}\/$ (that operate
on the unoccupied and doubly occupied states $|i0>\/$ and
$|i2>\/$, respectively, at the site $i\/$):

$$ \sip=\ciud \cidd\equiv|i2><i0|, $$ $$ \sib \equiv c_{i\downarrow}
c_{i\uparrow}\equiv|i0><i2|,\eqno(II.4) $$ $$ \siz \equiv 1/2 (n_{i} -
1)\equiv(|i2><i2|- |i0><i0|)/2.$$

\noindent The effective Hamiltonian is

$$ {\cal H}_{eff}=J \sum_{<ij>}(S_{i}^{+} S_{j}^{-}
+H.C.)+K\sum_{<ij>}S_{i}^{z} S_{j}^{z} -B \sum_{i}(2
S_{i}^{z}+1),\eqno(II.5)$$
 
\noindent where $J=2t^{2}/U,\, K=J+2V,\/$ and the pseudomagnetic field $B\/$ is
related to the chemical potential $\mu\/$ via $B=\mu +U/2 - zV.\/$ Note
that in the experimental system Ba$_{1-x}$K$_{x}$BiO$_{3}\/$,
$<S_{i}^{z}>\/$ is fixed at $-x/2;\/$  i.e., it corresponds to the
fixed-magnetization ensemble.

	An alternative representation, that is physically more transparent
than the pseudospin representation, uses the {\em hard-core boson\/}
operators $b_{i}^{+} = \sip\equiv \ciud \cidd$ and $b_{i} = \sib \equiv
\cid \ciu\/.$ Clearly these represent real-space, onsite-paired 6s$^{2}$
electrons, or Bi$^{3+}$ states, and have charge 2e. It is often useful to
think in terms of these hard-core bosons, so we will use both pseudospin
and hard-core-boson descriptions in our discussions below. The
correspondences between the pseudospin model (II.5) and the
extended-Hubbard model (II.2) are summarized in Table II.2. We will discuss
the viability of these models mainly for Ba$_{1-x}$K$_{x}$BiO$_{3}.\/$
To put our discussion in perspective, we review  earlier theoretical
treatments briefly.
\vspace{7.5mm}

\noindent{\large \bf II.2  Brief Review of Earlier Theories of Barium
Bismuthates}
\vspace{5mm}

\noindent{\bf II.2.1
 The Theory of Rice and Sneddon}
\vspace{3mm}

In one of the earliest attempts to understand the mechanism of
pairing in the bismuthates, Rice and Sneddon  \cite{TMR,EJ1} developed the
idea of the condensation of electrons, coupled strongly to the optic,
breathing-mode phonons, into a crystal of real-space pairs (CDW order).
These real-space pairs develop off-diagonal long-range order (ODLRO), if
the band width is finite, and thence superconductivity.

Rice and Sneddon worked with an effective, classical spin model.
They expressed the Hamiltonian (II.3), in the zero-bandwidth limit, in
terms of a spin variable $m_{i}\/$ (where $m_{i}=n_{i}-1\/$)
which assumes the 
values $1, 0,\/$ and $-1\/$ at each Bi site; $m_{i}= +1\/$ and
$m_{i}=-1\/$ correspond, respectively, to the configurations with two
electrons (Bi$^{3+}\/$) and no electron (Bi$^{5+}$) at the site $i,\/$
respectively, whereas $m_{i}=0\/$ corresponds to single occupancy
(Bi$^{4+}\/$) (the double degeneracy of this configuration 
is ignored).  At the BaBiO$_{3}\/$ end, the Bi\,6s band is half
filled, so the chemical potential is $-U/2$. The deviation of the chemical
potential from its value at half filling ($=-U/2\/$) acts as a magnetic
field in this effective spin model. Hence the effective, classical, spin
Hamiltonian (in the zero-bandwidth limit) is just the S=1,
Blume-Emery-Griffiths (BEG) model  \cite{MB},

$$ H=  U_{eff}/2 \sum_{i} m_{i}^{2}+V\sum_{<ij>}m_{i}m_{j},\eqno(II.6) $$

\noindent where $  U_{eff}/2=\frac{1}{2}(U_{0}-zg^{2}/C),\/$ 
$V=g^2/C + e^2\alpha/\epsilon r_{nn}\/$ (clearly $V > 0\/$), and
$\alpha\/$ is the Madelung constant (Sec.II.1).  The intersite term is
taken to be short ranged here and the effect of the longer-range part of
the interaction is included via the Madelung constant.  (Note the
difference between the derivation of this {\em classical spin model,}
obtained from model (II.2) by setting the hopping $t=0\/$ and
{\em ignoring} the the 
double degeneracy of the Bi$^{4+}$ configuration (up and down spin
states), from our earlier derivation in Sec.II.1 of the {\em quantum
spin model} in the limit of large-$U\/$, from a negative-$U\/$,
extended-Hubbard model.)

The Hamiltonian (II.6) has been studied extensively \cite{MB}
and its phase diagram is well known. For $ U_{eff} < zV\/$
(always satisfied if $U_{eff} < 0\/$), the ground state is
antiferromagnetic, with $m_{i}=\pm 1$ on the two sublattices
(i.e., the CDW phase of model (II.3)). If $U_{eff} > zV\/$, the
ground state has $m_{i}=0$ at all sites (i.e., a nonordered
phase with all Bi sites equivalent).

Rice and Sneddon use mean-field theory to obtain the phase
diagrams of model (II.3), with $t=0\/$, and the BEG model
(II.6).  Fig.II.1 depicts their phase diagrams in the $T -
U_{eff}/2V\/$ plane.  Though the mean-field phase boundaries
obtained for these two models do not coincide, the topologies of
the phase diagrams are the same.

At finite temperatures there are two types of excitations from
the CDW ground state: two-particle excitations
($m_{i}\rightarrow -m_{i}\/$); and single-particle excitations
($m_{i}=\pm 1 \rightarrow 0\/$), which break pairs.  In the
hypothetical limit $U_{eff}/ 2zV \rightarrow -\infty,\/$ a
transition from the CDW to the nonordered phase can occur only
via the two-particle excitations; this is a second-order
transition from a crystal of paired electrons to a liquid of
these pairs. As $U_{eff}/2zV\/$ increases and passes through
zero, the single-particle excitations become dominant and there
is a first-order transition, with a discontinuous jump in the
number of paired electrons (at $T=0\/$ this transition occurs at
$U_{eff}/2zV~=~1\/$).  For $U_{eff}/2zV > 1\/$ no phase boundary
is encountered, as $T\/$ increases; only the nature of the
disordered phase changes from a liquid of paired electrons to a
phase with no paired electrons.

Doping by Pb$^{4+}\/$ is modelled as in dilute magnetic alloys
since Pb$^{4+}\/$ behaves like a $m_{i} = 0\/$, nonmagnetic
impurity. In this picture  \cite{TMR}, the replacement of Bi by
Pb removes an electronically active site from the system.
Concentration fluctuations are assumed to be suppressed because
of long-range Coulomb forces. At this level of modelling, the
effect of doping by Pb is incorporated in the BEG Hamiltonian
(II.6) by the reduction of the local coordination number $z.\/$
For $U_{eff} < 0,\/$ both terms (site and bond)) in Eq. (II.6)
favour real-space pairing, and there is a second-order
transition to the nonordered state.  However, if $U_{eff} >
0,\/$ then these terms compete.The reduction of the average
cordination number $\bar{z}\/$ decreases the importance of the
bond term, so it is argued  \cite{TMR} that dilution can drive
the transition first order. Their phase diagram (Bethe-Peierls
approximation) in the $T/V-\bar{z}\/$ plane is shown in Fig.II.2
for three different values of $ U_{eff}/2V=-5, 1.2,\;{\rm
and}\;\; 2.3.\/$

The only ordered phase of the classical model (II.6) is the CDW
phase. In order to have a superconducting phase, one must
consider the quantum limit (with transverse spin fluctuations),
so hopping (i.e., $t_{ij}\/$) must be included. With nonzero 
$t_{ij},\/$ it is not clear whether this model has an
insulator-metal or an insulator-superconductor transition.
Superconductivity may develop  \cite{TMR} via a transition from
real-space, local pairing (a CDW) to momentum-space pairing (a
BCS superconductor), aided by dilution (i.e., an increase of the
effective, single-partcle band width); this scenario is
described below:

In the $x \rightarrow 1\/$ limit (the BaBiO$_{3}\/$ end), the
CDW is the ground state, and two-particle excitations propagate
only through high-energy intermediate states, resulting in a
very narrow band.  With dilution we get Pb$^{4+}\/$ sites, which
are $m_{i}=0,\/$ local, single-particle excitations over the CDW
ground state. As the concentration of Pb sites increases, these
excitations can move because of the hopping $t_{ij}\/$
[$=2(z-\bar{z})t\/$], so doping increases the band width for the
motion of single-particle excitations.  

In the $x \rightarrow 0\/$ limit (the BaPbO$_{3}\/$ end), there
are very few Bi atoms. They behave as impurities in
BaPbO$_{3}\/$.  If $U_{eff} > 0\/$ they prefer to be in the
$4^{+}$ state i.e., $m_{i}=0\/$ ($\bar{z} = zx \simeq 0,\/$ so the
bond term in Eq. (II.6) can be neglected) and act as donors
of electrons. As $x\/$ and hence $\bar{z}\/$ increases, a
metal-insulator transition becomes possible.  If, however, $
U_{eff} < 0,\/$ the Bi sites act as negative-$U\/$ centres and
charge disproportionation into Bi$^{3+}\/$ and Bi$^{5+}\/$ is
possible; i.e., an electron binds to the Bi$^{4+}\/$ site
because of the short-range, onsite, Coulomb potential.  This
cannot occur unless $|U_{eff}|/2 \gg 2(z-\bar{z})t.\/$ Hence
Rice and Sneddon argue that  \cite{TMR}, with decreasing
$\bar{z}\/$ (i.e., increasing the Pb concentration) and at $T =
0\/$, there can be a transition from a CDW insulator to a metal
only if $|U_{eff}|/2$ is not too large. If $t_{ij}\/$ increases
relative to the electron-phonon interaction 
(in the conventional BCS theory, $t_{ij}\/$ is much stronger than the
electron-phonon coupling), then the transition to an insulator from a
metal occurs at a larger value of $\bar{z},\/$  (i.e., a higher Bi
concentration is necessary to make the system insulating). If
$t_{ij}=0\/$, each electron is localized at a site and, for a 
system with $N\/$ electrons, there are $2^{N}\/$ states.  The spin
degeneracy  of the unpaired state is removed in the metallic state (when
$t_{ij}\neq 0\/$). As a result the entropic correction to the free energy
goes as $T^{2},\/$ in the metallic state, as opposed to $T\/$ in the
unpaired state (with $t_{ij}=0\/$). Thus the metal-insulator phase
boundary moves up perpendicular to the $T = 0\/$ axis in the
$T/V-\bar{z}\/$ phase diagram (Fig.II.2). At lower temperatures,
then, the metal can become a supeconductor. This is the usual
BCS (momentum-space) condensation of paired electrons into a
superconducting state. Since the superconductivity here is
mediated by optic phonons, the attractive interaction is
retarded and all energy integrals (e.g., in the $T_{c}\/$
equation) are cut off by $\omega_{0}$, the frequency of the
breathing-mode optic phonon.

\vspace{5mm}

\noindent{\bf II.2.2 The Theory of Jurczek and Rice}
\vspace{3mm}

The transition from real-space to momentum-space pairing as a
function of doping  \cite{TMR} described above was studied
further  \cite{EJ1} in 
the limit of finite and large bandwidth, to understand why the
CDW phase in the lead-doped system is so stable (till a Pb
concentration of $65 \%\/$). The mean-field study of a strongly
coupled electron-phonon system (with parameters taken from
band-structure calculations  \cite{LFM1,LFM4}) showed that the
``local" CDW state is indeed stable upto large doping levels.
(Band-structure calculations (Secs. I and II.3.1) had predicted
a strongly hybridized antibonding Bi/Pb\,6s-O\,2p$_{\sigma}\/$
band crossing the Fermi surface as in a metal, which was clearly
untenable since the lead-doped system shows a metal-insulator
transition at a Pb concentration of $65 \%.\/$) This theory
begins with the Hamiltonian (II.3) and an electron-phonon
coupling that is strong enough to make $U_{eff}=U_{0}-zg^{2}/C $
negative, which can lead to both CDW and pairing instabilities.
We outline this study below (first for the undoped and then for
the doped case.)

\noindent {\bf Undoped Case}

Here $x=1\/$ so the (uniform) CDW order parameter is:

 $$<\hat{n}_{i}> \equiv <\sum_{\sigma}c_{i\sigma}^{\dag}c_{i\sigma}>= 1 +
b_{c}\exp i {\bf Q.r}_{i},$$

\noindent where {\bf Q} = $\pi (1,1,1)$ and $b_{c}$ measures the deviation
of the valence at a Bi site from its nominal value (4+).  Bi sites with
$<\hat{n}_{i}> = 1 + b_{c}$ and $<\hat{n}_{i}> = 1 -  b_{c}$ are called
A and B sites, respectively. With this choice for the order
parameter and a conventional mean-field decoupling the
Hamiltonian becomes 

$$    H_{0}=t\sum_{ <ij> \sigma}c_{i\sigma}^{\dag}c_{j\sigma}+
\Delta_{c} \sum_{i}\hat{n}_{i}\exp i {\bf Q.r}_{i},\eqno(II.7)$$

\noindent where $\Delta_{c}=u_{p} b_{c}/2$ is the CDW gap parameter, 
$u_{p} \equiv U_{eff}+2ze^2\alpha/\epsilon r_{nn}$, and $\alpha$ is the
Madelung constant. The charge-disproportionation response-function is

$$R(b_{c}) \equiv 1/N \sum_{j=1}^{N} \exp i {\bf Q.r}_{j}
\int_{- \infty}^{E_{F}}dE \; n_{j}(E;b_{c}),\eqno(II.8) $$

\noindent and, at half filling, the Fermi energy is given by

$$ 1=1/N\sum_{j}\int_{- \infty}^{E_{F}}dE \; n_{j}(E;b_{c}),\eqno(II.9)$$

\noindent where $n_{j}(E;b_{c})$, the local density of states
at the site $j\/$ with energy $E\/$, obviously depends on the
value of $b_{c}\/,$ which must be determined self consistently.  If we
multiply both sides of 
the self-consistency  equation $<\hat{n}_{i}>=1 +  b_{c} \exp i{\bf
Q.r}_{i}\/$ by\,\, $ \exp i {\bf Q.r}_{j}\/$ and sum over all sites, it
becomes $R(b_{c}) = b_{c}\/$.  The local density of states (LDOS) that
satisfies this self-consistency equation can be obtained , as
usual, from the imaginary part of the Green function:

$$n_{j}(E) \equiv <j|\delta (E-H_{0})|j>=\frac{1}{\pi} \lim_{\eta
\rightarrow 0} Im <j|\frac{1}{E-i\eta-H_{0}}|j> $$
$$=\frac{1}{\pi} \lim_{\eta \rightarrow 0}Im G_{j}(z)|_{z=E-i\eta}.
\eqno(II.10)$$
\noindent Finally the self-consistency equation becomes the usual
mean-field gap equation:

$$1/u_{p} =\frac{1}{2N}\sum_{\bf k }1/(\Delta_{c}^{2}+t_{\bf
k}^{2})^{1/2},\eqno(II.11) $$
\noindent where the sum over momenta is restricted to half the Brillouin 
zone, $N\/$ is the number of sites in the lattice, $t_{\bf k}=t\sum_{\bf
k} e^{i{\bf k.a}} $, \, and ${\bf a}$ are nearest-neighbour lattice
vectors.
\vspace*{4mm}

\noindent {\bf Doped Case}

Jurczek and Rice accounted for dilution (i.e., Pb doping,
Sec.II.1) by replacing $\hat{n}_{j}\/$ with $p_{j}\hat{n}_{j},\/$ 
where $p_{j}=1,\/$ if site $j\/$ is occupied by a Bi atom, and
$p_{j} = 0,\/$ otherwise; the difference $\epsilon_{0,\/}\/$
($\simeq 2.5\/$ eV  \cite{LFM1,LFM4} between the onsite energies
of Pb and Bi, is an additional parameter. The following ansatz
is used for the inhomogeneous, CDW order parameter:

$$<\hat{n}_{j}>=p_{j}(1+ b_{c} \exp i{\bf Q.r}_{j}),\eqno(II.12)$$

\noindent i.e., for a site with a Pb atom $<\hat{n}_{j}> = 0.\/$
Mean-field decoupling yields the Hamiltonian

$$ H_{0}=t\sum_{<ij> \sigma}c_{i\sigma}^{\dag}c_{j\sigma}
+\epsilon_{0}\sum_{j}\hat{n}_{j}(1-p_{j})+\Delta_{c}(x)\sum_{j}\hat{n}_{j}p_{j}\exp
i{\bf Q.r}_{j},\eqno(II.13) $$

\noindent where $\Delta_{c}(x)={1\over 2}
(U_{eff}+2ze^2\alpha/\epsilon r_{nn}) b_{c}(x)\/$ and we have
displayed the
$x\/$ dependence of various quantities explicitly, as the order parameter
has to be obtained self consistently for each value of $x.\/$ For
BaBiO$_{3}\/$ there are two sets of inequivalent Bi sites, A and B. With
the introduction of Pb, sites in each of these sets can have either a Bi
or a Pb atom. Hence $R(b_{c})\/$ (Eq. (II.8)) becomes

$$R(b_{c} (x))=\frac{1}{2x(1-x)}[x \int_{- \infty}^{E_{F}}dE\;
\overline{n_{A-Bi}}(E;b_{c}) + (1-x) \int_{- \infty}^{E_{F}}dE\;
\overline{n_{A-Pb}}(E;b_{c})$$
$$-x \int_{- \infty}^{E_{F}}dE\;
\overline{n_{B-Bi}}(E;b_{c})-(1-x) \int_{ -\infty}^{E_{F}}dE\;
\overline{n_{B-Bi}}(E;b_{c})]
.\eqno(II.14)$$

The LDOS (Eq. (II.10)) must now be configurationally averaged. Even a
coherent-potential-approximation (CPA) calculation of this averaged LDOS
($\overline{n_{A-Bi}},\/$ etc.) is difficult since there are four
different averages. Hence an orthogonalized-moment method has been used  \cite{EJ1}.
This method uses the continued-fraction
representation of the one-particle Green function $G_{i}(z).\/$ Figures
II.3a and II.3b show the self-consistent LDOS for the Bi-A(B) sites  at Pb
concentrations of $20\%$ and $60\%,\/$ respectively. Note that, on doping
with Pb: (1) there is no real gap, but only a pseudogap, in the DOS at the
Fermi energy;  and (2) there are very few Pb states in the pseudogaps of
Fig.II.3. As the Pb concentration increases, the Bi subbands approach
each other: the upper Bi subband is pushed down because of repulsion by
the Pb band (the difference between the onsite energies of Pb and Bi is
$\epsilon_{0,\/} \simeq 2.5\/$ eV) but the lower Bi band is hardly
affected.

	It was  found   \cite{EJ1} that the local CDW is not
stable (i.e., $b_{c} = 0\/$) if the total coupling constant $\lambda(x)
\equiv {1\over 2}|U_{eff}(x)| N(E_{F},b_{c}=0)\/$ is less than a critical
value $\lambda_{c}(x).\/$ ($N(E_{F},b_{c}=0)\/$ is the DOS at the Fermi
energy in the absence of the CDW; and this positive, dimensionless, total
coupling constant is an effective measure of the onsite attraction
relative to the bandwidth in the noninteracting limit.) Also, the local
CDW in the BaPb$_{1-x}$Bi$_{x}$O$_{3}\/$ system was shown to be stable for
a wide range  ($ 0.35 < x < 1$) with reasonable values of $\lambda\/$.  In
this local CDW phase the DOS is nonzero in the pseudo gap (Fig.II.3), so,
to account for the semiconducting nature of this phase in
BaPb$_{1-x}$Bi$_{x}$O$_{3}\/$, Jurczek and Rice argued that the states in
the pseudo gap obtained here could well be localized because of diagonal
disorder (i.e., Pb doping).

An estimate of the electron-phonon coupling constant  has also been made
 \cite{EJ1}:  $U_{eff}\/$  can be separated into an
attractive part  $zg^{2}/C,\/$ from the electron-phonon interaction, and
$U_{0}\/$, the repulsive,  onsite electron-electron Coulomb interaction.
These two parts can be used to define the electron-phonon and
electron-electron coupling constants $\lambda_{ep}\/$ and
$\lambda_{ee}\/$, respectively ($\lambda\/$'s  defined as above). When
$U_{0}\/$ exceeds $U_{0c}=zg^{2}/C\/$ the effective, onsite, interaction
becomes repulsive and the ground state exhibits a spin-density wave, not a
charge-density wave (if near-neighbour repulsion is negligible
compared to the onsite 
term). Hence, in the CDW state, $\lambda_{ee}\/$ is maximum at
$U_{0}=U_{0c}$. This maximum value combined with typical values of
$U_{0}\/$  and the value obtained  for $\lambda_{c}\/$ gives an estimated
lower bound for $\lambda_{ep}\/$: it is found that $\lambda_{ep} \simeq\/$
 1 (for $\epsilon_{0} \simeq  1.7\/\/$ eV); i.e., according to this
theory, the system is just in the strong electron-phonon-coupling regime.
(This should be contrasted with the large values obtained in the
bipolaron-superconductivity theories, discussed below.) The values of both
$\lambda_{ep}\/$ and $\epsilon_{0}\/$ compare reasonably well with
experimental observations: $\epsilon_{0} \simeq 2.5\/$ eV (atomic estimate
given above) and the experimental value of $\lambda_{ep} \simeq 0.6\/$ in
the lead-doped system.  These values have been used  \cite{EJ1}
to obtain the superconducting $T_{c}\/$  via the
McMillan-Morel-Anderson Formula (McMillan, (1968; Morel \& Anderson, 1962)
and with $\lambda_{ep} (b_{c}= 0.65) = 0.7$: \/\/
$T_{c}={<\omega>\over 1.45} \exp
(\frac{1.04(1+\lambda_{ep})}{\lambda_{ep}-\mu^{*}
(1+0.62\lambda_{ep})}).$  
Here $<\omega>\/$ is the average phonon frequency (this is
normally taken to be the frequency of the phonon that couples  to the
electronic degree of freedom or the Debye temperature of the system),
$\mu^{*}$ is the Morel-Anderson Coulomb pseudopotential (Morel \&
Anderson, (1962).  Jurczek and Rice obtained $T_{c}\geq 7\/$ K  with
$<\omega>\simeq 150 cm^{-1}$, the frequency of the optic-mode phonon, and
$0.067< \mu^{*} < 0.116$.

	The main results of the above analysis is that the insulating,
local CDW phase of the lead-doped bismuthate is  quite robust against
doping.  There is a strong electron-phonon coupling in
BaPb$_{1-x}$Bi$_{x}$O$_{3}$ responsible for the attractive
electron-electron interaction that leads to the CDW and superconducting
instabilities in this system; and the  electron-phonon coupling constant
is fairly large.  We will briefly discuss the merits and shortcomings of
this model below.

\newpage
\noindent{\bf II.2.3 CPA On An Attractive Hubbard Model:
Semiconductor-Super-\newline conductor Transition}
\vspace{3mm}

	Yoshioka and Fukuyama  \cite{DY} have considered a
negative-$U\/$, Hubbard model with onsite disorder for the bismuthates.
They believe that, in BaPb$_{1-x}$Bi$_{x}$O$_{3}$, there is no
semiconductor-metal transition but only a transition  from a semiconductor
to a superconductor (at $T = 0\/$). Although such an assertion is not very
unreasonable (Sec.I.3), there is
no firm  experimental evidence for it.  In their model then, as long as
there is a gap in the DOS due to the difference in chemical potentials of
Bi and Pb, the system remains insulating. As soon as this gap vanishes,
the attractive onsite interaction leads to a pairing instability, which,
in the doped case, leads to a superconducting transition (at half filling
both the CDW and superconducting phases  are degenerate, but doping  favours
superconductivity). We review their theory briefly below:

In the negative-$U\/$ Hubbard model they consider,  disorder because of
the substitution of Bi by Pb is  modelled by a site-dependent chemical
potential.  Site energies  $\epsilon _{A}$ and $\epsilon _{B}$ are
associated with the atoms A and B (see below) respectively.   There is an
onsite, attractive interaction mediated by a strong, electron-phonon
interaction.  They further assume that atom A has no conduction electron
while B has two (see below).  So at $x\/$=1 the conduction band is full
and at $x\/$= 0, it is empty .  Now the connection between this picture of
a binary alloy and the experimental system BaPb$_{1-x}$Bi$_{x}$O$_{3}$ is
made in the following manner:

	 The hybridization of Bi/Pb\,6s and O\,2p orbitals creates ten
bands near the Fermi level (Sec.I, Fig.I.11 and I.13).  The
Fermi level falls in 
the highest $\sigma^{*}\/$ subband which is half-filled at $x = 1.\/$ A
unit-cell-doubling  distortion  creates a gap along the entire
non-interacting  nested Fermi surface (Sec.I) crossing this subband.
Hence this subband splits; and the lower split subbands is
filled at $x = 1$ and 
empty at $x=0$. Such a distortion corresponds to the frozen-in,
breathing-mode phonon observed in this system.  If the difference between
Bi and Pb  site energies were much smaller than this gap, then for
discussing the low-energy physics of this system it would
be sufficient to concentrate on the lower split subband only.  

In the bismuthates, actually this is
not the case; the Pb site energy is higher and the difference between
atomic levels of Pb and Bi is $\simeq 2.5$eV) Then the mapping of the
real system  onto the alloy is done by assigning the  PbO$_{3}$ cluster to
A atom and the BiO$_{3}$ cluster to B atom (and ignoring the existence of
the CDW for other purposes).

\noindent The resulting model  is given by the Hamiltonian

$$ H=  \sum_{<ij>\sigma}(t_{ij}+\frac{1}{2} \epsilon_{i}\delta_{ij})
c_{i\sigma}^{\dag}c_{j\sigma} -\sum_{i\sigma}  U
\hat{n}_{i\sigma}\hat{n}_{i - \sigma},\eqno(II.15)$$

\noindent with $\epsilon _{i} = \epsilon_{A(B)}$  for A (B) atoms.  The
randomness is treated in the CPA, whereas the attractive interaction is
treated via standard,  mean-field  approximation, it is i.e., assumed that
$<\hat{n}_{i}>=n_{A(B)}$ for i belonging to A(B);  n$_{A}$ and n$_{B}$ are
determined self-consistently.

\noindent In this approximation the mean-field  Hamiltonian  becomes

$$ H=  \sum_{<ij>\sigma}(t_{ij}+{\tilde \epsilon}_{i\sigma}\delta_{ij})
c_{i\sigma}^{\dag}c_{j\sigma} .$$

\noindent where\,\, ${\tilde \epsilon}_{i\sigma}=
\frac{1}{2}(\epsilon_{i}-U<\hat{n}_{i -\sigma}>) $

\noindent To study the effect of disorder, we need 
 $\varepsilon $,  the difference between the chemical potentials of the A
and B site in presence of the mean-fields, i.e.,

$$\varepsilon =\varepsilon^{(0)}-U(n_{A}-n_{B})$$

\noindent where $\varepsilon^{(0)}=\epsilon_{A}-\epsilon_{B}$. 
 The local densities of states at A and B sites are given in the CPA by

$$\rho_{A,B}=-\frac{1}{\pi}Im\left(\frac{F(z)}{1-(\epsilon_{A,B}-\Sigma
(z))F(z)} \right)_{z=E+i0} ,\eqno(II.16)$$

\noindent where the self energy calculated in CPA, 
is $\Sigma (z)=z-1/F(z)-F(z)/4$ and $$F(z)=1/N \sum_{k} G(k,
i\omega_{n}).$$

\noindent Now $n_{A}$ and $n_{B}$ are determined self consistently via

$$n_{A(B)} =\int_{ - \infty}^{E_{F}}dE\; \rho_{A(B)}(E)\;\;$$

\noindent and the lever rule $ (1-x)n_{A}+xn_{B}=x\/$. Finally,
assuming that there is a gap in the density of states in the
semiconducting phase, we get (at T = 0)

$$n_{A}=\frac{x}{4} w^{2}/\varepsilon^{2}\;\;{\rm and}\;\;
n_{B}=1-\frac{1-x}{4} w^{2}/\varepsilon^{2}.\eqno(II.17)$$

\noindent  where, $w$ is half  the noninteracting bandwidth, and thence 

$$\varepsilon=\varepsilon^{(0)}+U(1-\frac{w^{2}}{4\varepsilon^{2}})$$

\noindent which gives the condition that $\varepsilon^{(0)}$
should satisfy to produce a gap in the density of states.  Figures II.4a and
II.4b  show, respectively, the boundary between semiconducting and
metallic phases and the DOS.  The total DOS $\rho(E)$ (Fig.II.4b) is
obtained from $\rho (E) = (1-x) \rho_{A} (E) + x \rho_{B} (E)$, where
$x\/$ is determined self consistently from  Eqs. (II.17) and the lever
rule.  With the DOS thus obtained, the superconducting T$_{c}$ is
calculated from a Cooper instability calculation in the framework of the
CPA .  The expression for $T_{c}$ is formally the same as for a pure
system.  $$1=\frac{U}{2} \int_{-\omega_{D}}^{\omega_{D}} dE\; \rho (E)
\frac{tanh[(E-E_{F})/2T_{c}]}{(E-E_{F})},$$ 

\noindent while  $\rho(E)$ is determined in the manner described above.

Fig.II.5 shows  T$_{c}$ versus $x\/$  for different values of  $U$.  The
cut off $\omega_{D}$ is some phonon frequency whose value is taken to be
larger than the bandwidth in this calculation.  but the phase diagram
qualitatively the same. If the cut-off is reduced, T$_c$ decreases. Hence
depending on the parameter values of the system ($w,\, \varepsilon^{0}$
and $U$) the model system becomes either a superconductor or a
semiconductor as $x\/$ is varied. Note that the phase is symmetric about
$x=0.5$.

	The results obtained here do not really conform to the
experimental observations in the BaPb$_{1-x}$Bi$_{x}$O$_{3}$ system.  The
insulating phase here has no CDW order as observed in the system. This is
a very simple-minded model only  explore the possibility of a
semiconductor-superconductor transition but it may not be relevant for the
BaPb$_{1-x}$Bi$_{x}$O$_{3}$ system.
\vspace{5mm}

\noindent{\bf II.2.4  Ba$_{1-x}$K$_{x}$BiO$_{3}$  As A Doped
Peierls-Insulator}  
\vspace{3mm}

The models we have discussed so far are for the lead-doped bismuthate.
They were all put forward before the discovery of superconductivity in
Ba$_{1-x}$K$_{x}$BiO$_{3}$. Subsequent theoretical sork on this system has
been hampered by  the limited amount of experimental data that is
available.  We discuss below one of the few theoretical  attempts
 \cite{MJR} to understand the Ba$_{1-x}$K$_{x}$BiO$_{3}$ system.  The
objective of this study is limited to understanding the insulating
behaviour in BaBiO$_{3}$ system and its disappearence on  doping with
potassium.

	Rice and Wang \cite{MJR}  interpret the BaBiO$_{3}$ system as a Pierls
insulator which undergoes a semiconductor-metal transition on doping
because of the loss of the commensurate Peierls condensation energy. They
assume that the ground state at $x=0$ is a bond-ordered, CDW rather than
an onsite  CDW  i.e., a Peierls insulator in which there
is an alternation of bond lengths.  Experiments indicate the
possibility of 
 \cite{DEC,SP1,SP2,BB5} alternating Bi-O bond length in  BaBiO$_{3}$
(though this bond-length alternation can equally well arise from a site
diagonal CDW, as in the theories of Rice and Sneddon and Jurczek and
Rice). While considering the effect of doping,
and the consequent disappearence of the insulating, bond ordered phase,
they consider only the  Ba$_{1-x}$K$_{x}$BiO$_{3}$, as the lead-doped
system is diagonally disordered and a consideration of disorder is
absolutely essential for its  description. On the other hand, in the
Ba$_{1-x}$K$_{x}$BiO$_{3}$, potassium replaces barium and hence the effect
of disorder is probably less important.

They start with the tight-binding model  \cite{LFM1,LFM4}(\&
Sec. I.3)  involving the Bi\,6s conduction band  with $(1-x)\/$
electrons  per unit cell in the
Ba$_{1-x}$K$_{x}$BiO$_{3}$ system, hybridizing strongly with the three oxygen
2p$_{\sigma}$ bands (with six electrons  per unit cell ). The Hamiltonian
they use is: 

$$ H= \frac{\epsilon_{p}}{2}
\sum_{j\sigma}(b_{j\sigma}^{\dag}b_{j\sigma}-\frac{1}{2} 
\sum_{\delta} a_{j+\delta \sigma}^{\dag}a_{j+\delta \sigma})-\sum_{j\sigma
\delta} V_{j,j+\delta} (a_{j+\delta \sigma}^{\dag}b_{j\sigma}+H.c.)+C/2
\sum_{j \delta}u^{2}_{j+\delta,j} . \eqno(II.18)$$

\noindent Here $ b_{j\sigma}^{\dag}$ creates an electron in a Bi\,6s orbital
with energy $ \epsilon_{p}/2$ (the zero of energy is set midway between the
energies of Bi\,6s and O\,2p levels) at the site j of the reference cubic
lattice (noncubic distortions are ignored);\, $a_{j\sigma}^{\dag}$
is a similar operator for the O\,2p$_{\sigma}$ orbital (at energy
-$\epsilon_{p}/2$) and  
\,$j+\delta$ denotes one  of the six nearest neighbour sites of $j$; C
denotes the spring constant for the stretching of the ($j,\; j+\delta$)
band, described by $u_{j,j+\delta}$ the deviation from the equilibrium
separation

$$ V_{j,j+\delta}=sign(\delta)(V- \gamma_{c} u_{j,j+\delta}) \eqno(II.19)$$

\noindent is the hopping matrix element between Bi and O where  
-$\gamma_{c}$ is the derivative of $V_{j,j+\delta}$ with
respect to $u$. The $sign$ on 
the right hand side of Eq.(II.19) reflects the symmetry of the
$\sigma\/$ bonds between Bi\,6s and O\,2p.  
Here the phonon couples to the hopping of the electron, whereas in
the model introduced by Rice and Sneddon and Jurczek and Rice (Sec.II.2.2)
the phonon couples to the charge difference on neighbouring Bi
sites.  In the case $u_{j,j+\delta} = 0\/$ for  
all  $j,\/$ the alloy is a metal and H is easily diagonalized to give two
$\sigma\/$ bands and two non bonding dispersionless oxygen bands. In
BaBiO$_{3}\/$ (i.e., at $x=0\/$) the uppermost tight binding band
(which is half filled) is unstable with respect to the  commensurate Peierls
distortion 

$$ u_{i,i+\delta}=(-1)^{i}\Delta / \gamma_{c},\;\; {\rm for\; all} \;
i .\eqno(II.20)$$   

\noindent Here $\Delta$ is the amplitude of the dimerization and has  the
dimensions of energy. This dimerization leads to a splitting of both the  
original $\sigma$ bands and we get four subbands with energies
$E_{k,n=1 \cdots 4}$, given by

$${\rm E}_{k,n }=\pm
[\frac{\epsilon_{p}}{2}^{2}+6(V^{2}+\Delta^{2}) \pm 
[(12V\Delta)^{2}+ (V^{2}-\Delta^{2})^{2}\epsilon_{k}^{2}/V^{2}]]$$

\noindent where $\epsilon_{k}=zV\sum_{\bf a} e^{i{\bf k.a}}$\,\,\,
${\bf a}$ being the nearest-neighbour lattice vector.

 The Peierls gap $E_{p}$ is the gap between the upper two bands (or,
equivalently, the gap between the two lower bands) and

$$
E_{p}=[ \frac{\epsilon_{p}}{2}^{2}+6(V+\Delta)^{2}]^{1/2}-[
\frac{\epsilon_{p}}{2}^{2}+6(V-\Delta)^{2}]^{1/2} 
.\eqno(II.21)$$

\noindent At half filling, the three lower subbands are filled and the
topmost subband (from the splitting of the $\sigma^{*}$ band) is empty.
The total energy is  then given by 

$$ E_{tot}=3NC(\Delta / \gamma_{c})^{2}+2 \sum_{k}E_{k,1}.\eqno(II.22)$$

The Peierls distortion parameter $\Delta$ is determined from the
self-consistency condition obtained by setting 
$\partial E_{tot}/\partial \Delta=0$. A second derivative of $E_{tot}$ with
respect to $\Delta$ gives the restoring force for small displacements of
$u_{j,j+\delta}$ 
from its equilibrium value (as obtained from the self-consistency equation).

According to this theory, then, the insulating gap of
BaBiO$_{3}$ has to be equal to this Peierls gap. So the Peierls
gap is fitted to the optical gap (2eV) of BaBiO$_{3}$ to obtain
estimates for the parameters of this theory.  Using $
\epsilon_{p}/2=-1.1 $ eV and $V=2.2$ eV (values taken from the
band-structure results of Mattheiss 
and Hamann \cite{LFM3}, as discussed earlier) these authors
estimate the Bi-O spring constant $C=15.7$ eV/A$^{2}\/$ and the dimensionless
electron-phonon coupling constant $\lambda_{ep}={2\gamma_{c}^{2}}{3\pi CV}
=0.16\/$. With these choices they get 
the bond-alteration amplitude at $x=0\/$ to be $u\simeq 0.08$\AA which is
quite close to the experimental value \cite{RJC1,RJC2,BB1,BB2} in 
BaBiO$_{3}\/$. 

	Next, an estimate can be made for the critical concentration
$x_{0}$ of potassium at which the Peierls distortion vanishes and hence a
semiconductor-metal transition takes place via the vanishing of $\Delta$.
This is easily done by partially emptying the second top-most (filled at
x=0)  subband and
self-consistently calculating the value of $\Delta\/$.  $x_{0}\/$
turns out to be $11 \%\/$, considerably smaller than the experimental value
of $x=0.40\/$ for the Ba$_{1-x}$K$_{x}$BiO$_{3}\/$  system.

\vspace{5mm}

\noindent{\bf II.2.5  Local-Pair Superconductivity and Barium Bismuthates}
\vspace{3mm}

	The local-pair model for superconductivity has been studied in
detail by Micnas et. al \cite{RM} and several other people,  mainly in 
the context of the possiblity of bipolaron superconductivity  \cite{BKC,AA}
for a variety of systems (like Ti$_{4}$O$_{7}$) which show a CDW
ordering. They studied the negative-$U\/$, extended-Hubbard model in the
large\/$U\/$ limit. We do not review here the extensive literature on  negative-$U\/$
models which are not directly related to the bismuthates.

The possible application of this local-pairing idea to the barium
bismuthates has been pointed out by Varma \cite{CMV}. He proposed that 
the  barium bismuthates be modelled by a 
negative-$U\/$,  extended-Hubbard model with strong onsite interaction  $U$.
In this limit, the model becomes identical to the local-pair model 
as mentioned above.   He then noted that the
phase diagram of the negative-$U\/$, extended-Hubbard model 
is qualitatively  similar to that of the barium bismuthates.

In the early local-pairing papers the  negative-$U\/$ was
attributed to the electron-phonon interaction 
(as in Rice and Sneddon); however, the local
pairing and bipolaron mechanism requires an extraordinarily strong
coupling between electrons and phonons to produce an adequately large and
negative-$U$.  Subsequently Varma has pointed out the possiblity of an electronic 
origin for the negative-$U\/$ and, in a recent article  \cite{RM} 
on local pair superconductivity, this  alternative has also been alluded to.

The Hamiltonian studied  is  the extended Hubbard Hamiltonian:

$$ H=t\sum_{<ij>\sigma}c_{i\sigma}^{\dag}c_{j\sigma}-\frac{U}
{2}\sum_{i\sigma}\hat{n}_{i\sigma}\hat{n}_{i-\sigma}+\frac{V}
{2}\sum_{<ij>\sigma\sigma'}\hat{n}_{i\sigma}\hat{n}_{j\sigma'}.
\eqno(II.23)$$ 

	In the local-pairing model one takes $U\/$ to be larger than the
band width, so the negative-$U,\/$ extended-Hubbard model (II.2) can be
mapped onto the pseudospin model (II.5)  and the phase diagram  obtained
via mean-field theory.  Typical phase diagrams are
qualitatively similar to those observed for the lead- and potassium-doped
bismuthates.  There are three distinct phases: (pseudospin)
antiferromagnetic (CDW for model II.2), spin flopped (superconductor for
(II.2)), and nonordered (metallic for model II.2).  A first-order boundary
separates the superconducting phase from the CDW phase; all other
transitions are continuous.   In the $T - x\/$ plane the first-order
boundary should appear as a region of two-phase coexistence (although
reported earlier  \cite{AWS1}, such a region has not been
confirmed in later experiments with high-quality samples).  The
superconducting phase of this model consists of strongly bound,
onsite-paired electrons. A condensation of these {\em bosons}  gives rise
to supeconductivity.   The system is, of course, diamagnetic as the spin
degree of freedom is suppressed by the attractive interaction.  We will
discuss this model later in Section III and show that
superconductivity and a host of other phenomena in
the barium bismuthates cannot be described in terms of local pairing.

\vspace{7.5mm}
\noindent{\large \bf II.3 Other Related Work}
\vspace{5mm}

	There have been some other  recent attempts to understand the
superconducting phase and  high superconducting 
T$_c$ in the barium bismuthates.  An attractive interaction between
doped carriers in a CDW background can  be generated, in certain
circumstances, by the exchange of  CDW amplitude modes \cite{DG}.
The local depression of the CDW amplitude because of doping
leads to an attractive interaction for the doped carriers and mediates
pairing in $k\/$-space.  The superconducting transition is of the BCS type,
with the phonon cutoff frequency $\omega_{D}$ replaced by a much higher
scale of energy   (the frequency of the CDW amplitude
mode).  The formalism is analogous to its positive-$U\/$ counterpart, namely,
the spin-bag mechanism \cite{kampf} for high-T$_{c}$ Cuprates.

Some recent work \cite{DNM} has tried to understand the
semiconducting behaviour of the Ba$_{1-x}$K$_{x}$BiO$_{3}$
system starting from the model of Jurczek and Rice (Jurczek and
Rice \cite{EJ1}, see Sec.II.2.2). Their motivation 
was to estimate the various parameters of this model from the known
experimental results on Ba$_{1-x}$K$_{x}$BiO$_{3}.\/$

 They use the Jurczek-Rice Hamiltonian (II.3) and  assume that this system
is in a CDW ordered state in its semiconducting phase. There are two types
of sites in the system: Bi$^{3+}\/$ and Bi$^{5+}\/$
at half
filling with concentrations $c_{3+}=c_{5+}=0.5\/$ each. Doping by
potassium changes them to $(1-x)/2\/$ and $(1+x)/2,\/$ respectively. In
this picture the origin of the semiconducting phase is quite simple: The
Bi$^{3+}$ and Bi$^{5+}$ sites give rise to the lower and upper bands
containing $2c_{3+}\/$ and $2c_{5+}\/$ states and if there is a gap
between them, then the Fermi level is in the gap as the filling of the
lower band is 2$c_{3+}=1-x\/$. The system is then an insulator. Manh,, et
al. calculate the band structure of the system described above where they
account for the disorder due to potassium doping in the usual
Bethe-Peierls approximation and the interaction is treated in the
mean-field approximation.

	The local density of states is calculated by these authors in the
alloy approximation: They use a continued fraction representation for
one-particle Green function with three unknown parameters in the theory:
onsite attraction $U,\/$ nearest-neighbour Coulomb repulsion $V$, and the
effective electron-phonon coupling constant $\lambda_{ep}$ ($=gN(E_{F}$),
where $g\/$ is the electron-phonon coupling strength (Eq. (II.3)) and $
N(E_{F})\/$ is the DOS at the Fermi level.  All the energies
are scaled by the band width (the value used by the authors is the width
of the $\sigma$-antibonding band in Ba$_{1-x}$K$_{x}$ BiO$_{3},\/$ which
is nearly 6 eV from band structure calculations).  These authors try to
fit the experimental values of the optical gap $E_{G}\/$ at  $x= 0\/$
 where $E_{G} = 2\/$eV and at $x= 0.4\/$  where, $E_{G}  = 0,\/$
(i.e., where  Ba$_{1-x}$K$_{x}$BiO$_{3}\/$ 
becomes  metallic).  With the gap in the DOS obtained by them at those values
of $x,\/$ they conclude that for  $ -3eV <  U\/ < -2eV\/$, and
$\lambda_{ep} < 0.7\/$ produce a good fit for the optical gap.

\sloppypar
	All the theoretical studies discussed so far are based on models
where the onsite, Coulomb interaction is either suppressed
(Rice-Sneddon case), or the onsite $U_{eff}$ is either very
small or even negative and attractive, and the nearest-neighbour
Coulomb repulsion produces a CDW state.  There is,
however, a suggestion 
 \cite{GB} that these systems are positive-$U\/$ systems and their
insulating phase is be a spin-Peierls insulator, melting at
a high temperature into a resonating-valence bond liquid: One starts with a
strongly coupled spin and lattice system plus a Heisenberg-Hubbard model
for the spin and charge degrees of freedom.  Now such a spin-Peierls state
involving spins at each Bi$^{4+}$ ion will lead to a modulation in the Bi-Bi
distance. As far as we know there is no
indication of an alternating  Bi-Bi distance  in the structural work on
these systems. We do not present a detailed discussion of this model,
in this  review.

There is some work by Jin et al. \cite{jin} where a
molecular dynamics simulation of phonon density of states is
used to study the superconductivity in barium postassium
bismuthate within a strong coupling Eliashberg formalism. These
authors used a moderate electron-phonon coupling strength
($\lambda \simeq 1$) and assumed that strongth electron-phonon
coupling exists even at high phonon energies (30-60$meV$). The
\stc that they could reach are fairly high. 

\vspace{7.5mm}

\noindent{\large \bf II.4  Discussion }
\vspace{5mm}

Most of the theoretical work on the bismuthates starts with: the
negative-$U\/$, 
extended Hubbard model to account for  the experimental observation
 \cite{RJC1,RJC2,AWS2} of a  CDW
phase close to half filling and the occurrence of superconductivity on
doping.  The diamagnetic behaviour of these systems 
can  also be explained by using this model since the attractive interaction
suppresses the free-electron moments. 

	The theory of Sneddon and Rice was aimed at motivating the
attractive $U\/$ model for the barium bismuthate system. They did not try
to obtain experimental fits from this model: A qualitative description of
the pairing was their objective. In the model of Jurczek and Rice the more
realistic case of finite band-width and disorder was discussed. The
important result obtained from this work is that the CDW can survive
rather large doping. The T$_{c}$ obtained from this model was low for
realistic electronic DOS at $E_{F}\/$. Their model produces a pseudo-gap at
the Fermi surface, instead of a real gap as seen in the experiments, even
in the insulating side and does not account for the observed transport
properties in the insulating regime.

	The  model of Yoshioka and Fuku\-yama focus\-ses at\-ten\-tion on the
in\-sula\-tor-super\-conduc\-tor tran\-si\-tion; The con\-nec\-tion bet\-ween their
``alloy'' and the real BaPb$_{1-x}$Bi$_{x}$O$_{3}$ system is rather
far-fetched. The phase diagram obtained by them is symmetrical about $x\/$
and $1-x\/$, which is not the case for the real system.  The theory of
metal insulator transition in potassium doped system put forth by Rice and
Wang \cite{MJR}, although interesting, is not quite satisfactory as it does not
consider the effect of Coulomb interactions. Also the eventual development
of superconductivity at low temperatures in the metallic regime of these
systems is completely ignored.   They obtaine a semiconductor-metal
transition at a very low value of K concentration ($11
\% \/$) compared  to the experimental value ($\simeq 40 \%\/$).

	The local pairing mechanism  for the superconductivity in these
systems essentially involving the large\/$U\/$ limit of the
negative  $U\/$,
extended Hubbard model, can not produce reliable experimental fits of both
the superconducting and the normal properties.  It produces too small a
superconducting coherence length and too large an optical gap. The theory
of charge-bag can produce a superconducting T$_{c}$ that is large but does
not account for the observed unusual transport properties in the CDW or
semiconducting phases of these systems. The positive $U\/$ model
has not so far been used to calculate the normal and superconducting 
properties.  


	The unusual experimental observation of the wide separation
between the  optical and transport gaps  \cite{SU1} has not been addressed
by any of the theories we discussed.  We find that this two-gap feature,
coupled with other experimental observations, both in the superconducting
and the normal phase of the bismuthates, puts quite stringent constraints
on the models for these systems.  We have recently explained this
experimental observation  and developed a theory for the bismuthate
systems \cite{AT2,AT3} {\em which can explain many of the experimental
observations in these systems} on the basis of the negative-$U,\/$
extended-Hubbard model;  we discuss this in the following sections.

The negative-$U$   generated by an electron-phonon coupling is
retarded (and hence cut off in energy), whereas the negative-$U\/$
generated by electronic processes acts over the entire non-interacting
band-width. This makes the two mechanisms quite different, with
different experimental manifestations. This is another major issue we
focus on We argue that the two well-separated gaps can be understood only
if the attraction is electronic in origin and suggest experiments to
verify this assertions. We also calculate various normal-state and
superconducting properties from the negative-$U\/$ extended Hubbard model
and the values of the parameters of this model that  best fit experimental
data.

Though there are strong phenomenological grounds (Sec.
II.1) for using  the negative-$U\/$ extended Hubbard model for the
bismuthates,  there is no  clear derivation  of the negative-$U\/$.   In 
our work, it is used as a parameter of the theory. In Sec.III we
discuss the large-$U$ limit of this model where a mapping onto a
pseudospin, antiferromagnetic, anisotropic Heisenberg model (II.5) allows
us to investigate the spin model. We discuss briefly the
mean-field theory for this model and obtain various correlation functions in the
random phase approximation (RPA).  We also suggest a novel mechanism of
charge transport via pseudo-spin exchange. We discuss the limitations of
the large$-U\/$ limit and motivate the more interesting and appropriate
limit of intermediate $U\/$ which we take up in Section IV.

\vspace{1.5cm}

\noindent {\large \bf III. The Negative-$U\/$, Extended-Hubbard Model:}
{\large \bf Large-$U\/$ Limit} 

\vspace{5mm}
{\bf III.1: Summary of Results}
\vspace{5mm}

In Sec.II we motivated the use of the
negative-$U\/$, extended-Hubbard model (II.2) for the barium
bismuthates. It has been suggested  \cite{CMV} that, for a
preliminary understanding, these bismuthates be modelled by the
simpler Hamiltonian (II.5), since theoretical studies of this model
are far easier than those of model (II.2).  Model (II.5), an anisotropic,
pseudospin $S = 1/2,\/$ Heisenberg antiferromagnet in a pseudomagnetic
field, is the large$-U\/$ limit ($U\gg zt$) of the negative-$U\/$ (i.e.,
onsite attraction) Hubbard model (II.2) with nearest-neighbour
repulsion $V.\/$  (We follow the convention that $U\/$ and $V\/$ are
always positive; the sign of the interaction is displayed when
needed.)\,\, It has been studied via different approximations by many
workers  \cite{AT1,AT2,CMV,EJ1,RM}.
Here we examine 
the viability of this limit for the barium bismuthates by doing
detailed calculations for the phase diagram of this model and the 
properties -- thermodynamic functions, correlation functions, and response
functions -- of its metallic, semiconducting, and superconducting phases
 \cite{AT1,AT2} and comparing them with experimental observations.
Our overall conclusions are: (a)  On the positive side,
even the anisotropic-Heisenberg limit of the extended-Hubbard model
leads to various results which are either in qualitative accord with
experiments or which suggest new experiments for the elucidation of the
properties of these alloys. (b) However, it is not
good enough to obtain many important properties of these systems; 
its most serious limitation is that it leads to coherence lengths
in the superconducting phase that are far too small compared to
those obtained experimentally ($\simeq 70-80 A\/$).  We are thus led
to use the intermediate$-U\/$ limit of model (II.2), which we use in
Sec.IV to obtain better agreement with experiments.  We summarize
our large-$U\/$ results below: 

\begin{description}
\item[(1)] In the simplest, mean-field approximation, the anisotropic,
    $S = 1/2,\/$ Heisenberg antiferromagnet exhibits, typically,
    three phases, which, in the language of the extended-Hubbard model,
    are (seeSec.III for the correspondences between the two models):
    a nonordered metal (NO), a charge-ordered semiconductor (with a
    CDW), and an s-wave superconductor (SC).  Phase transitions
    between all excepting superconducting and CDW phases are
    continuous (second order); 
    the superconductor-CDW transition is first order.  Slight 
    perturbations of the model (say by the inclusion of
    next-nearest-neighbour 
    couplings) can yield an intermediate phase that exhibits
    both superconducting and CDW ordering  \cite{KSL}.
    At special values of coupling constants, bicritical, tetracritical,
    and tricritical points or critical end points can be obtained.
\item[(2)] A random-phase-approximation (RPA) calculation of the excitation 
    spectra in these phases shows the existence of a gap ($\simeq V$)
    in their semiconducting, CDW phase, which is different from 
    the gap ($\simeq U$) for the breaking of pairs in 
    the superconducting phase, which should appear in 
    the optical spectrum.
\item[(3)] It is useful to think of the phases in the following way:
    Bi$^{3+}\/$ (6s$^{2}\/$ configuration) corresponds to a
    real-space pair of two electrons and constitutes a charge-2e,
    hard-core boson. 
    The CDW phase is a crystal of these bosons, the superconducting
    phase a superfluid resulting from their bose condensation, and
    the nonordered-metallic phase their normal-liquid phase. 
\item[(4)] Charge conduction occurs by the motion of these charged
    bosons, or equivalently, by the exchange of pairs of Bi$^{3+}\/$
and Bi$^{5+}\/$ states. For this conduction mechanism the
    current operator has been constructed \cite{AT1} and thence, using the 
    RPA, the frequency-dependent conductivity $\sigma (\omega)\/$ can 
    be obtained \cite{AT2}. It is shown that:

\item[(a)] In the nonordered-metallic phase the conductivity $\sigma
(\omega)$ has  a Drude form, i.e.,\, $
\sigma(\omega)=n(2e)^{2}\tau/m^{*}(1-i\omega\tau),$ with $ \tau $ a
phenomenological transport lifetime that accounts for scattering, $m^{*}$
the effective mass, and $n$ the number of paired carriers; in the
grand-canonical ensemble, i.e., with fixed chemical potential,  goes to
zero exponen\-tially as the temperature goes to zero;  with  fixed (i.e.,
at  fixed filling), this means that the metallic phase does not extend
down to zero temperature at any filling (See Sec.IV).
\item[(b)] In the CDW semiconducting phase the conductivity shows an
activated  behaviour, 
       i.e.,\,$ \sigma (\omega\rightarrow 0 )\simeq \exp
(-\Delta/K_{B}T)$,\,where  $ \Delta$  is the gap (measured from the
chemical potential) in the spectrum of the hard-core bosons ( or, equivalently,
	the pseudospin-wave excitation spectrum ) in the CDW phase; this
	is in accord with experiments. 
\item[(5)] Exactly as in the case of the bismuthates, the CDW phase of the
        model exhibits two gaps - a charge gap
        associated with the breaking of\, Bi$^{3+}$  pairs, which costs an
        energy$\simeq U$  and another ( $ \Delta\simeq V $,  the
nearest-neighbour repulsion) associated with the activated behaviour of
$\sigma(\omega  \rightarrow 0 $). We calculated \cite{AT1} the optical
conductivity in the single- and two-particle channels using the
Kubo formula and the RPA in the large-$U$ limit. The result is
shown in Fig.III.4. It is interesting to note that even at a
simple level of approximation \cite{AT2}, the peak in the
two-particle channel (at $0.24$ eV) is down by a factor of 9
compared to that in the one-particle channel (at $2$ eV).  
\item[(6)] The metallic phase of this model is not a conven\-tional Fermi
liquid, but rather a charged, hard-core, bose liquid, and must have exotic
    properties. For example, the specific heat is certainly not a linear
    function of the temperature T. (A crossover to a conventional Fermi
    liquid occurs as $U\/$  decreases.) 
\end{description}

	We now show explicitly the calculations that lead to
the above results. The calculations are
described in two parts: The first part (III.2.1) is devoted mainly to 
to the mean-field theory of the anisotropic, pseudospin Heisenberg
model (II.5). In the following section (III.2.2) we show the
calculations that lead to the 
excitation spectra in RPA  in each of these mean field phases. We use the
RPA excitation spectra then to obtain various correlation functions. We
discuss the novel conductivity mechanism that we suggest for the
bismuthates in their insulating phase and calculate the corresponding
current operator in the large$-U\/$ limit.

\vspace{7.5mm}
\noindent {\large \bf III.2 Calculations: An Outline}
\vspace{5mm}

\noindent{\bf III.2.1  Mean-Field Phase Diagram}
\vspace{3mm}

	The phase diagram of the pseudospin model (II.5) is well known at the
level of the mean-field approximation; however, some
recent treatments are misleading  \cite{RM,SR1}, so we outline 
briefly how these phase diagrams are obtained  \cite{AT1}.

	The order parameters chosen are $<S_{i}^{x}>$ and $<S_{i}^{z}>$ 
where $i$ is the site index and can have only two labels A and B (indicating 
the two sublattices ). Also the rotational symmetry in pseudospin space
about the $z$-axis (the magnetic field  picks out the $z$ direction)
allows one to choose the spin to be in the $xz\/$ plane, i.e., $<S_{i}^{y}>=0$,
without any loss of generality. A straightforward mean-field
decoupling scheme is used to reduce the Hamiltonian (II.5) 
to the following site-diagonal mean-field Hamiltonian:

$$ {\cal H}_{mf}=J \sum_{<ij>}S_{i}^{x} <S_{j}^{x}> 
+  \sum_{<ij>}S_{i}^{z}(K <S_{j}^{z}>-B).
$$

This Hamiltonian is diagonalized by the following eigenfunctions
 
$$|i+>=\cos \frac{\theta_{i}}{2} |\uparrow> 
+ \sin \frac{\theta_{i}}{2} |\downarrow>$$ 
and 
$$|i->=\cos \frac{\theta_{i}}{2} |\downarrow> 
 - \sin \frac{\theta_{i}}{2} |\uparrow>,$$
where
$$\tan \theta_{i}=J_{0}<S_{j}^{x}>/(B-K_{0}<S_{j}^{z}>) $$

\noindent and $|\uparrow>$ and $|\downarrow>$ are the eigenvetors of 
$S^{z}$ operator with eigenvalues +1/2 and -1/2, respectively.
\noindent Here $J_{0}=zJ$ and  $K_{0}=zK$ and $\beta=(K_{B}T)^{-1}$.
The corresponding eigenvalues are 
$$\Lambda_{i}^{\pm}=-B \mp R_{j}$$
\noindent with
$$ R_{j}^{2}=(J_{0}<S_{j}^{x}>)^{2}+(B-K_{0}<S_{j}^{z}>)^{2}.
$$

The mean-field free energy per site is 

$$f=-\frac{1}{\beta}[\log (2 \cosh(\beta R_{A}))+\log (2 \cosh(\beta R_{B}))]
$$
$$+ J_{0}<S_{A}^{x}><S_{B}^{x}>-K_{0}<S_{A}^{z}><S_{B}^{z}>-B.$$

(This can also be obtained by using Peierls or Gibbs-Bogoliubov
variational method.  \cite{MG}
Finally the self-consistency equations follow easily by minimising the 
free energy with respect to the order parameters:

$$<S_{i}^{x}>=(J_{0}<S_{j}^{x}>/2R_{j})\tanh (\beta R_{j}).\eqno(III.1)$$

$$<S_{i}^{z}>=[(B-K_{0}<S_{j}^{z}>)/2R_{j}]\tanh (\beta R_{j});\eqno(III.2)$$

The subscripts $i$ and $j$ label nearest-neighbour sites; i.e.,\, they
belong to the  two different sublattices A and B (or B and A), 
respectively. If more than one solution is obtained 
for these self-consistency equations, then the one which yields the  
lowest value of the free energy  is picked.

Note that we can work with either fixed pseudomagnetic field (the analogue
of the grand-canonical or fixed-chemical-potential ensemble for the original,
extended-Hubbard model) or fixed $<S^{z}>$ (the analogue of the canonical
or fixed-filling ensemble);
the results obtained from both approaches have to be the same by
virtue of the equivalence of the two ensembles (see Note 2).

	The phase diagram in both canonical and 
grand-canonical ensembles have been obtained. A typical phase diagram 
that results from this mean-field 
calculation is shown, in Fig.III.1a, in the $B-T\/$ plane and, in
Fig.III.1b, in the $< S^{z} >-T\/$ plane. (Parameters for this phase
diagram are: $J_{0} = 1, K_{0} = 1.301$.) There are three phases: a
nonordered metal at large $B\/$ or $T\/$ (the pseudospin-model analogue is
a paramagnet with some alignment 
of the pseudospins by virtue of the finite magnetic field), a CDW\,
semiconductor (the pseudospin-model analogue: a ferrimagnet (antiferromagnet 
at\linebreak $B =0$)), and an s-wave superconductor (pseudospin-model
analog: a spin-flopped phase, with nonzero $x\/$ component of
pseudospins). The pseudospin-model analogues can be obtained easily
by consulting Table II.2.  In terms of the hard-core bosons
introduced above, these three phases correspond,  respectively, to a
bose liquid, a bose crystal, and a bose-condensed superfluid (note
3) phase.

Full lines in Fig.III.1a indicate continuous, second-order
transitions, 
whereas the dashed line in Fig.III.1a indicates a first-order phase boundary;
this  boundary shows up as a region of two-phase coexistence if $< S^{z} >$
(i.e.,\, the filling) is held fixed (Fig.III.1b). If the temperature and 
filling are  such that  the system lies in this two-phase 
region, then, in equilibrium, the  model system phase separates into two 
regions - one superconducting and the other semiconducting - with one 
interface between them. This two-phase region is not a new mixed phase with
both superconducting and CDW\, order. To obtain such a phase, 
next-nearest-neighbour interactions are required in the pseudospin model (II.5) 
(see below). Note that at no value of the filling does the metallic phase
extend down to $T = 0\/$. (This manifests itself in an
exponential depletion of the carrier density if one tries to approach 
$T = 0$ at fixed B\,.) The two second-order lines meet the first-order
boundary (Fig.III.1a) at a bicritical point. The positions of these
second-order lines can be obtained analytically  as described in Appendix B.
As Varma has noted\cite{CMV}
in this pseudospin-model limit the maximum $T_{c}$ for the
superconducting phase occurs at the value of filling for which the
metal-semiconductor transition first appears, i.e., at the filling at
which the bicritical point occurs.\, (Strictly speaking, this maximum
occurs at a slightly higher temperature.  The two critical lines come
in tangentially at the bicritical point, but this becomes evident
only from a renormalization-group calculation.)

It is well known from the work of Liu and
Fisher \cite{KSL} and  Matsuda and Tsuneto \cite{MS}  that 
slight modifications of the pseudospin model (II.5) can lead to more
complicated phase diagrams, at least at the level of mean-field theory. In
particular, if we allow for next-nearest-neighbour interactions of the
form $\sum_{<<ij>>} J^{\prime}_{ij} S_{i}.S_{j}$, where $<<ij>>$ indicates
that $i$ and $j$ are next-nearest neighbours, then, depending on the
relative sizes of the different interactions in the model, phase diagrams
of the sort shown schematically in Fig.III.2 can be obtained. (Such
next-nearest-neighbour interactions could arise from
next-nearest-neighbour repulsion $V^{\prime}$ or hopping $t^{\prime}$ in
the original extended-Hubbard model.) With such couplings it is possible
to get a bona fide intermediate phase in which both CDW and
superconducting order are present (Fig.III.2). As in Fig.III.1
solid lines denote 
continuous transitions and dashed ones first-order transitions.  Note that
a variety of multicritical points can be obtained now  \cite{DRN2,ADB}:
bicritical, tricritical, and tetracritical
points, and also a critical end point.  We mention these multicritical
points here and discuss them briefly in Appendix D\, mainly  for the sake
of completeness.  It is quite likely of course that the precise
multicritical behavior that obtains in the barium bismuthates depends
sensitively on effects that are not included in our model, such as
disorder, long-range Coulomb interactions and deviations from a cubic
crystal structure, etc.  Also, it will not be very easy to determine the
nature of such multicritical behavior experimentally, for that will
require very accurate control of both the temperature and the
concentration of the dopant.

 	The  phase boundary between the CDW and nonordered metallic phases 
belongs to the three-dimensional Ising universality class and the
superconductor-nonordered metal transition to the three-dimensional
$XY$ class. The transitions from the intermediate phase (Fig.III.2)
to the CDW and superconductor phases are, respectively, in  the
three-dimensional $XY$ and Ising universality classes. In the pseudospin
model,  the coherence length in the
superconducting phase is very small (see Appendix F), so apparent
mean-field exponents will not be seen here as they might well be in
the barium bismuthates, in which coherence lengths are of the order
of 80 \AA. 
\vspace{5mm}

\noindent{\bf III.2.2  Excitation spectra, Correlation function and the
Optical Conductivity in the Random-Phase Approximation}
\vspace{3mm}

	The excitation spectra in all the phases of the pseudospin
model can be obtained in a straightforward manner by using the 
equation-of-motion method followed by the random-phase approximation
 \cite{SBH} as shown in Appendix C. 

We use the Hubbard operators \cite{SBH} $\hat{P}^{i}_{\alpha
\beta}\equiv |i\alpha><i\beta|$,  where the state $|i\alpha>$ refers to
the states $|+>$ \/ or \/ $|->$ at the site $i$, in the
mean-field Hamiltonian described above. The double-time retarded
Green functions for these operators are denoted by
\mbox{$G^{m,n}_{\alpha\beta,\gamma 
\delta}(t-t^{\prime})$} or, equivalently, $<<\hat{P}^{m}_{\alpha\beta}(t)
|\hat{P}^{n}_{\gamma \delta}(t^{\prime})>>$. The Green function is defined
in the standard way as

$$ G^{m,n}_{\alpha \beta,\gamma \delta}(t-t^{\prime})=-i\theta(t-t^{\prime})
<[\hat{P}^{m}_{\alpha\beta}(t)
 ,\hat{P}^{n}_{\gamma \delta}(t^{\prime})]_{-}>\eqno(III.3)$$
\noindent where [P,Q]$_{-}$ denotes a commutator between the operators P
and Q.

The equations of motion for the Green functions are easily obtained in the
random-phase approximation  \cite{AT1,RM}.
The poles of these Green functions can be obtained simply from the
solutions to the equations of motions.  From these poles we can 
obtain the excitation spectra in the various phases of 
Fig.III.1. The details are discussed in the Appendix C.

	In Fig.III.3 we show the pseudospin-wave excitation spectra
in the three phases (respectively, CDW,\, superconductor, and nonordered
metal) of the phase diagram of Fig.III.1a. The spectra are drawn along the
symmetry directions of a cubic Brillouin-zone.  Data for these
spectra are given in the figure captions.  Note that there is a gap
as expected in the CDW\, (i.e.,\, ferrimagnetic in the pseudospin
language) phase.  This gap is obtained from the dispersion relation
in the CDW phase: 

$$\omega^{2}=2B-K_{0}m \pm [K_{0}^{2} \eta^{2}+(m^{2}-\eta^{2})J^{2}({\bf q})]
.\eqno(III.4)$$
\noindent Here $\eta=<S^{z}_{i}>-<S^{z}_{j}>$  is the staggered magnetisation.
The $\pm$ sign indicates that there are two branches in the spectrum, as 
should be the case in an antiferromagnetically ordered spin system. 
There is a gap in the spectrum of the non-ordered metallic
phase also at any finite magnetic field which
is merely the chemical-potential gap of the charged bose
liquid. (In the pseudospin language this is the gap
due to the aligning pseudomagnetic field.)  
From this expression it is easy to see that the gap at $T = 0\/$ 
in the CDW phase is 
$$ \Delta=[K_{0}^{2}-J^{2}(0)]^{1/2},\eqno(III.5)$$
\noindent Where

$$J({\bf q})=J_{0} \gamma_{\bf q},$$

$$\gamma_{\bf q}=\frac{1}{z} \sum_{\bf a} \exp (i{\bf q.a}), $$

\noindent with $\{{\bf a}\}$ is any one of the nearest-neighbour lattice
vectors, 

\noindent and
$$m\equiv n-1=<S_{i}^{z}>+<S_{j}^{z}>$$

The gap $\Delta$ vanishes (\,as does the CDW\, phase) as the anisotropy goes 
to zero, i.e.,\, as {\bf K}\, starts to become equal to {\bf J}. This is 
the transport gap $E_{A}$ that shows up in measurements of the dc
conductivity in the CDW\, phase see below.

	The excitation spectrum for the metallic phase also shows a 
gap. However, this does not lead to activated behaviour in the conductivity:
This gap is merely the chemical-potential gap of the charged bose
liquid that constitutes the metallic phase and is a manifestation of the
finite charge susceptibility at $q = 0$ and $T > 0$ (see Note 1).

	Various correlation functions can be obtained from the Green
functions that have been calculated.  For example, in the nonordered phase

$$<S^{+}({\bf q},\omega)S^{-}({\bf q},\omega)> =
\frac{1}{2\pi}(n-1)/(\omega-\epsilon({\bf q})),$$   
\noindent where
$$ \epsilon({\bf q})=2B-(n-1)K_{0}-(n-1)J({\bf q})
=\beta^{-1} \log (n/(2-n))-(n-1)J({\bf q}).$$

\noindent Similarly in the CDW phase

$$<S^{+}({\bf q},\omega)S^{-}({\bf q},\omega)>
=\frac{1}{\pi} \frac{[\omega-2B-2K_{0}(m+\eta)(m+\eta)]}
{(\omega-2B-K_{0}m)^{2}-[K_{0}^{2} \eta^{2}+J^{2}({\bf q})(m^{2}-\eta^{2})]}.
\eqno(III.6)$$

It is easy to check that at $\omega = 0$ these correlation functions 
diverge as $q\rightarrow0$ at the critical and multicritical points
of the phase diagrams of Figs. III.1 and III.2.  Here this divergence
is characterized, of course, by mean-field exponents; corrections to these
exponents and to the scaling forms for various thermodynamic and correlation
functions can be obtained from renormalization-group calculations as described 
in Appendix D.

	Given the Green functions that have been calculated, we can easily
obtain the frequency-dependent conductivity by noting that the only mode of
conduction in the limit of large $U\/$ is that in which the
charged boson comprising Bi$^{3+}$ moves or, equivalently, pairs of
Bi$^{3+}$ and Bi$^{5+}$ sites are exchanged.  The current operator for this
mode of conduction, (Appendix E and  \cite{AT1}), is given by:

$${\bf j}=-i(2e)J \sum_{<ij>}(\sip.\sjm-H.C.) {\bf a}.$$

	By using the Kubo formula  and an RPA-decoupling 
scheme we can evaluate the conductivity in all the 
phases (\,all effects of scattering
are contained in a phenomenological scattering time $\tau$). The result of
such a calculation for the metallic phase is of 
the Drude form

$$\sigma(\omega)=\frac{n(2e)^{2}\tau/m^{*}}{(1-i\omega\tau)},$$

\noindent where $n$ is the number of charge-2e carriers and $m^{*}$ is 
the effective mass.  

\noindent For the CDW\, phase,  as $\omega\rightarrow 0$ we get  
$ \sigma(\omega \rightarrow 0) \approx \exp(-\Delta/K_{B}T) $.
(This is the anisotropy gap in the pseudospin-wave excitation spectrum in  the 
antiferromagnetic phase.)
\vspace{7.5mm}

\noindent{\large \bf III.3 Remarks on the viability of the Large$-U\/$ limit for the
bismuthates}
\vspace{5mm}

	The phase diagram and the other results obtained above for the
pseudospin model have many features in common with the observed properties
of the barium bismuthates (Sec. III.2.1).  However, there are serious
limitations which, in our opinion, make this large-$U$ model not viable.
The negative-$U\/$  Hubbard model may be a more appropriate  limit  for
the barium bismuthates and may offer a more realistic description of these
systems if we choose intermediate values of $U$.

	The most serious limitation of the large-$U$\, pseudospin model is that
the coherence length in its superconducting phase is about one lattice
spacing.  This is obvious since the model corresponds to hard-core, charge-2e
bosons which are onsite-paired electrons. An explicit calculation
of this coherence length is  given in the Appendix F.  In contrast, the 
experimentally measured coherence length is much larger ($\simeq$ 80 A).

	One of the attractive aspects of the large-$U$\, pseudospin model
is that it yields the two gaps seen experimentally (see Secs. I and II) --
an optical gap $\simeq U$ and a transport gap $\simeq 2zV$.  From the
experimental data for BaBiO$_{3}$ one would therefore have to infer that
$U \simeq 2.0$ eV and $2zV \simeq 0.24$ eV.  However, from the DOS
 \cite{BB5} it can be estimated that (see Note 4) $zt = 0.8eV$, so
the large-$U\/$ assumption (i.e., $U\gg 2zt$) is clearly not true.  Also the
charge contrast (relative to uniform Bi$^{4+}$) implicit in the CDW\,
phase of the large-$U\/$ model is too large compared to its value in these
bismuthates. (The charge contrast has not been  
measured precisely in any experiment.  However, the value of the charge 
contrast that can be inferred from the relative sizes of the enlarged and 
contracted oxygen octahedra is much smaller than 
$\pm 1$, which the large-$U\/$ model predicts).

Both the shortcomings discussed above show that, if the
negative-$U\/$  model is at all viable for these bismuthates, it is so only
if the large-$U\/$ limit is abandoned.  Thus we must look at
the intermediate- and small-$U\/$ limits, which we discuss next
in Sec.IV. 


\newpage
\noindent{\large \bf IV. The Negative-$U\/$, Extended-Hubbard Model At
Intermediate $U$}

\vspace{5mm}

\noindent{\bf IV.1 Summary of Results}

\vspace{5mm}

In this Section we continue our discussion  of the viability  of
modelling barium bismuthates 
using the negative-$U\/$, extended-Hubbard model (with a 
repulsive interaction $V\/$ between nearest-neighbour sites) in the 
intermediate-U case. We obtain the phase diagram  of this model via
a mean-field theory and the compute the effect of fluctuations 
at $T = 0\/$ in the RPA. The details of our calculations are discussed
elsewhere \cite{AT1,AT2,AT3}. We discuss only the principal 
results here.   Furthermore, we comment on  the feasibilty of 
using the negative-$U\/$, extended-Hubbard model for these bismuthates  by
fitting a variety of experimentally measured quantities, such as the
optical and transport-activation gaps in the semiconducting phase, the
superconducting $T_c$ and  coherence length \cite{AT1,AT2,AT3}.

\noindent Our  principal results for the intermediate-$U\/$ are:

\begin{description}

\item[(1)] A phase diagram  (Figs. IV.1a and IV.1b) for the
intermediate-$U$ case, extended-Hubbard model by using a mean-field
approximation which is can distinguish between a variety of
phases: a semiconducting, charge-density-wave (CDW) phase, superconducting
phases with s-wave, extended-s-wave, or both of these types of ordering,
intermediate phases with both superconducting and CDW order,  and  a
nonordered,  metallic phase. This phase diagram has the same topology as
that  of the large-$U\/$ limit (Sec.III). Furthermore, it is in
qualitative accord with experimental phase diagrams in so far as it
exhibits three distinct thermodynamically stable phases:  a CDW phase, a
singlet superconducting phase, and a nonordered metallic phase.
Phase transitions between all excepting CDW and superconducting phases are
continuous; the superconductor-CDW transition is first order. 

\item[(2)] If the negative-$U\/$ is electronic in origin (see point (6)
below), then in the CDW phase there are two-particle and two-hole bound
states, i.e., {\em Cooperons}, with energies in the gap of the
two-particle spectrum, (Fig.IV.2b). 
We have calculated  the pairing susceptibility within 
the RPA in the
particle-particle channel \cite{AT1,AT2} to obtain these
bound-state (Cooperon)  energies in the CDW phase.In the limit of large $U,\/$
this energy evolves smoothly into  
the anisotropy gap in the pseudospin-wave spectrum (Sec.III).
The bound-state energy moves continuously towards
the chemical potential as $V\/$ approaches zero. 

\item[(3)] Charge transport in the CDW semiconducting phase is dominated
by the motion of these Cooperons. The large-$U\/$ limit  of this process
reproduces the conductivity mechanism in the pseudospin model (Sec. III).
Within the RPA the conductivity shows an activated behaviour with an
activation energy equal to the energy of the particle-particle bound
state. This is the transport gap.  It is thus clearly different from the
optical gap. For the latter is equal to the gap in the single-particle
spectrum, i.e., it is twice the CDW  gap parameter in the small-$U\/$
regime.  At large values of $U,\/$ this gap is $\simeq U$; note that this
is not the gap in the pseudospin-wave spectrum, but is the energy required
to break a pair.

\item[(4)] At large $U\/$ the pairs are tightly bound in real space and can be
thought of as a quantum lattice gas of hard-core bosons; at small $U$
k-space pairing is the proper description of these pairs.  This calculation
yields a natural interpolation between these two limits at $T=0$.

\item[(5)] The conductivity in the nonordered, metallic phase is of the Drude
form.  At large $U\/$ the carriers are charge-2e, hard-core bosons 
(Sec. III); at small $U\/$ the metal is, of course, a conventional
Fermi liquid.

\item[(6)] For the above explanation for the two gaps to be valid, the
attractive, onsite interaction must be electronic in origin: An 
attraction mediated by phonons is retarded and operates only over some
characteristic phonon frequency. Such an interaction cannot produce the
necessary binding to form the two-particle bound state as the CDW gap is
$\simeq$ 1ev, and any phonon-mediated attraction must be cut off well
below this energy range. 

\end {description}

	The remaining part of this Section is organized as follows: In
Section IV.2  we describe our mean-field and RPA  calculations in the
intermediate-$U$ range of the negative-$U\/$ Hubbard model very briefly.
In Sec.IV.3 we conclude with a discussion of the strengths and
weaknesses of the model, suggest future directions of study, and  propose
some experiments which can be used to verify our theory.

\vspace{7.5mm}
{\large \bf IV.2 Outline of the Calculation}

\vspace{5mm}

{\bf IV.2.1  Mean-Field Theory and The Phase Diagram}

\vspace{3mm}

The one-band, negative-$U\/$, extended-Hubbard Hamiltonian, 
defined on a simple cubic lattice, is given by

$$H = -t\sum_{<ij> \sigma}c_{i \sigma}^{\dag} c_{j \sigma} - \frac{U}  {2} 
\sum_{i \sigma} \hat{n}_{i \sigma} \hat{n}_{i -\sigma} + \frac{V} {2}
\sum_{<ij> \sigma \sigma'} \hat{n}_{i\sigma} \hat{n}_{j\sigma'} - \mu
\sum_{i\sigma} \hat{n}_{i \sigma}.\eqno(IV.1)$$ 

Though the qualitative features of the phase diagram of model (IV.1)
are known  \cite{SR2}, some earlier mean-field
treatments  \cite{SR2} of it contain some errors (see Note 5). 
The qualitative  features of the phase diagram of model (IV.1) remain the
same as those of the large-$U\/$ model;  however, there are quantitative
differences.  

 	The phase diagram (Figs. IV.1a and IV.1b) is obtained  
by using a mean-field approximation which is capable of distinguishing
between a variety of phases: a semiconducting, charge-density-wave (CDW)
phase, superconducting phases with s-wave, extended-s-wave, or both of
these types of ordering, intermediate phases with both superconducting and
CDW order,  and  a nonordered,  metallic phase.  The order parameters used
are:

$$<c_{i \uparrow}^{\dag} c_{i \downarrow}^{\dag}> \equiv b_{s}^{*},$$
 
$$<c_{i \uparrow}^{\dag} c_{j \downarrow}^{\dag}> \equiv b_{e}^{*},$$

$$<c_{i \sigma}^{\dag} c_{j \sigma}> \equiv C_{0},$$
\noindent and  
$$<\hat{n}_{i \sigma}> = {n \over 2} + {b_{c} \over 2}
e^{i{\bf Q}.{\bf r}_{i}}, \eqno(IV.2) $$

\noindent where ${\bf Q}\equiv\pi(1,1,1)$, $b_{c}$ is the CDW
order parameter, $b_{s}$ and $b_{e}$ are, respectively, onsite and
extended-singlet-superconducting order parameters, and C$_{0}$ a
self-consistent contribution to the band width. Only
translationally invariant order parameters are considered. A standard, 
Hartree-Fock decoupling,  yields the
mean-field Hamiltonian (we take the order parameters to be real)   

$$H_{mf} = -\sum_{{\bf k} \sigma}(\tilde{\epsilon}_{\bf k} + \bar{\mu})
c_{{\bf k} \sigma}^{\dag}
c_{{\bf k} \sigma} - {u_{p} \over 2} \Delta_{c} \sum_{{\bf k} \sigma}
c_{{\bf k + Q} \sigma}^{\dag}
c_{{\bf k} \sigma} - \sum_{\bf k} (\Delta_{s}({\bf k}) c_{-{\bf k}
\downarrow} c_{{\bf k} \uparrow} +
 c^{\dag}_{{\bf k} \uparrow}c^{\dag}_{-{\bf k}\downarrow})$$
$$ + N [U b_{s}^{2} - V z b_{e}^{2} + \frac{1} {4} n^{2} u_{m} 
+ \frac {1} {4} b_{c}^{2} u_{p} + V z C_{0}^{2}].\eqno(IV.3) $$

\noindent Here $N$ is the total number of lattice sites, $c_{{\bf
k}\sigma}$ is the Fourier transform of $c_{i\sigma}$, $u_{m}\equiv U -
2zV$, \, $u_{p} \equiv U + 2zV$, $\bar{\mu} \equiv \mu + u_{m}n,$
$\tilde{\epsilon}_{{\bf k}} = z \tilde{t} \gamma_{{\bf k}};\; \gamma_{{\bf
k}}\equiv  \frac {1} {z}
\sum_{{\bf a}}\exp(i{\bf k}. {\bf  a})$,      where
\{${\bf a}$\} are the nearest neighbour lattice vectors;$ \,
\Delta_{c}\equiv u_{p}b_{c}/2$  and  $\Delta_{s}({\bf  k})\equiv
Ub_{s}-zVb_{e}\gamma_{\bf k}$ are, respectively, the gap 
parameters in the CDW and superconducting phases.

This mean-field Hamiltonian (IV.3)  can be easily diagonalized and the four
eigenvalues are $\pm$ $E_{{\bf k}}^{\pm}$, where 

$$E_{{\bf k}}^{\pm} = [R_{\bf k} \pm 2 P_{\bf k}]^{1/2},$$

\noindent 
$$R_{\bf k} = \Delta_{c}^{2} + U^{2}b_{s}^{2} +
 V^{2}z^{2}b_{e}^{2} \gamma_{\bf k}^{2} + {\bar{\mu}}^{2} +
\tilde{\epsilon}_{\bf k}^{2},$$ 
\noindent and
$$P_{\bf k} = [(Ub_{s} Vz \gamma_{\bf k} b_{e} + {\bar{\mu}}
\tilde{\epsilon}_{\bf k})^{2} + 
 \Delta^{2}_{c} (u^{2}b_{s}^{2} + {\bar{\mu}}^{2})]^{1/2}.$$

With these four eigenvalues the mean-field free energy per site becomes

$$ f= - \frac {1} {N\beta} \sum '_{\bf k} \left[\ln  4 \cosh^{2}
\left( \frac {\beta E_{\bf k}^{+}} {2} \right) +\ln 4 \cosh^{2}
 \left( \frac{\beta E_{\bf k}^{-}} {2} \right) \right]$$ 
$$ + Ub_{s}^{2} - V z
b_{e}^{2} + \frac {1} {4} n^{2} u_{m} + \frac {\Delta_{c}^{2}}
{u_{p}} + z V C_{o}^{2} + {\bar {\mu}} (n-1)$$
 
  The order parameters
are determined  from the self-consistency 
equations that follow from a minimization of this free energy.
The prime over summation indicates a reduced brillouin zone.
A numerical, iteration method is used to solve the self-consistency
equations for the order parameters for different values of $U/zt$ and $V/zt$
($z$, the number of nearest-neighbour sites, is 6 in three dimensions) and 
at various values of filling $\delta=n-1$. 

The phase diagrams in both filling-temperature and chemical
potential-temperature planes are shown in Figs. IV.1a and IV.1b.
The topology of 
the phase diagram remains the same as in the large-$U\/$ case, but the
numbers obtained (for the positions of phase boundaries, etc.) are quite
different. There are three thermodynamically stable phases: A CDW phase
characterized by $\Delta_{c}\neq0$ and $b_{s}=b_{e}=0$; separated from
this CDW phase by a first-order line in the $\mu-T$ plane is the
superconducting phase (characterized by $b_{s}\neq0, b_{e}\neq0$ and
$\Delta_{c}=0$); and then there is a nonordered phase (all the order
parameters are zero in this phase) at high temperature (at $\delta=1$ this
phase goes down to $T=0$).  The phase transitions between the ordered phases
and the nonordered phase are all continuous. The two lines of continuous
transitions meet at a bicritical point. The first-order line in the
grand-canonical (i.e., fixed $\mu$, Fig.IV.1b) ensemble opens out into a
region of two-phase coexistence (CDW and superconducting phases) in the
canonical (i.e., fixed $\delta$, Fig.IV.1a) ensemble. $C_{0}$ is nonzero
in all the phases.  The hatched region in Fig.IV.1a is the two-phase
coexistence region.  In this two-phase region (and in equilibrium) the
system phase separates into semiconducting (CDW) and superconducting
phases, with one interface between them.

For $V=0$ and at half-filling (i.e., $\delta = 0$) in model 
(IV.1) the CDW and 
superconductor solutions of the self-consistency equations are connected
by a pseudospin-rotation symmetry (see Note 6) of the Hamiltonian and
hence are degenerate; their free energies are equal and they coexist. Away
from half filling this degeneracy is removed and the superconducting phase
becomes the thermodynamically stable one:  In the large-$U\/$ limit this
corresponds to the (pseudo-spin) antiferromagnetic phase giving way to the
(pseudo)spin-flop phase in a (pseudo)magnetic field.  The degeneracy
between CDW and superconducting phases is also lifted when $V$ becomes
nonzero (the pseudospin analogue, described in Table II.2, is
that the pseudospin model of Sec.III becomes anisotropic for
$V\neq 0$): at half filling the CDW phase is the 
thermodynamically stable one; of course, sufficiently far away from half
filling, the superconducting phase becomes stable.

The gap equation for the CDW gap parameter is of the usual BCS
form, from which the  CDW-nonordered transition temperature follows.
The  gap equation   is 

$$\frac {1} {u_{p}} = \frac {1} {2} \int_{-W}^{+W} \rho (\epsilon) \frac
{\tanh 
{\beta \over 2}[(\epsilon^{2}+\Delta_{c}^2)^{1/2}-\bar{\mu}]} {(\epsilon^{2}+
\Delta_{c}^2)^{1/2}}  d\epsilon,\eqno(IV.4) $$
 
\noindent where 2$W$ is the noninteracting band width.  At half filling
$\bar{\mu}=0$ and, if we take the  density of states $\rho(
\epsilon)$ to be a constant, this  becomes the 
usual BCS gap equation. 
At $T\equiv 1/\beta=0$ and with a constant DOS this gives the CDW 
gap parameter $\Delta_{c}= W/\sinh(2W/u_{p}).$ 

For the  
superconducting phase the calculations are  also of the BCS type. The
gap equation  now involves the  2$\times$2 matrix equation:

$$\tilde{M} = \left[{\bf 1} - \frac {1}{2N} \s'_{k} \frac {\tanh
({\tilde{\epsilon}}_{{\bf k}} + 
 {\bar {\mu}})} {(\tilde {\epsilon}_{\bf k}+{\bar {\mu}})} 
\left( \begin{array}{cc}  U & -Vz \gamma_{\bf k}\\
U\gamma_{\bf k} & -Vz \gamma^{2}_{\bf k} \end{array}
\right)\right].\eqno(IV.5) $$ 

\noindent where ${\bf 1}$ is the 2 $\times$ 2 unit matrix.
The determinant of the matrix $\tilde{M}$ should vanish at $T_{c}$. 
 Note that, in this mean-field theory, the maximum superconducting $T_{c}$
occurs  at  the bicritical point.
\vspace{5mm}

\noindent {\bf IV.2.2 The Cooperon Bound State In The Random
Phase Approximation} 
\vspace{3mm}

We showed in Sec.III that the experimental observation \cite{SU1}
of well-separated optical and transport gaps can be explained by
identifying the optical gap with the 
pair-breaking energy and the transport gap with the pseudospin-wave gap in
the CDW semiconducting phase.  {\it This explanation remains viable even in
case of   the  \negu, \exhb model for intermediate $V$}.

	As in the large-$U\/$ case the CDW phase continues to show two
well-separated gaps  for intermediate values of $U$ and $V$.  The physics
of these two gaps  is  as follows: Our mean-field
theory yields quasiparticle valence and conduction bands with a gap
separating them, (Fig.IV.2a). In ${\bf k}$-space  this gap  opens up
along the entire noninteracting  Fermi surface.  In the ground state, the
valence band is full whereas the conduction band is empty (\,at half
filling, i.e.,\, for pure BaBiO$_{3}$) and this system is
insulating.  Now consider putting two electrons into the conduction band.
They still feel the residual attraction  $U$ (modified by appropriate
coherence factors in general).  Hence they  form a bound state (a
Cooperon, i.e.,\,a charge-2e boson) in the gap in the two-particle
excitation spectrum, (Fig.IV.2b).  Two holes in the valence band form a
similar bound state.  This bound state forms provided the source of the
attractive $U$ is electronic   \cite{CMV} and not phononic since $U\/$ must
remain attractive over energy scales larger than the CDW gap $\Delta_{c}$,
which is 1 eV, as measured from the Fermi level. Also a phononic mechanism
cannot give rise to a binding of nearly 0.8 eV (required to fit
experimental observations; see below).  It is easy to see that charge (and
all other) transport in this CDW\, (semiconducting) phase will be {\em
dominated by these Cooperons}.  The thermal activation energy required for
exciting these Cooperons is the gap for exciting two quasiparticles minus
the binding energy of these Cooperons (i.e., the energy of the bound state
measured from the chemical potential) (Fig.IV.2b).  Hence it is lower
than (half) the semiconducting (\,CDW\,) gap 2$\Delta_{c}$. On the other
hand, optical experiments, which excite single particles across the\, CDW
gap, measure this larger gap, $\Delta_{c}$.

The existence of a two-particle bound state in the CDW phase is 
indicated by the appearence of a pole in the pair susceptibility
in the Cooper channel (at wave vector ${\bf q}=0$) below the
bottom of the continuum of the 
two-particle excitation spectrum  (the two-particle continuum starts at
2$\Delta_{c}$ (Fig.IV.2b).  

The (singlet) pair
susceptibility   in the Cooper channel is defined by
$$\chi_{p,ij;i'j'}({\bf q}, \nu) = \frac {1}{2\pi N^{2}}$$
$$\times \int_{-\infty}^{+\infty} d
\tau \exp(i\nu\tau) 
\sum'_{\bf k,k'} \langle T_{\tau} [ c_{{\bf k+q}\uparrow,i'}(\tau)c_{{\bf
-k}\downarrow,j'}(\tau) 
c_{{\bf k'}+q\uparrow,i}^{\dagger}(0) c_{{\bf
-k'}\downarrow,j}^{\dagger}(0) ] \rangle,$$ 

\noindent where $T_{\tau}$  represents time-ordering,  $i,j$, etc. are band
indices (in the CDW ground state there are two bands), the subscript $p$
stands for pairing, and the expectation value is calculated with 
respect to the interacting ground state.  We restrict ourselves to T=0.

We show in Fig.IV.3 the RPA ladder diagrams for the pair
susceptibility $\chi_{p,ij;i'j'}({\bf q}, \nu)$. We have
calculated the sum of these diagrams  \cite{AT1,AT2,AT3} in the
presence of a CDW background and starting from 
a negative-$U\/$, extended-Hubbard model at half filling.  From this the
two-particle (hole) excitation spectrum is obtained, and thence the
location of the Cooperon bound state.  It is shown that, as long as $V\/$
(more precisely, $\Delta_{c}$), is small  which is satisfied for the barium
bismuthates because the $\Delta_c \simeq$ 1eV), the bound state
falls within the 
gap of  the two-particle spectrum (Fig.IV.2b) and, for any nonzero $V$,
this state lies above the chemical potential (all energies are measured
from the chemical potential). Hence, so long as $V\neq$ 0, the CDW phase
is not unstable with respect to the superconducting phase at half filling.

We  show that the resulting two-particle bound-state energy in
the intermediate-$U\/$ regime evolves into  the pseudospin-wave gap of the
large-U model.  It is interesting to note  that the RPA excitation
spectrum calculated at intermediate values of $U\/$ evolves into the RPA
pseudospin-wave excitation calculated in the large-$U\/$ limit (for it is
not evident a priori that the  two RPA's are equivalent in all respects).

	In the presence of a CDW  background we have done  the analogue of
the Cooper-instability  \cite{LNC} calculation by following the
t-matrix approach  \cite{ABM}. This requires the solution of a secular
equation involving a 14$\times$14 matrix \cite{AT1,AT2}.  Had it been
possible to cast this equation into an equation of the type
$1/\lambda=\Phi(\nu)$ (where $\lambda$  is some effective, attractive
interaction), then its solutions could have been portrayed as shown
schematically in Fig.IV.4, i.e., as the intersections (empty circles) of
the horizontal line ( $y = 1/\lambda$, where $\lambda$ is negative since
the residual interaction is attractive) with the curves (thick lines)
representing $\Phi(\nu)$. The vertical lines (thin lines) are the
unperturbed CDW eigen-energies starting from $ \pm 2\Delta_{c}$. We have
taken the system to be  finite system for the purposes of this figure,  so
the energy levels are discrete.As the system size goes to infinity, the
eigen-energies of the interacting system which fall above  $ 2\Delta_{c}$
or below  -$ 2\Delta_{c}$ form the two-particle and two-hole continua
respectively, and correspond to scattering states; however, the two
eigenenergies inside the gap remain separated from these continua, and
correspond to the bound states (the one below the chemical potential is
the two-hole bound state and the one above is the two-particle bound
state).

	The energy of the bound state depends on $U,\/$ $V\/$ and $zt$. In
order to pin down at least some of the parameter values, we proceed as
follows:\newline 
(1) Set  $\Delta_{c}$ = 1eV, since the optical 
gap, which is twice the CDW gap parameter, is 2eV for BaBiO$_{3}$.  (We use
the mean-field gap equation and a constant density of states which makes
$\Delta_{c}=z\tilde{t}/\sinh(2z\tilde{t}/u_{p}),$ \, where $u_{p}=U+2zV$
and $z\tilde{t}$ is half the band width.)\newline
(2) The Cooperon bound-state energy (measured from the chemical
potential) is set equal  
to the transport gap of BaBiO$_{3}$, i.e., 0.24eV. The values of
$U$ (both screened and unscreened; see below)  and $V$
required to satisfy these two conditions for  different choices of the
values of $zt$ are plotted in
Figs. IV.5a and IV.5b. The main results obtained from  the above
calculations, relevant to the modelling of the barium bismuthates,  are
summarized below:

\begin {description}

\item[(1)] For the potassium-doped system the band width, close to the 
 metal-semi\-conductor transition  is roughly 1.6eV, as obtained from the
DOS (Batlogg, et al \cite{BB5}).  Given this value we obtain,
from Figs.IV.5a,b $U = 1.9$eV and $2zV = 0.13$eV. Clearly then
we must consider the intermediate-$U\/$ and not  the large-$U\/$ case.
\item[(2)] The two-particle bound-state energy goes to zero
continuously as $V$ 
approaches zero. At $V$=0 there is a gapless two-particle mode that
signals the instability of the CDW state towards superconducting order. In
Fig.IV.6 we show the values of the interaction-parameters obtained from
the two conditions described above,  for a fixed value of $W\/(=zt)$, as a
function of the bound-state energy. Note that  the extended-Coulomb term
$V\/$ goes to zero as the energy of the bound state approaches the
chemical potential.
\item[(3)] The bound-state energy  goes to $2zV$  as the band width approaches
zero (Fig.IV.5b). Note that in the large-$U\/$ limit the
anisotropy gap is exactly equal to $2zV$, a feature that is reproduced
correctly here as $ zt\rightarrow 0$.  This shows that the above  results
interpolate smoothly between the small- and large-$U\/$ results for the
extended-Hubbard model at $T = 0\/$.

\end {description}

	We now compare our results with experimentals. However, we cannot
naively use our estimates for the values of $U\/$ and $V\/$ to calculate
the superconducting and CDW transition temperatures for the following
reasons: Our estimates of $U\/$ and V have been made at the extreme
insulating limit of BaBiO$_{3}$, whereas superconductivity and the
metal-semiconductor transition occur after significant doping ( $\delta
\simeq  0.4$ for the potassium-doped system).  We must, therefore, try to
incorporate the effects of metallic screening on the strength of the
Coulomb interaction. In addition, doping has another  important
effect  on these systems. The arguments used for an attractive,
onsite interaction were based on nonlinear screening of the ionised
configurations of the Bi atoms by the highly polarizable O octahedra.
It is conceivable that, with increased doping, as the system moves towards
the metallic side and the lattice distortions and the
charge-disproportionated, highly ionized Bi configurations begin to
disappear, the mechanism which leads to the attractive interaction gets
suppressed rapidly. At this stage it is difficult to calculate these
effects, as we do not have a first-principles theory for the
negative value of $U$.

	In order to account for at least some of the physical processes
that result from doping, we have included the effect of metallic screening
at the simplest level via the RPA. The corresponding RPA screening
diagrams are shown in Figs.IV.7a and  IV.7b (repeated excitations of
particle-hole or particle-particle pairs, out of the Fermi sea ).  Note
that screening operates on both the particle-particle and particle-hole
channels. Hence the effective interactions responsible for CDW and
superconductivity are both reduced. Using the standard RPA procedure to
calculate the t-matrix we write the corresponding effective (screened)
interaction as ${\cal V}_{sc}({\bf q})={\cal V}({\bf q})/\left[1-{\cal
V}({\bf q}) \chi_{0}({\bf q})\right]$ where ${\cal V}$ is 
the bare interaction and $\chi_{0}({\bf q})$ is the appropriate bare 
susceptibility (particle-particle or particle-hole). Now the
particle-hole susceptibility at wave vector ${\bf Q}$ is the same as
particle-particle susceptibility at zero wave vector. Hence, in
the simplest approximation, we can replace $\chi_{0}$ by the density of states
$N(0)$ at the Fermi level and get the effective screened interaction
governing the CDW or the superconducting transition. For
$N(0)\/$ we use the experimental value at the metal-semiconductor boundary
of the potassium-doped system. 


By using the above procedure to account for screening, we plot the results
for the superconducting and CDW transition temperatures in Figs.IV.8a 
and IV.8b calculted via the BCS formula discussed earlier (with
both bare and 
screened interactions),  as a function of the band width. The results for 
the phase boundaries obtained using this procedure are discussed in
Appendix C and shown in Fig.IV.9.  Note that we have used
screening corrections only for the on-site 
Coulomb term. A calculation of the t-matrix for the
momentum dependent $V\/$ term 
is rather complicated and is not done here. Since the values of $V\/$ we are 
interested in are quite low (Fig.IV.9) compared to $U,\/$ it is
reasonable to neglect the screening corrections to $V.$ Indeed the 
dashed  line in Fig.IV.9 shows the superconductor-nonordered phase boundary
with $V$=0 (we show the results for two different set of parameters values,
and note that the parameter values that are appropriate  for the bismuthates 
correspond to Fig.IV.9). This shows the negligible effect $V\/$
has on $T_{c}$ in the range 
of interactions  that we are interested in. In passing we note that 
the effect of $V\/$ on
the superconducting  \stc vanishes as $\delta$ approaches zero \cite{AT1,AT2}.


	In any case, our results clearly show that the values obtained for the
superconducting and CDW transition temperatures are too high compared to
the experimental values. The effects of fluctuations neglected in our
mean-field theory and the effects of disorder could reduce these transition
temperatures by as much as a factor of 4, but probably not much more. It
is likely that the reduction of the strength of attraction with doping (as
described earlier) must be incorporated in our theory order to get better
experimental fits for the $T_c$. (This especially for the
superconducting transition temperature;  the mean-field, CDW transition
temperature is not orders of magnitude off from the experimental value,
lending support to our hypothesis that, as 
doping increases, the negative-$U\/$ is reduced.)

	In spite of this limitation of  large transition temperatures, our
theory yields quite a few interesting results.  Firstly, the large
difference between  the transport gap  the optical and gap can be
understood from the existence of the two-particle bound state in the CDW
phase. The conductivity is  given by\,
$\sigma(\omega)=(n{e^{*}}^{2}\tau/m^{*}) / (1-i\omega \tau) $, where
$m^{*}$ is the effective mass of the charge carriers, i.e., the Cooperons
and $e^{*}=2e$ is their charge.  
$n(T) \approx \exp(-E_{A}/K_{B}T)$ is the number of charge-2e carriers at
temperature $T$, where $E_{A}$ is the transport gap. Hence
$\sigma(\,\omega\rightarrow 0)$
shows an activated behaviour with an activation energy $E_{A}$. Secondly,
this theory clearly reproduces some of the qualitative features of the
phase diagram for the bismuthates (see the concluding discussion below).
In addition, the value of 2$\Delta/K_{B}T_{c}$ is 3.5, and the
system is
diamagnetic at low temperatures in our model. We have taken the standard
mean-field expression for the superconducting coherence length and used
the observed values of the Fermi velocity to obtain a coherence length of
nearly $30\AA$. Since our calculation grossly overestimates the
superconducting \stc, it is not surprising that our estimate for the
coherence length is quite small compared to the experimental value.
Nevertheless, it is much better than the extremely short coherence length
predicted by the large-$U\/$ (Sec.III) where the superconductivity arose
from onsite paired electrons.  Here, in the intermediate-U case, the
pairing is in momentum space and the pairs are spatially extended.

\newpage

\noindent{\large \bf IV.3 Concluding Discussions} 
\vspace{5mm}

In this and the preceding Sections we have investigated the
negative-$U\/$, extended-Hubbard model to model the barium bismuthate
superconductors. In our study we have  assumed that the mechanism that
leads to the negative value of $U\/$ is electronic rather than
phonon mediated. At the end of studying both large- and
intermediate-$U\/$ ranges we have been successful in
understanding some of these features associated with the
bismuthate systems.

Perhaps the most important aspect of our treatment is the
detailed theory we have provided for the occurence of two gaps
in the semiconducting phase of the barium bismuthates. We have
identified the transport gap as the energy of the two-particle
bound state ({\em Cooperon}) and the optical gap as twice the CDW gap.
This picture of the two gaps is consistent with the fact that
the transport gap is not easily accessible to optical
reflectance, photo-acoustic or photoconductivity studies, all of
which involve single particle excitations.  Note that {\em the
above picture implies that the semiconducting phase of the
bismuthates is unique, with transport properties dominated by
charge-2e bosons\/}.

Indeed, by fitting the theoretical gaps to the experimental ones
(Sec.IV.2)  we have estimated  the interaction parameters of the system. With
these parameter values  we have  calculated other normal-state and
superconducting properties. We have set up the formalism for the
calculation  of the particle-particle excitation spectrum  in the RPA for
the extended-Hubbard model in the CDW phase. In addition, we have
demonstrated that a smooth interpolation between the large- and
small-$U\/$ theories is possible at $T = 0\/$, and that the
particle-particle excitation spectrum we calculate at intermediate $U,\/$
evolves  smoothly  into the correct pseudospin-wave spectrum in the
large-$U\/$ limit. For typical intermediate values of $U\/$, and $V\/$
(i.e., when $U \leq 2zt$ and $V < U$)\,    our phase diagram in the
$T-x\/$ plane (Fig.IV.1a), has the same qualitative features as the
large-$U\/$ phase diagram (Fig.III.1) and shares many features with
experimental phase diagrams.  Furthermore, the charge-contrast in the
\cdw\, phase  gets reduced and the superconducting coherence length
enhanced over the  values obtained in the large-$U\/$ limit, in slightly
better accord with the experiments.   We have also argued that the
processes that give  rise to the attractive interaction, if electronic
in origin, must become less and less effective as
the system is doped towards the metallic side, so the strength of the
attractive interaction ought to be reduced in order to obtain the
experimentally observed superconducting $T_c$.


It is worth stressing that our theory disfavours a purely phonon-mediated
mechanism for the negative $U\/$ (unless an alternative explanation can be
found for the well-separated optical and transport gaps). A phonon
mediated-attraction is necessarily retarded (cut off at  characteristic
phonon frequencies $\simeq 100$ K); the  minimum energy range over which
the attraction must operate,  in order to produce the two gaps, is about
1eV. In addition, the large binding required to produce the two-particle
bound state at about 0.24eV above the chemical potential cannot come from
phononic mechanism.  It has been observed recently (Hinks, et
al., \cite{DGH2,DGH3}) that 
calculations based on realistic phonon  these systems can, at least at the
level of molecular-dynamics simulations, produce  rather large
superconducting $T_{c}$, without requiring a very strong electron-phonon
coupling  ($\lambda_{e-p}$ around 0.6-0.8) or a large DOS at the Fermi
level. The correct scenario could be that in the semiconducting phase of
these systems, the attraction is mainly electronic in origin;  however,
this electronic component of the attraction becomes weak as the system is
driven towards the metallic side (see discussion above),  and the
electron-phonon induced attraaction also becomes important in the
superconducting phase.

	According to our theory the carriers responsible for charge
transport  in the semiconducting phase of the bismuthate systems are
charge-2e bosons (Cooperons). Hence  experiments that can pick up the
charges of  carriers in the semiconducting phase would be of great value
in verifying our theory. Unfortunately the determination of the
charge of  carriers in  pure transport measurements is very difficult;
they generally yield only the {\em ratio} of the effective
charge and effective mass. 

However, the following is worth looking into. Stanton and
Wilkins \cite{CJS} have observed that the 
noise spectrum of the high-field transport in a semiconductor is
$S(\omega, E)=(n{e^{*}}^{2}\tau_{0})/(1+(\omega \tau_{0})^{2})
\times 
(A/2L)[k_{B}T_{0}/m^{*} +(e^{*}E \tau_{0}/m^{*})^{2}]  $, where $E\/$ is the
external electric field, $T_{0}$ is the equilibrium lattice temperature,
$\tau_{0}$ is the phenomenological relaxation time of the charge carrier
of effective charge $e^{*}$ and mass $m^{*}$, and $A$ and $L$ are
respectively the area and length (in the direction of transport) of the
sample. Now if the thermal- and electric-field dependent  components  of
the noise spectrum could be separated, the above expression for  $S(\omega
,E)$ would allow us to calculate the charge of the carriers and their
effective mass separately. To measure the noise spectrum, however,   one
must do a very controlled experiment at low temperatures on a pure,
single-crystal sample of BaBiO$_{3}$. A measurement of the phase change on
passing through a magnetic field (Bohm-Aharanov effect) can also, in
principle, detect the charge of the carrier, but such an experiment on
mesoscopic samples of  BaBiO$_{3}$ may be difficult to perform and hard to
interpret.  Even more likely possiblilties for observing the charge-2e
Cooperon may be in tunnelling  measurements from a superconductor into the
BaBiO$_3$ semiconductor, since pair tunnelling should now become viable.
 
We believe that such experimental work that confirms or rules
out this picture of the semiconducting phase of the barium
bismuthates would be of great interest.  A confirmation of this
transport mechanism in the semiconducting phase of the barium
bismuthates would confirm the electronic origin, as opposed to
the phononic origin, for the negative $U$.

We note that the phase diagram (Fig.IV.1), obtained for the
negative-$U\/$, extended Hubbard model in our mean-field theory
does not contain any intermediate phase (i.e., one with both CDW
and superconducting order), unlike the mean-field phase diagram
of Micnas, et al \cite{RM}. In the large-$U$ case earlier work (Liu and
Fisher \cite{KSL} and Matsuda and Tsuneto \cite{MS}) has shown clearly that, in
mean-field theory, such a phase is possible only in the presence
of longer range of interactions.

Needless to say, in spite of the attractive features discussed
above, the negative-$U\/$ extended-Hubbard model, as applied to
these bismuthates, has its limitations. The main shortcomings
arise because we have neglected electron-phonon interactions,the
long-range part of the Coulomb interactions, and disorder
(charge disorder in the case of Ba$_{1-x}K_{x}$BiO$_{3}$).  We
comment on these shortcomings below.

The neglect of Coulomb interactions is probably the main reason
for one important discrepancy between the experimental phase
diagram (Fig.I.2b) and the one obtained (Fig.IV.1) from the
model discussed above: The former has no two-phase region for $
x < 0.4$, unlike the latter.  Since the CDW\, and the
superconducting phases have different average Bi-occupancies (or
charges), long-range, Coulomb interactions strongly disfavour
any bulk phase separation. Instead they might favour the
formation, for small deviations from half-filling, of a new phase
consisting of a dispersion of globules of superconducting
material (which in the large-$U$\, limit and with small enough
globule sizes can be thought of as discommensurations,cite{CMV})
inside a semiconducting \cdw\,\, background. If the background
compensating charge (because of potassium ions) were smeared
uniformly, one might expect this dispersion to be a periodic
array. However, in the actual system the potassium ions are
distributed randomly, so the dispersion might well be random.
Such a phase would be semiconducting above a new critical
temperature (much lower than the critical temperature for the
bulk superconductivity) corresponding to the destruction of the
phase coherence between the different superconducting globules (
this temperature would be determined by the weak Josephson
coupling between the globules). It is interesting to speculate
that the semiconducting region, observed experimentally 
(Fig.III.2b) away from half-filling, corresponds to such a
dispersed phase.

Another possibility that one ought to explore, both
experimentally, and theoretically, even in the nearest-neighbour
model (IV.2) and the mean-field framework, is whether the
first-order coexistence region Fig.IV.1b gets replaced by an
incommensurate phase, or a succession of commensurate phases.

We must also account for the effects of disorder: In the
potassium-doped case since K$^{1+}$ replaces Ba$^{2+}$, in
addition to one electron being removed from the system (which we
have considered in our model), a random Coulomb potential is
also introduced (K is effectively a 1$^{-}$ impurity).  In the
simple theory presented here, we have neglected this random
potential.  The effect of randomness is even more serious in
lead-doped BaBiO$_{3}$ since the Pb atoms remove the
negative-$U\/$ centers of Bi randomly. Only very simple studies
of the effects of this randomness on CDW ordering have been
carried out so far (Jurczek and Rice \cite{EJ1} and Jurczek \cite{EJ2}). 

Since BaPbO$_{3}$ is a metal, it is clear that for a proper
study of the entire range of lead doping, the one-band model
(II.2) is inadequate and the more complex model (II.1) has to be
used. Also the electron-phonon interactions are strong in these
systems and has to be treated in a better way than has been done
so far, especially because of the structural transitions in
these systems. Clearly more detailed experimental and
theoretical studies are called for to address these issues.

It is worth noting in the end that there are some other
theoretical attempts recently using negative-$U$, extended
Hubbard model to understand the bismuthates. Ohkawa \cite{ohkawa}
has studied a negative-$U$, extended-Hubbard model using $1/d$
expansion in the slave boson approach \cite{barnes}. The
leading order effect in $1/d$ produces a "kondo" like "charge
fluctuation" regime discussed earlier in the context of an
attractive impurity model \cite{AT4}. A
mean-field analysis of this model produces, at a lower temperature, both CDW
and superconducting phases as we have shown
above \cite{AT1,AT2}. Ohkawa argues that the normal state of
barium potassium bismuthate represents a diamagnetic heavy
electronic system with a large ($\simeq 3-4\times 10^3$K)
charge-kondo temperature $T_K$. Superconductivity occurs in a band
of width 4$T_K$ in the usual BCS route (via Cooper pairing),
with the enhanced mass of the electrons leading to high $T_c$. 
A three band, negative-$U$ extended Hubbard model has also been
considered by Bala and Oles \cite{bala} recently leading to
a phase diagram similar to ours (as described in Sec.IV).
\vspace{5mm}

\noindent {\large \bf Acknowledgements} 
\vspace{3mm}

We thank the University Grants Commission and the Council of Scientific and
Industrial Research in India for research support. The
Supercomputer Education and Research Centre at Indian Institute
of Science, Bangalore is acknowledged for computational support.
Two of us (TVR and AT) would like to thank IFCPAR-408-2 for
partial support for work done in France. AT acknowledges helpful
discussions with  B. K. Chakraborty, F. Cyrot-Lackmann, C.
Escribe-Filippini, D. M. Gaitonde and S. Sarker. AT also
acknowledges support from Jawaharlal Nehru Centre for Advanced
Scientific Research, Bangalore.

\pagebreak
\centerline{\large \bf Appendices }
\vspace{7.5mm}

\noindent{\large \bf Appendix A}
\vspace{5mm}

\noindent{\bf The Pseudospin Hamiltonian as the large-$U\/$
limit of the Hubbard Model}

There are various ways of transforming the Hubbard model to a spin model
in the large-$U$ limit.
We use most straightforward route via second-order, degenerate,
perturbation theory, which is also physically transparent. Other methods
use a canonical transformation  \cite{CKi,DGr}.

\noindent Consider the negative-$U\/$  Hubbard model\hfill

$$
 H= -t \sumij c^{+}_{i\sigma}c_{j\sigma}-U\sum_{i} (n_{i}-1)^{2}.\eqno(A1)
$$

In the  large-$U\/$ limit, the first term can be taken as a perturbation 
and one can construct an effective Hamiltonian describing the low-energy 
physics.
    
	The full Hilbert space consists of four states per site: the
zero-occupancy, double-occupancy  and single-occupancy (with up and down
spins) states denoted by  $|0>$, $|2>$, $|\uparrow >$ and $|\downarrow>$,
have energies $-U$, $-U\/$ and 0 respectively.  The degenerate,  lower
energy states with double  and zero occupancy are separated from the
higher energy doublet of singly occupied states by the energy $U$. A
chemical potential term lifts the degeneracy between the low energy
doublet\, $| 0 >$ and $| 2 >$.  However, as long as $ U \gg \mu$, one can
do a degenerate perturbation calculation involving the nearly degenerate
lower energy manifold of unoccupied and doubly occupied states (see
below).

	We now use degenerate perturbation theory to obtain the
lowest-order (second order in $t$, as the first-order term vanishes)
effective Hamiltonian in the subspace of states\, $|0>$ and $|2>$,\, with
the singly occupied states $|\uparrow>$ and $|\downarrow>$  being virtual
(intermediate) states.

	It is clear that at this order, only the nearest-neighbour sites
will be connected by the perturbation and so we choose a
direct-product basis in the site representation: 
$|i0;j0>,$\,$ |i0;j2>, |i2;j0>, |i2;j2>$ 
where $|i0;j0>$ is the state with zero occupancy at 
both $i$ and $j$ sites, etc. The projected Hamiltonian in this subspace is

$$
  H=4\frac{|t|^{2}}{U}(| i 0;j 0 >< i 0;j 0|+| i 2;j 2 >< i 2;j 2|)
    +4\frac{t^{2}}{U}(| i 0;j 2 >< i 2;j 0| + h.c.).\eqno(A2)
$$

\noindent Now we define the operators 

$$ \sip \equiv|i2><i0|, $$
$$ \sib \equiv|i0><i2|,\eqno(A3) $$
$$ \siz \equiv\frac{(|i2><i2|- |i0><i0|)}{2},$$

\noindent where $|i2>$ represents the state with  2 electrons at the $i$-th. 
site and $  \sib=(\sip)^{\dagger}.$

These operators obey the commutation relations of  spin operators
and hence are called pseudospin operators. Note that $ \siz=\frac{1}{2}$
implies that the i-th site is a doubly occupied site while
$ \siz=- \frac{1}{2}$ implies it is unoccupied.

 	In terms of these operators the effective Hamiltonian becomes
$$ H_{pseudospin}= J \sum_{<ij>}({\bf S}_{i}.{\bf S}_{j}-\frac{1}{4}),
\eqno(A4)$$
\noindent where  $J=4t^{2}/U$.

This is an isotropic Heisenberg Hamiltonian. Deviations away from half
filling, modelled by a chemical potential $\mu \neq U$, will break the
pseudospin-rotation invariance  \cite{AT4} and lift the  degeneracy between
the states $|0>$ and $|2>$. So long as  $U \gg \mu$, the corrections to
the results of degenerate perturbation theory because of $\mu$ are $
\frac{t^{2}}{U}.\frac{\mu}{U}$ i.e.\, they are clearly higher order in $
\frac{1}{U}$. The term $V \sum_{<ij>} n_{i}n_{j} $, when projected onto
the subspace of $|0>$ and $|2>$, takes the form $4V \sum_{<ij>}
(\siz+{1\over 2})(\sjz+{1\over 2})$, which also breaks the pseudospin rotation
symmetry.

\noindent Thus the final effective Hamiltonian in the projected subspace is

$$
{\cal H}_{pseudospin}= J \sum_{<ij>}({\bf S}_{i}.{\bf
S}_{j}-\frac{1}{4})+4V\sum_{<ij>} S^{z}_{i} S^{z}_{j}
-B\sum_{i}(2S^{z}_{i} +1),\eqno(A5) 
$$

\noindent where $B=\mu+U-2zV$ and $z$ is the coordination number of the
lattice. 

\vspace{7.5mm}

\noindent{\large \bf Appendix B}
\vspace{5mm}

\noindent {\bf The Relation Between Pseudomagnetic Field and Magnetization in the
Different Mean-Field Phases and the Phase Boundaries Across Them}

	To obtain the pseudomagnetic field (and hence the chemical
potential) in the different mean-field phases obtained by the mean-field
theory  of the pseudospin model (Sec. III.2),  we use the following
thermodynamic identity 

$$
-\partial f /\partial B=n=<S_{i}^{z}>+<S_{j}^{z}>+1.\eqno(B1)
$$
\noindent The subscripts  $i$ and $j$ label sites on the two different
sublattices A and B  (or B and A).  This equation, in conjunction with the
self-consistency equations (Sec.III.2) yield the pseudomagnetic field in
in terms of the magnetisation $m$.

\vspace{5mm}
\noindent {\bf B.1 Pseudomagnetic Field}
\vspace{3mm}

\noindent {\bf Superconducting Phase}
\vspace{2.5mm}

	In the superconducting phase $<S_{A}^{z}>=<S_{B}^{z}>\equiv
<S^{z}>$
 and $|<S_{A}^{x}>|=|<S_{B}^{x}>|$. 
\,\,  Hence  $ R_{A}=R_{B}\equiv E $  (See Section III). Thus equation
(III.2) gives 

$$B=m/(\frac{\tanh^{-1}(\beta E)}{E}) + K_{0}<S^{z}>,\eqno(B2)$$

\noindent where $m=n-1$,\, is the magnetization, 
 $J_{0}=zJ$ and $K_{0}=zK$.

\noindent From the self-consistency equation (Sec.III)
$$<S_{A(B)}^{x}>=\frac{J_{0}}{2R_{B(A)}}<S_{B(A)}^{x}>\tanh (\beta R_{B(A)}).$$

\noindent For the superconducting phase

$$J_{0}^{-1}=\frac{\tanh (\beta E)}{2E}.\eqno(B3)$$

\noindent By combining equations (B2) and (B3), we get

$$B=(K_{0}+J_{0})m/2\eqno(B4)$$

\noindent in the superconducting phases.

\vspace{3mm}

\noindent {\bf Nonordered Phase}
\vspace{2.5mm}

Similarly for the nonordered phase ($<S_{A(B)}^{x}>=0$ and
$<S_{A}^{z}>\/ = \/<S_{B}^{z}> \equiv <S^{z}>$ ) whence
$$B=1+\beta^{-1}\tanh^{-1}m +\frac{k_{0}}{2}m.\eqno(B5)$$

\vspace{3mm}
\noindent {\bf Charge-Density-Wave Phase}
\vspace{2.5mm}

Here $ <S_{A}^{x}>=<S_{B}^{x}>=0.$ So, following
the procedure outlined above  we get

$$m=1/2[\tanh \beta (B-K_{0}/2(m-\eta ))+\tanh \beta (B-K_{0}/2(m+\eta ))],
\eqno(B6)$$
\noindent where $\eta=<S^{z}_{A}>-<S^{z}_{B}>$

\vspace{5mm}
\noindent{\bf B.2 Phase Boundaries}

\vspace{3mm}
\noindent {\bf Charge-Density-Wave -- Nonordered}
\vspace{2.5mm}

 The CDW-nonordered phase boundary is obtained by noting that the
staggered magnetisation\, $\eta=<S_{A}^{z}>-<S_{B}^{z}>$ goes to zero
continuously as this boundary is approached from the CDW side.  By
expanding the self-consistency equations about this boundary ($\eta
\rightarrow 0$ ) we get
 
$$k_{B}T_{CN}=\frac{K_{0}}{2}(1-m^{2}).\eqno(B7)$$

\noindent where $T_{CN}$ is the transition temperature.

\vspace{3mm}
\noindent {\bf Superconductor-Nonordered}
\vspace{2.5mm}

Here the order parameter $<S^{x}> \rightarrow 0\/$ as the phase
boundary is approached from the low-temperature side. An
expansion of the self-consistency equations yields

$$B=K_{0}m/2+J_{0} \tanh \beta(B-K_{0}m),\eqno(B8)$$

which along with Eq.(B4) gives 

$$T_{SN}=J_{0}m/(k_{B} \log [(1+m)/(1-m)]).\eqno(B9)$$

\vspace{7.5mm}

\noindent{\large \bf Appendix C}
\vspace{5mm}

\noindent {\bf The Equation of Motion for the Green Function in the RPA}
\vspace{3mm}
               
To obtain the excitation spectra in different mean-field phases,
we used a Random-Phase-Approximation technique \cite{AT1,RM} for the
closure of the equations of motion for the Green function (Sec. III).  We
use the Hubbard operators \cite{SBH} $\hat{P}^{i}_{\alpha
\beta}\equiv |i\alpha><i\beta|$. In the models we discuss $\alpha\/$ and
$\beta\/$ can take two values (+ and --) corresponding to the two
mean-field eigenstates
\,$|+>=\cos \frac{\theta}{2} |\uparrow> + \sin \frac{\theta}{2} |\downarrow>$ 
and  $|->=\cos \frac{\theta}{2} |\downarrow> - \sin \frac{\theta}{2}
|\uparrow>\/$.  The indices $i\/$ label sites on a bipartite 
lattice; with nearest-neighbour interactions, the only relevant 
indices are the two sublattice labels (A and B).  The commutation 
relation between these Hubbard operators is 

$$
[\hat{P}^{i}_{\alpha\alpha^{\prime}},\hat{P}^{i}_{\beta\beta^{\prime}}]_{-}=
\delta_{ij}(\hat{P}^{i}_{\alpha\beta^{\prime}} \delta_{\alpha^{\prime}\beta}
 -\hat{P}^{i}_{\alpha^{\prime}\beta} \delta_{\alpha\beta^{\prime}}),\eqno(C1)
$$

\noindent where [$\hat{R}$,$\hat{S}$]$_{-}\/$ represents a commutator between 
$\hat{R}\/$ and $\hat{S}\/$. In terms of these projection operators
the double-time, retarded Green function 
$G^{m,n}_{\alpha\beta,\gamma\delta}(t-t^{\prime})\/$ (also denoted by
\mbox{$<<\hat{P}^{m}_{\alpha\beta}(t)
 |\hat{P}^{n}_{\gamma \delta}(t^{\prime})>>,\/$}  \cite{DNZ} is

$$ G^{m,n}_{\alpha\beta,\gamma\delta}(t-t^{\prime})=-i\theta(t-t^{\prime})
<[\hat{P}^{m}_{\alpha\beta}(t),
\hat{P}^{n}_{\gamma \delta}(t^{\prime})]_{-}>.\eqno(C2)
$$

\begin{sloppypar}

	This Green function gives the probability amplitude for
the process in which, at time $t^{\prime},\/$ the $n^{th}\/$ site
makes a transition from  the state $|n \delta>\/$ to the state $|n
\gamma>\/$ followed by another transition, at site $m\/$ and a later
time $t,\/$ from the state $|m \beta>\/$ to the state $|m \alpha>\/$.
The Fourier transform of this Green function in time is 

\end{sloppypar}

$$G^{m,n}_{\alpha \beta, \gamma \delta}( \omega)
=\frac{1}{2 \pi}\int_{-\infty}^{\infty}
 dt G^{m,n}_{\alpha \beta, \gamma \delta}(t-t^{\prime})\exp
i\omega(t-t^{\prime}).\eqno(C3) 
$$

Correlation functions of two projection operators are  expressed
in terms of the Green function in the usual way:

$$<\hat{P}^{m}_{\gamma \delta}(t^{\prime}) \hat{P}^{n}_{\alpha \beta}(t)>=
\int_{-\infty}^{\infty} d \omega [G^{m,n}_{\alpha \beta, \gamma
\beta}( \omega-i0)  - G^{m,n}_{\alpha \beta, \gamma \beta}(
\omega+i0)] f_{B}( \omega),\eqno(C4)$$ 

\noindent where $f_{B}(\omega)\/$ is the Bose occupation factor
$(\exp \omega/k_{B}T-1)^{-1}.\/$ 

Since any local operator can be represented in terms of the
projection operators as $ A_{i}=\sum_{\alpha \beta}(A_{i})_{\alpha
\beta} \hat{P}_{\alpha \beta}^{i},\/$ the pseudospin Hamiltonian can
be written as 

$$
H=-\sum_{i,\alpha \beta} u_{\alpha \beta}^{i}\hat{P}_{\alpha \beta}^{i}
-\sum_{ij,\alpha \alpha^{\prime}, \beta \beta^{\prime}} W^{ij}_{\alpha
\alpha^{\prime}, \beta \beta^{\prime}}
\hat{P}_{\alpha \alpha^{\prime}}^{i}\hat{P}_{\beta
\beta^{\prime}}^{j},\eqno(C5) 
$$
\noindent where
$$
u_{\alpha \beta}^{i}=B[2(S_{i}^{z})_{\alpha \beta}+1] 
$$
\noindent and

$$
W_{\alpha \beta, \gamma \delta}^{ij}=
\frac{J}{2}[(\sip)_{\alpha \beta} (\sjm)_{\gamma \delta}
+(\sib)_{\alpha \beta} (\sjp)_{\gamma \delta}]
-K(\siz)_{\alpha \beta}(\sjz)_{\gamma \delta}.\eqno(C6)
$$

An obvious symmetry of $W_{\alpha \beta, \gamma \delta}^{ij},\/$ which 
we use, is 
$W^{ij}_{\alpha \alpha^{\prime}, \beta \beta^{\prime}}=W^{ji}_{\alpha
\alpha^{\prime},  \beta \beta^{\prime}}.\/$  The equation of motion
is, therefore 

$$
i\partial G^{ij}_{\alpha \alpha^{\prime},
 \beta \beta^{\prime}}(t-t^{\prime}) / \partial
t=\frac{1}{2\pi}\delta (t-t^{\prime}) 
 <[\hat{P}^{i}_{\alpha\alpha^{\prime}}, \hat{P}^{j}_{\beta\beta^{\prime}}]_{-}>
+<<[\hat{P}^{i}_{\alpha\alpha^{\prime}},H]_{-} 
| \hat{P}^{j}_{\beta\beta^{\prime}}>>.\eqno(C7)
$$

If we substitute Eq. (C5) in Eq. (C7), rearrange dummy   
indices , and use Eq.(C1) and the symmetry  of $W^{ij}_{\alpha
\alpha^{\prime}, \beta\beta^{\prime}}\/$, we obtain the following 
equation of motion for the  Fourier transform (i.e., in ${\bf
q},\, \omega$-space) of the Green function.  

$$
\noindent \omega  G^{ij}_{\alpha \alpha^{\prime},
 \beta \beta^{\prime}}({\bf q}, \omega)=\frac{1}{2\pi}
<\delta_{ij}(\delta_{\alpha\beta^{\prime}}\hat{P}^{i}_{\alpha\beta^{\prime}}
-\delta_{\beta\alpha^{\prime}}\hat{P}^{i}_{\beta\alpha^{\prime}})>$$
$$-\sum_{\mu \mu^{\prime}}<u^{i}_{\mu
\mu^{\prime}}(\hat{P}^{i}_{\alpha \mu^{\prime}} 
\delta_{\alpha^{\prime} \mu}-\hat{P}^{i}_{\mu \alpha^{\prime}}
\delta_{\alpha \mu^{\prime}}) \hat{P}^{j}_{\beta \beta^{\prime}}>
-\sum_{k,\mu \nu \nu^{\prime}}W^{ik}_{\alpha^{\prime} \mu,\nu\nu^{\prime}}
<<\hat{P}^{i}_{\alpha \mu} \hat{P}^{k}_{\nu \nu^{\prime}} 
 | \hat{P}^{j}_{\beta \beta^{\prime}}>>$$
$$ +\sum_{k,\mu \nu \nu^{\prime}}W^{ik}_{\mu \alpha, \nu \nu^{\prime}}
<<\hat{P}^{i}_{\mu \alpha^{\prime}} \hat{P}^{k}_{\nu \nu^{\prime}} 
 | \hat{P}^{j}_{\beta \beta^{\prime}}>>.\eqno(C8)
$$

In terms of the Hubbard operators, the RPA is defined by the
decoupling in which a three-operator Green function is factorised into 
products of two-operator Green functions and expectation values of
the remaining operator, as shown below:

$$
<<\hat{P}_{\alpha \alpha^{\prime}}^{m} \hat{P}_{\gamma \gamma^{\prime}}^{l}
| \hat{P}_{\beta \beta^{\prime}}^{n}>>_{\omega}=\delta_{\alpha \alpha^{\prime}}
<\hat{P}_{\alpha \alpha^{\prime}}^{m}>
G^{ln}_{\gamma \gamma^{\prime}, \beta \beta^{\prime}}(\omega) 
+\delta_{\gamma \gamma^{\prime}}
<\hat{P}_{\gamma \gamma^{\prime}}^{l}>
G^{mn}_{\alpha \alpha^{\prime}, \beta \beta^{\prime}}(\omega)$$
$$+\delta_{lm} [\delta_{\alpha^{\prime} \gamma}G^{mn}_{\alpha \gamma^{\prime},
 \beta \beta^{\prime}}(\omega)-\delta_{\gamma^{\prime} \alpha}
 G^{mn}_{\gamma \alpha^{\prime},
 \beta \beta^{\prime}}(\omega)].\eqno(C9)$$

In our model (C5), the last term on the right hand side of
Eq.(C9), arising out of the commutation of $ \hat{P}_{\alpha 
\alpha^{\prime}}^{m}\/$ and $ \hat{P}_{\gamma
\gamma^{\prime}}^{l},\/$  vanishes as there is no 
interaction connecting the sites on the same sublattice
(the delta-function $\delta_{lm}\/$  requires this but
$W^{ij}_{\alpha\alpha^{\prime},\beta\beta^{\prime}}=0\/$ for
$i=j\/$). In a model with interactions connecting 
the same sublattices (e.g., a model with further than nearest-neighbour 
interactions) this term has to be considered.

By using the RPA factorisation in Eq.(C8) we obtain 

$$
(\omega-\varepsilon^{i}_{\alpha \alpha^{\prime}}) G^{ij}_{\alpha
\alpha^{\prime},  \beta \beta^{\prime}}({\bf q},
\omega)=L^{i}_{\alpha^{\prime} \alpha} \sum_{k,\mu \nu} 
W^{ik}_{\alpha\alpha^{\prime}, \nu \mu} G^{kj}_{\mu \nu,
 \beta \beta^{\prime}}({\bf q}, \omega)
+\frac{1}{2\pi} L^{i}_{\alpha \alpha^{\prime}}
\delta_{ij} \delta_{\alpha \beta^{\prime}}\delta_{\beta \alpha^{\prime}}
\eqno(C10)$$

\noindent where
 
$$\epsilon_{\alpha}^{i}=u_{\alpha \alpha}^{i}+\sum_{l,\gamma} 
W_{\alpha \alpha, \gamma \gamma}^{il}({\bf q}=0)
<\hat{P}^{l}_{\gamma \gamma^{\prime}}>,\eqno(C11)
$$

$$\varepsilon^{i}_{\alpha \alpha^{\prime}}=\epsilon_{\alpha}^{i}
-\epsilon_{\alpha^{\prime}}^{i},\eqno(C12)
$$

\noindent and

$$
L^{m}_{\alpha \alpha^{\prime}}=<\hat{P}^{m}_{\alpha \alpha}>
-<\hat{P}^{m}_{\alpha^{\prime} \alpha^{\prime}}>.\eqno(C13)
$$

Since the site indices can have 2 values (denoting the 2 
sublattices), and since $\alpha\/$ and $\alpha^{\prime}\/$  can have
only two values, the most general form of the above equations of
motion here is a $4\times 4$  matrix equation  given by 

$$
 \left(  \begin{array}{cccc}
           \omb      & 0         & \paqp  & \paqm   \\
           0         & \opb      & -\paqm & -\paqp  \\ 
           \pbqp     & \pbqm     & \oma   & 0	    \\
           -\pbqm    & -\pbqm    & 0      & \opa    
          \end{array}  \right)
	 \left(  \begin{array}{c} 
	    \gabpm \\    
	    \gabmp \\
	    \gbbpm \\
	    \gbbmp 
	         \end{array}  \right) $$

$$ = \frac{1}{2\pi} \lpm \left( \begin{array}{c}
		           0    \\
		           0    \\
		           \rpp \\
		           -\rmm 
		         \end{array}  \right),
\eqno(C14)$$

\noindent where $\beta \beta^{\prime}$ take values $+- $ or $-+$ and

$$ \Pi^{i}_{\pm {\bf q}}={1\over 2} [\pm J({\bf q})(1+\cos \theta_{A}
\cos \theta_{B})-K({\bf q})\sin \theta_{A} \sin
\theta_{B}] L^{i}_{+-}.\eqno(C15)
$$ 

The sublattice indices can be dropped in the superconducting and
nonordered phases (i.e., there is no broken lattice-translational
symmetry). This matrix equation (C14) reduces to a 2$\times 2$ matrix 
equation in the CDW and the superconducting phases and to a
scalar equation for the nonordered 
metallic phase because of the symmetry of these phases. 

\vspace{7.5mm}
 
\noindent {\large \bf Appendix D }
\vspace{5mm}

\noindent {\bf Scaling forms for Thermodynamic Functions near
Multicritical Points} 
\vspace{3mm}

In the phase diagrams of Figs.III.1 and III.2 the following
multicritical points occur: bicritical points, tricritical points,
tetracritical points and critical end points. For the sake of completeness
we give the scaling form  \cite{AAA} for the singular part of the free
energy in the vicinities of these points.

\vspace{3mm}
\noindent {\bf Bicritical and Tetracritical Points}
\vspace{2.5mm}

Bicritical and tetracritical points are studied easily using the same
model, namely, the Landau-Ginzburg free-energy functional
 \cite{DRN2,ADB,JMK}.

$$
{\cal F}_{LG}=\int d^{d}x [\frac{1}{2}( (\nabla {\bf \psi})^{2}+
 r_{0}|{\bf \psi}|^{2})+u|{\bf \psi}|^{4}
-\frac{1}{2} g(\psi_{1}^{2}-\frac{1}{2}(\psi_{2}^{2}+
\psi_{3}^{2}))+v\sum_{i=1}^{3} \psi_{i}^{4}],\eqno(D1)$$

\noindent where ${\bf \psi}$ is a three-component, vector order parameter with
 components
$ \psi_{i},\,\, i=1\dots 3 $.  A bicritical point is obtained if $v<0$ 
and a tetracritical point if $v>0$. The temperature-like scaling
 variable $t$ depends linearly on the deviation of $r_{0}$ from
its value at the multicritical point.

The singular part of the free energy ${\cal F}_{s}$
assumes the following scaling form in the vicinity of a bicritical point
which occurs at the temperature $T_{b}$:

$$
 {\cal F}_{s}=(T-T_{b})^{2-\alpha_{b}} {\cal
F}_{b}(\frac{g}{(T-T_{b})^{ \phi_{g}}}, 
 \frac{v}{(T-T_{b})^{ \phi_{v}}}),\eqno(D2)
$$

\noindent where the mean-field values of the exponents are
$\alpha_{b} = 0 ,\phi_{g} 
= 1$  and  $\phi_{v} = 0$.  The specific-heat exponent $\alpha_{b}$ 
is related to the 
correlation length exponent $\nu_{b}$ via the hyperscaling relation
$d\nu_{b}=2 -\alpha_{b}$, where $d$ is the dimension ( the mean-field 
value of $\nu_{b} = 1/2$ ).\, To leading order in $\epsilon = 4 - d$
the bicritical exponents are

$$\nu_{b}=1/2(1+5 \epsilon/22),$$
$$\phi_{g}=1+3\epsilon/22,\eqno(D3)$$
$$\phi_{v}=-\epsilon/22.$$  

Note that, even though $v$ is irrelevant, it should be displayed
explicitly since it is a {\em dangerous\/}
irrelevant variable.     More details regarding
this point and a comprehensive discussion of the (similar) scaling 
in the vicinity of tetracritical points can be found elsewhere
 \cite{DRN2,AAA,MEF1}.

For the anisotropic antiferromagnets, that we are interested in,
the Landau-Ginzburg free energy is an expansion in powers of {\bf M} and
{\bf M$_{s}$}, the magnetization and the staggered magnetization. However,
in the vicinity of the bicritical point, such a free energy can always be 
put in the form (C1) given above.

\vspace{4mm}	
\noindent {\bf Tricritical Point}
\vspace{2.5mm}

In the vicinity of a tricritical point the Landau-Ginzburg 
free-energy functional can be written as  \cite{MEF1,DMu,DRN1,ED1,ED2}

$$
{\cal F}_{LG}=\int d^{d}x [\frac{1}{2}( (\nabla \psi)^{2}+
a_{2} \psi^{2})
+a_{4} \psi^{4}+\frac{1}{6} a_{6} \psi^{6}
-H \psi], \eqno(D4)
$$

\noindent where $\psi$ is a scalar order parameter. The tricritical point
occurs at $H=0, a_{2}=a_{4}=0$, and $a_{6}>0$ ( $a_{2}$ depends linearly
on temperature ).  The scaling form for the singular part of the free
energy ${\cal F}_{s}$ has  the following form in the vicinity of the
tricritical point $T_{t}$:

$$
{\cal F}_{s}=(T-T_{t})^{2-\alpha_{t}} {\cal
F}_{t}(\frac{g}{(T-T_{t})^{ \phi_{t}}}, 
 \frac{H}{(T-T_{t})^{\phi_{v}}}),\eqno(D5)
$$

\noindent where, in mean-field theory, the exponents are $ \alpha_{t}=1/2 $,\, 
 $\phi_{t}=1/2$, $ \Delta_{t}=5/4$, and the scaling field $g$ is defined
to be proportional to $a_{4}$. Since the upper-critical dimension for
tricritical behaviour is 3, corrections to mean-field theory are
weak logarithmic ones  \cite{AAA,MEF1}. 

\noindent {\bf Critical End Point}

The simplest Landau-Ginzburg theory which yields a critical end
point is defined by the functional  \cite{AAA,TALZ,JRB,MEF2}

$$  {\cal F}_{LG}=\int d^{d}x [\frac{1}{2}( (\nabla \psi)^{2}+
 r\psi^{2})+u\psi^{4}+v\psi^{6}+w\psi^{8}],\eqno(D6)
$$

with $v<0, w>0,$ and $\psi$ a scalar order parameter.  (For the critical
end point shown in Fig.III.2 a more complicated Landau-Ginzburg functional
would be required, but Eq.(D6) suffices for the purpose of illustrating
the essential features associated with this point.)  If $\psi$ is uniform
and we use mean-field theory (i.e., we minimise ${\cal F}_{LG}$ with
respect to $\psi$), then a critical end point is encountered at
$u=\frac{v^{2}}{4w}$  and $r=0.$  As this point is approached from the
disordered phase, thermodynamic functions, such as the order-parameter
susceptibility $\chi$, diverge as they do along the critical line which
terminates at the critical end point (Fig.III.2); i.e., at the
mean-field level $\chi \sib r^{-1}$ as $r\rightarrow0$ (as usual r
depends linearly on the deviation of the temperature from the temperature
at which the critical end point occurs).  However, the order parameter
jumps from zero to $(|v|/2w)^{1/2}$ as we go from the disordered phase to
the ordered phase through the critical end point. This behavior does not
depend on the mean-field approximation. Even renormalization-group
calculations  \cite{TALZ,JRB,MEF2} yield the same qualitative behavior:
Thermodynamic functions like $\chi$ diverge with the same exponent at a
critical end point as they do along the critical line which terminates at
this point  and the order parameter jumps at the critical end point.
\vspace{7.5mm}

\noindent{\large \bf Appendix E}
\vspace{5mm}

\noindent {\bf The Current Operator in the Pseudospin Model}

 The effective Hamiltonian in the pseudospin model is (see Appendix A) 

$$ H=4{|t|^{2}\over U} \sum_{<ij>}\siz.\sjz
+4\frac{t^{2}}{U} \sum_{<ij>}(S_{i}^{+} S_{j}^{-}
+H.C.)+4V\sum_{<ij>}\siz.\sjz$$
$$-B\sum_{i}(2\siz +1).\eqno(E1)$$

To couple this to an electromagnetic field we use the Peierls
substitution  $ t \rightarrow t \exp -i\int_{i}^{j}{\bf A}.d{\bf l}$
where ${\bf A}$ is the vector potential ($\hbar=c=1$). Such
a substitution will obviously leave the first term
in H unaffected and the only contribution to the current comes from the 
transverse part of the pseudospin-pseudospin interaction term.

	By expanding $H$ to linear order in ${\bf A}$, and noting
that a minimal coupling to the current is of the form ${\bf j.A}$, we
can easily identify the current operator for the above model as

$$
 {\bf j}=-i(2e) \sum_{<ij>} J{\bf a}(\sip.\sjm-H.C.),\eqno(E2)
$$

\noindent where ${\bf a}$ is the nearest-neighbour vector and 
$J=4{|t|^{2}\over U}$.

	There are alternative ways of 
deriving the current operator (of course with identical results) 
for the projected Hamiltonian . For example, we can construct the 
dipole operator ${\bf P}$ for this model and then use the relation 
${\bf j}= -i[{\bf P},H]_{-}$ to get the current operator. 	
\vspace{7.5mm}

\noindent{\large \bf Appendix F}
\vspace{5mm}

\noindent {\bf The Ginzburg-Landau Equation and the Super\-conduc\-ting
Co\-herence Length  for the Pseudo\-spin Model}

	The Ginzburg-Landau  free-energy functional for the pseudospin
model in terms of the order parameters $<\sip> = \psi_{i}$ and
$2<S^{z}_{i}> = (n-1)=m$ is:

$${\cal F}[\psi]=-J/N\sum_{<ij>}(\psi^{\dagger}_{i} \psi_{j}+H.C.)+K_{0}m^2$$
$$+(\beta N)^{-1}\sum_{i}[(S_{i}-1/2) \log (S_{i}-1/2)
+(S_{i}+1/2) \log (S_{i}+1/2)],\eqno(F1)$$ 
where N is the number of sites, and
$$S_{i}^{2}=|\psi_{i}|^{2}+m^{2}/4.\eqno(F2)$$

	It is then easy to show that an expansion to order $\psi^{2}$
gives the phase boundary between the superconducting and nonordered
phases. To this order the free energy per site is

$${\cal F}[\psi]= -J/N\sum_{<ij>}(\psi^{\dagger}_{i} \psi_{j}+H.C.)
+(\beta N)^{-1}\sum_{i}|\psi_{i}|^{2}/(m \log n/(1-m)). \eqno(F3)$$

\noindent which of course gives the same expression for $T_{CN}$ as
Eq.(B8)

	Note that the coupling
to the electromagnetic field is expressed in terms of the following gauge
transformation on the bond $(ij)$,

$$\psi_{i}\psi_{j} \rightarrow \psi_{i}\psi_{j} \exp -i(2e) \int_{i}^{j}
{\bf A}.{\bf dl}.\eqno(F4)$$ 

\noindent where ${\bf A}$ is the vector potential.
Note that the current operator (Appendix E) follows from a 
functional derivative of $\cal F$ with  respect to ${\bf A}$

	To calculate the coherence length, we take the magnetic field $H$
to be in the $z$ direction, so that, in the Landau gauge, the vector
potential is ${\bf A}=(0,-Hx,0)$. Then 
$$\psi_{l}\psi_{m} \rightarrow
\psi_{l}\psi_{m} \exp [-i\phi_{lm}]\eqno(F5),$$ where

$$\phi_{lm}=2eHl_{x}a^{2}.\eqno(F6)$$
\noindent Here $a$ is the lattice constant and $l_{x}$ is the
$x$-coordinate of the $l$-th\, site;

 Close to the transition, by expanding the Ginzburg-Landau free
energy  and retaining terms upto order of $\psi^{2}$, we obtain

$${\cal F}[\psi]= -J/N\sum_{<ij>}(\psi^{\dagger}_{i} \psi_{j}+H.C.)
+T/T_{SN}(0)\sum_{i}|\psi_{i}|^{2},\eqno(F7)$$

\noindent where T$_{SN}$(H) is the transition temperature in a magnetic
field H.
	
An expansion of the first term on the right hand
side gives the $(\nabla \psi)^{2}$ term. By equating the coefficient of
this term to $\xi^{2}$  (where $\xi$ is the superconducting
coherence length) we obtain $$\xi^{2}=a^{2}/\tau,\eqno(F8)$$ 
where $\tau=1-T/T_{SN}(0)$. (Equivalently, one can write down the
equation for  $\psi_{i}$ and compare it with the eigenvalue equation 
of a quantum oscillator.)

Hence, as we have already mentioned in the text, the coherence
length well away from the transition is about one a lattice
constant. This result is meaningful from a physical point of
view as, in the large-$U\/$ theory, there is no length scale in 
the problem other than the lattice constant.

\newpage


Note 1. Some other correlation functions are gapless in the superconducting
    phase reflecting merely the XY-model nature of the ordering; however, 
    they do not correspond to any physically measurable function.

Note 2. In an earlier treatment  \cite{RM} it was claimed
	 that a
      mixed phase with both CDW  and                
      superconducting order
      could be obtained within mean-field theory for model    
       by working
      with fixed $<S^{z}>$ (the analogue of the canonical ensemble).
       This is
      not correct. The phase diagrams obtained from both
       fixed-$B$ (i.e.,\, 
      grand canonical ensemble) and fixed-$<S^{z}>$
       (i.e.,\, canonical 
      ensemble) calculations have to be the same by the 
      equivalence of the
      two ensembles. The first-order boundary of Fig.III.1a
      (obtained by 
      controlling B) appears as the two-phase coexistence region
       of Fig.III.1b
       (obtained by controlling $<S^{z}>$); and the discontinuity
        of $<S^{z}>$
      across the the first-order boundary of Fig.III.1b yields
       directly the 
      boundaries of the coexistence region of Fig.III.1a. 
      The only difference
      between the two ensembles is the following: At 
      coexistence and with
      fixed $<S^{z}>$ , the system phase separates into 
      two phases--one 
      superconducting and the other a semiconducting, 
      CDW                       
      phase--with one interface between the two. 
      (The fractions of the two 
      phases follow from the lever rule.) However, 
      with fixed $B$ on
      the first-order boundary), either one or the other 
      phase fills the
      whole system (since $<S^{z}>$ is not fixed) and there 
      is no interface;
      formally one or the other can be picked out by allowing 
      $B$ to approach
      the first-order boundary from either the 
      superconducting or the
      CDW side. As we have discussed in 
      the text, to obtain\linebreak[2]
      a bona fide mixed phase with both\linebreak supercoducting and 
      charge-density-wave order, next-nearest\--neighbour 
      interactions must be added to the pseudospin model. This has also
been pointed out recently by  \ldots

Note 3.  The bose condensation of the hard-core bosons is observed 
	from the
    behavior of the chemical potential as one approaches the
    nonordered--singlet superconductor phase boundary from the 
    high-temperature side. The chemical potential loses  explicit
    dependence on temperature on entering the superconducting phase from
    the high-temperature disordered phase 
    signifying the onset of a bose condensation of the hard-core bosons.
    The pseudomagnetic field in the singlet superconducting
    phase is given by [Eqn (B4)] \mbox{$B=(K_{0}+J_{0})<S^{z}>$} and 
    is clearly independent of temperature.

Note 4. From the width ($\simeq 6$ eV) of the antibonding band of the LAPW,
      band-structure calculation  \cite{LFM1,LFM2}. The
value of the hopping 
      matrix element is obtained from the  experimental values of the
      (renormalized) DOS at the Fermilevel  \cite{BB5}  when
fitted to a 
      tight binding band with square density of states, it is less
      ($\simeq 0.20$ eV) than the value (0.3 eV) obtained from the 
band-structure calculation.

Note 5. In earlier work  \cite{SR2} the
 only a singlet superconducting order parameter
b$_{s}$. In a negative-$U\/$, extended-Hubbard model-context, 
choice is naive as we have shown here; both b$_{s}$ and b$_{e}$ are
nonzero in the superconducting phase.  

Note 6: This symmetry is quite easily understood: The
negative-$U\/$ Hubbard model can be transformed into the corresponding
positive-$U\/$ model,  as described earlier. (If the
chemical potential of the negative-$U\/$ model is non zero (i.e., the system
is away from half filling) this transformation leads to a positive-$U\/$,
Hubbard model with a magnetic field coupling to the spin.)  The positive-$U\/$ 
Hubbard model has a global SU(2) spin rotation symmetry that mixes the up and 
down spins  (Affleck, et al. (1988)), whence the $x(y)$ ordered  and 
the $z$ ordered antiferromagnetic 
(SDW) states are degenerate. In the negative-$U\/$ context, this
symmetry is a pseudospin rotation symmetry  \cite{AT4}
which means that the superconducting state is degenerate with the CDW
state at all filling. [However, as the negative-$U\/$ model away from half
filling maps onto a positive-$U\/$ Hubbard model with a magnetic field, this
symmetry between the $x(y)$ and the $z$-ordered states is lifted, i.e., 
the corresponding symmetry in the negative-$U\/$ model,
between CDW and superconducting states
is destroyed away from half filling.
\newpage
\vspace{2cm}
\begin{center}
{\bf Table I.1 }\\ 
Normal State Properties of BaPb$_{1-x}$\/Bi$_x$O$_3$ \\ 
\begin{tabular}{|l|r|r|r||c|}
\hline 
\multicolumn{4}{|c|}{Metallic} & Semiconducting \\ \hline
$x$ & $0.0$ & $0.12$ & $0.2$ &  $1.0$ \\ \hline\hline 
$\rho (\Omega {\rm cm}) \times 10^{-4}$ & $2.10$ & $4.15$ & $5.40$ & \\
\hline 
$n ({\rm cm}^{-3})$ & $1.42 \times 10^{20}$ & $16.0 \times
10^{20}$ & $20.0 \times 10^{20}$ &
\multicolumn{1}{|p{4cm}|}{Prefactor $110 \times 10^{20}$ (activa\-ted
with gap $0.24$ eV\/)} \\  \hline  
$\mu ({\rm cm}^2 {\rm v}^{-1} {\rm s}^{-1})$ &  $210.0$ & $9.40$ &
$5.80$ & \\ \hline
$\lambda_{\rm mf} (A)$ & $220.0$ & $22.0$ & $15.0$ & \\ \hline 
$m^{\star}/m_e$ & $0.16$ & $0.32$ & $0.23$ & \\ 
\hline 
\end{tabular}

\vspace{2cm}
{\bf Table I.2}\\
Superconducting Properties of BaPb$_{1-x}$O$_3$ 
and Ba$_{1-x}$K$_x$BiO$_3$ \\
\begin{tabular}{|l||c|c|}
\hline 
& Ba$_{0.6}$K$_{0.4}$BiO$_3$ & BaPb$_{0.75}$Bi$_{0.25}$O$_3$ \\
\hline \hline
$ dH_{C_2}/dT\,\bigm |_{T_c}$ & $-5 \pm 0.5 \,{\rm KOe}/^{\circ}{\rm K}$ &
$-5.3 \,{\rm KOe}/^{\circ}{\rm K}$ \\  \hline 
$ dH_C/dT\,\bigm |_{T_c}$ & $-75.0 \pm 8.0 \,{\rm Oe}/^{\circ}{\rm K}$ & $
-50 \pm 5 \,{\rm Oe}/^{\circ}{\rm K}$ \\ \hline
$-\chi_{\rm GL}$  & $45.0 \AA$ & $80.0 \AA $ \\ \hline 
$\xi_{\rm GL}$ & $70.0 \AA $ & $80.0 \AA $ \\ \hline
$\Delta C/T_c$ & $2.1 \pm 0.4 \,{\rm mJ/mole K}^2$ & $0.94 \,{\rm mJ/mole
K}^2$ \\ \hline
$N(0)$ & $0.42 \,{\rm states/eV}$ & $0.42 \,{\rm states/eV}$ \\ 
$N^{\star}(0)$ & $\le 0.64 \,{\rm states/eV}$ & $\le 0.64 \,{\rm
states/eV}$ \\ \hline 
$\lambda_{\rm el-ph}$ & \multicolumn{1}{p{4cm}}{$\cong 0.5$ \/, (but
less than $0.8$)} & \multicolumn{1}{|p{4cm}|}{$\cong 0.7$\/,
(but less than $0.8$)} \\ \hline 
$\gamma(=\Delta C/1.43T_c)$ & $\cong 1.5 \/{\rm mJ/mole K}^2$ & $0.8 \pm
0.2 \/{\rm mJ/mole K}^2$ \\  \hline
$\lambda_{\rm penetration}$ & $\cong 4000 \AA$ & $5000 - 6000 \AA$
\\ \hline
\end{tabular}

\newpage
\vspace{2cm}
{\bf Table II.1} \\[1cm] 
\begin{tabular}{|cl|cl|}
\hline\hline
   & ${\rm U}_{3+} = 19.7$ &    & ${\rm U}_{3+} = 18.8$ \\
Bi & ${\rm U}_{4+} = 19.7$ & Sb & ${\rm U}_{4+} = 11.9$  \\
   & ${\rm U}_{5+} = 32.3$ &    & ${\rm U}_{5+} = 52.0$ \\
\hline
\end{tabular}
\begin{tabular}{|cl|cl|cl|}
\hline
   & ${\rm U}_{2+} = 16.9$ &    & ${\rm U}_{2+} = 15.9$ &    & ${\rm
U}_{2+} = 18.2$ \\ 
Pb & ${\rm U}_{3+} = 10.4$ & Sn & ${\rm U}_{3+} =  9.1$ & Ge & ${\rm
U}_{3+} =  9.5$ \\  
   & ${\rm U}_{4+} = 25.5$ &    & ${\rm U}_{4+} = 31.6$ &    & ${\rm
U}_{4+} = 48.7$ \\ 
\hline
   & ${\rm U}_{1+} = 14.3$ &    & ${\rm U}_{1+} = 13.1$ &    & ${\rm
U}_{1+} = 14.6$ \\ 
Tl & ${\rm U}_{2+} =  9.4$ & Sn & ${\rm U}_{2+} =  9.2$ & Ga & ${\rm
U}_{2+} =  9.2$ \\  
   & ${\rm U}_{3+} = 20.9$ &    & ${\rm U}_{3+} = 26.4$ &    & ${\rm
U}_{3+} = 26.4$ \\ 
\hline
   & ${\rm U}_{3+} = 13.3$ &    & ${\rm U}_{3+} = 18.7$ &    & ${\rm
U}_{2+} = 13.9$ \\ 
Nb & ${\rm U}_{4+} = 11.7$ &  V & ${\rm U}_{4+} = 17.0$ & Ti & ${\rm
U}_{3+} = 15.8$ \\  
   & ${\rm U}_{5+} = 53.0$ &    & ${\rm U}_{5+} = 64.0$ &    & ${\rm
U}_{4+} = 56.6$ \\ 
\hline\hline
\end{tabular}

\vspace{2cm}
{\bf Table II.2}\\
Correspondence between Pseudospin model \& Extended Hubbard model \\
\begin{tabular}{|p{6cm}||p{6cm}|}
\hline
Pseudospin model & Extended Hubbard model \\ \hline\hline
Up spin & Vacancy \\ \hline
Down spin & Double occupancy \\ \hline
Magnetisation $m$ & Deviation from $\frac{1}{2}$ filling \\ \hline
Magnetic field $B$ & Chemical Potential $\mu$ \\ \hline
Antiferromagnetic (AFM) order & Bose Crystal of Cooperons (CDW phase)
\\ \hline
Spin flopped phase (i.e. AFM with nonzero XY component of spins) &
Bose condensate of Cooperons (S-wave superconductivity) \\ \hline
Non-ordered phase (with partial alignment for $B \neq 0$ & Bose
liquid of the Co\-ope\-rons (normal metal) \\ \hline 
\end{tabular}
\end{center}

\newpage
\centerline{\large \bf Figure Captions}

\vspace*{1em}

\begin{description}
\item[Fig.I.1a] The underlying cubic perovskite structure of BaBiO$_3$
with oxygen atoms at the corners of the octahedra centered around
Bi-atoms.  Bi-atoms occupy the vertices of a simple-cubic lattice; Ba
atoms are at the body centres of this cubic lattice.

\item[Fig.I.1b] Possible distortions of this underlying cubic perovskite
structure that lead to the various lower-symmetry structures found in
Ba$_{1-x}$K$_{x}$BiO$_{3}$ and BaPb$_{1-x}$Bi$_{x}$O$_{3}$.

\item[Fig.I.2a] Schematic phase diagram of the
BaPb$_{1-x}$Bi$_{x}$O$_{3}$ system in the $T-x$ plane (after
Sleight \cite{AWS1}). 
 The orthorhombic(O$_{II}$) and tetragonal (T) phases are metallic at high
temperatures and superconducting at low temperatures.  The orthorhombic
(O$_{II}$) and monoclinic(M) phases are semiconducting.  The precise
nature of the metal-semiconductor is not clear, but here it is shown as a
first-order transition with the hatched region indicating a region of
two-phase coexistence.  All the four phases undergo high-temperature
structural transitions to a cubic phase.

\item[Fig.I.2b] Schematic phase diagram of the
Ba$_{1-x}$K$_{x}$BiO$_{3}$ system in the $T-x$ plane after
Pei, et al. \cite{SP1,DGH2} The entire cubic phase is metallic and
the hatched, low temperature region is a superconducting phase.
The monoclinic(M), orthorhombic (O), and rhombohedral (R)
regions are semiconducting. 

\item[Fig.I.3a] Temperature dependence of the resistivity of
BaPb$_{1-x}$Bi$_{x}$O$_{3}$ on a log-log scale in the metallic regime and
across the metal-semiconductor transition regime.  Note that the
temperature-coefficient of resistivity is negative at $x$ = 0.3, but
resistivity goes to zero at low temperatures. (Afte Uchida et
al. \cite{SU1}) 

\item[Fig.I.3b] The temperature dependence of the electrical resistivity of
BaPb$_{1-x}$Bi$_{x}$O$_{3}$ in the semiconducting regime (after
Uchida et al. \cite{SU1}). 

\item[Fig.I.3c] Temperature dependence of the electrical
resistivity of Ba$_{1-x}$K$_{x}$BiO$_{3}$ (for $x=0.45$) thin
films for two different samples $i$ and $j$ prepared under
slightly different annealing conditions(see Sato, et
al. \cite{sato}). 

\item[Fig.I.4] Reflectivity spectra in the metallic regime $0\leq x \leq
0.2$ \/ of BaPb$_{1-x}$Bi$_{x}$O$_{3}$ alloys.  The solid curves are
results of a Drude fit.  Note the deviation of the Drude fit a
$x=0.2$ (after Uchida et al. \cite{SU1}). 

\item[Fig.I.5] Variation of the optical gap $E_g$ (i.e., the position of
the peak in the optical absorption spectrum) and the transport gap $E_A$
(i.e., transport activation energy) in eV, of BaPb$_{1-x}$Bi$_{x}$O$_{3}$
with concentration $x$, in the semiconducting phase.  The optical gap does
not vanish at the semiconductor-metal transition.  Two different scales
have been used to show the wide separation of the energies of the two
gaps (after Uchida, et al. \cite{SU1}).

\item[Fig.I.6a] Optical conductivity spectrum (in a linear scale) in the
semiconducting phase of BaPb$_{1-x}$Bi$_{x}$O$_{3}$  alloys.  Note the
gradual shift of the conductivity peak towards high frequency with
increasing $x$ (after Uchida, et al. \cite{SU1}).

\item[Fig.I.6b] Optical conductivity of
Ba$_{1-x}$K$_{x}$BiO$_{3}$ with $x=0.0$ (solid), $x=0.1$ (dot),
$x=0.2$ (dash), $x=0.33$ (dot dash) and $x=0.4$ (solid) from
Blanton, et al. \cite{blanton}. 

\item[Fig.I.7] (a) The composition dependence of the mag\-ne\-tic
sus\-cepti\-bi\-lity at 150K in\/\break \mbox{BaPb$_{1-x}$Bi$_{x}$O$_{3}$}.
The solid line is the dia\-magne\-tic con\-tribu\-tion to
sus\-cepti\-bi\-lity from the ionic core electrons (after Uchida,
et al. \cite{SU1}). In (b) is shown the normal state magnetic
susceptibility of both lead and potassium doped systems as a
function of temperature. The diamagnetic core contributions for
the lead and potassium doped systems are, respectively,
8.6$\times 10^{-5}$ emu/mole and 7.8$\times 10^{-5}$ emu/mole
(after Batlogg, et al. \cite{BB5}). 

\item[Fig.I.8a] The temperature dependence of the resistivity of a single
crystal thin-film sample of\/ BaPb$_{0.7}$Bi$_{0.3}$O$_{3}$\/ at different
levels of oxygen deficiency.  The labels 5 to 1 are in the increasing
order of oxygen deficiency (after Enomoto et al. \cite{YE}).

\item[Fig.I.8b] The temperature dependence of the carrier concentration
for a BaPb$_{0.7}$Bi$_{0.3}$O$_{3}$ single-crystal thin filsm. The labels
5 to 1 are in order of increasing Oxyen deficiency  (after
Enomoto et al. \cite{YE}). 

\item[Fig.I.9] Temperature dependence of the upper critical
field (in Tesla): 
(a) For BaPb$_{1-x}$Bi$_{x}$O$_{3}$ at different filling $x$. S
and P denote, respectively, single crystal and polycrystal
samples. Doping $x$ and the superconducting $T_c$ of the samples are
given with each curve (after Kitazawa, et al. \cite{KK1}).
(b) For Ba$_{1-x}$K$_{x}$BiO$_{3}$ thin films at $x=0.45$ (after
Sato, et al. \cite{sato}).

\item[Fig.I.10] A spherically symmetric bismuth $ 6s$ orbital surrounded by
the oxygen p orbitals in the BiO$_6$ octahedron.  Only the $\sigma$
orbitals of each of the six oxygen atoms are shown.

\item[Fig.I.11] LAPW band-structure results for
BaPb$_{1-x}$Bi$_{x}$O$_{3}$ alloys. The effect of Pb doping is
shown from (c) through (a); 
$\sigma$ and $\sigma^*$ denote the bonding and antibonding $\sigma$ bands
respectively.  The antibonding band that crosses the Fermi level
$E_F\/$ comes from the Bi/Pb6s-O2$_p$\/\/$\sigma$ bonding.  Lead
doping adjusts 
the filling of the $\sigma^{*}$ subband and also introduces a noticeable
chemical shift in the position of the $6s$ band (open circles) (after
Mattheiss \cite{LFM3}).  

\item[Fig.I.12] Comparison of schematic bands for the high- and low-$T_c$
\/oxide superconductors. (a) The Fermi level lies in the antobonding band
of predominantly p character for the high-$T_{c}$ materials. (b) For the
low $T_c$ materials the Fermi level is in the degenerate $\pi^{*}$
manifold.  The quantities $\epsilon_p,\; \epsilon_d,\;$  and
$\epsilon$, are the energies of the ligand p-level and cation d and s
levels, respectively (after Mattheiss \cite{LFM3}).

\item[Fig.I.13] Band-structure results for the lead-doped system with a
tight-binding fit. (a) LAPW  energy bands (solid lines) - for cubic \/
BaPb$_{0.7}$Bi$_{0.3}$O$_{3}$\/ compared with those derived from simple
tight-binding models involving three (dotted lines) and five (dashed
lines) tight-binding parameters respectively \cite{LFM2}.  Note the ten
bands are the high-lying Ba-derived bands.
(b) Sketch of the electron-like F-centered simple cubic Fermi surface
predicted by the three-parameter tight-binding model for BaBiO$_3$.  Note
that the electron surface nests with an identical hole surface centered at
the Brillonin-zone corner (R) (after Mattheiss and Hamman \cite{LFM1}).

\item[Fig.I.14] Model tight-binding conduction bands for cubic (dashed
line) and distorted (solid line) BaPb$_{1-x}$Bi$_{x}$O$_{3}$ alloys.  The
downward movement of the Fermi level as a function of doping is also
shown.  Note that at half filling the gap is at the Fermi
surface (after Mattheiss \cite{LFM3}). 

\item[Fig.I.15] (a) LAPW band-structure results for the
Ba$_{1-x}$K$_{x}$BiO$_{3}$ system.  The wide bands of Bi6s-O2p are bounded
above and below by the $\sigma^* $\/ and  \/ $\sigma$ bands.  The Fermi
level lies in the $\sigma^*$ band.  Note the similarity between the band
structures of the K-doped and Pb-doped systems  (after
Mattheiss  and Hamman \cite{LFM1,LFM3}).  (b) The full potential LMTO band
structure near fermi level for 2BaBiO$_3$. Solid line is for $t=0^{o}$ and
$b=0\AA$, dashed line is for $t=10^{o}$ and $b=0\AA$, dot-dashed line is for
$t=10^{o}$ and $b=0.09\AA$. The brillouin zone has been folded
(after Liechtenstein, et al. \cite{liech}).

\item[Fig.I.16] $T_c\/$ versus the Sommerfeld constant $\gamma$ for various
superconductors.  bismuthates and cuprates have the highest $T_c$'s on an
absolute scale (and especially given their low DOS at $E_F\/$).
The extraordinary character of these two types of systems relative to all
other superconductors, including oxides, is highlighted in the above plot.
The dashed line demarcates low-$T_c$ superconductors from the high-$T_c$
superconductors (after Batlogg \cite{RJC2}).

\item[Fig.II.1] $T_c$ as a function of $U_{eff}$/2V calculated in the
Hartree approximation for the BEG model (lower curve).  Solid lines
denote second order and dashed lines denote first-order transitions and
the dot, a tricritical point. In the Hartree calculations $U_{o}$ and
$e^{2}/\alpha a$ were kept fixed at the values 2 and 1, respectively, and
$g^2$/C is varied (after Rice and Sneddon \cite{TMR}).

\item[Fig.II.2] $T_c$ as a function of $\overline{z}$, the average
coordination number in the BaPb$_{1-x}$Bi$_{x}$O$_{3}$ \/ system for three
different values of $U_{eff}$/2V.  $T_c$ is calculated in the
Bethe-Peierls approximation.  The inset on the top right corner shows the
effect of a nonzero band with ($t_{ij} \neq 0$) in this phase diagram, on
the region near zero temperature, as described in the text, for
$U_{eff}/2V > 0$.  In this diagram M is a metallic phase and S stands for
BCS superconductor  (after Sneddon and Rice \cite{TMR}).

\item[Fig.II.3a] The DOS of the Bi A-site (solid line) and Bi B-site
(dashed line) at 20\% Pb doping using $\epsilon_o$ = 2eV and $\lambda =
0.8$.  The Bi subbands approach each other because of a level repulsion
between the Bi and Pb single bands.  The total DOS is shown in the inset
at the top right corner.  The different level shifts of the Bi subbands
produce a pseudogap around the Fermi energy.  In (b) is shown the case of
60\% Pb doping using the same parameter values as in (a).  The Bi subbands
continuously approach each other with increasing Pb doping.  The states
around the Fermi energy are still almost composed of Bi sites (after
Jurczek and Rice \cite{EJ1}).

\item[Fig.II.4a] The boundary between semiconducting and metallic phases
in the $\epsilon^{(0)}-x$ \/ plane for several choices of $U$.  The
symbols are explained in the text.  In (b) is shown the DOS obtained in
the CPA at two different values of U (after Yoshioka and
Fukuyama \cite{DY}). 

\item[Fig.II.5] The superconducting transition temperature as a function
of filling $x$ for different values of $U$ with $\epsilon^{(0)}$ kept fixed.
 Notte that the phase diagram is symmetric with respect to $x
\longrightarrow 1-x$ (after Yoshioka and Fukuyama \cite{DY}).

\item[Fig.III.1a] A typical phase diagram of the pseudospin model in the
pseudomagnetic field ($B$) and temperature ($T$) plane.  The full lines are
second-order phase boundaries and the dashed lines indicate first-order
phase boundaries.  The two continuous lines and the first-order line meet
at a bicritical point (the full circle).  The first-order line separates
the charge-density-wave (CDW) phase (antiferromagnet in the spin language)
from the singlet superconducting (SS) (spin flopped in the spin language)
phase.  The two continuous lines separate the non-ordered metallic (NOM)
phase (paramagnet in the spin language, with a net moment because of the
aligning field) from the CDW and SS phases.  The arrows show the
arrangement of the spins on nearest-neighbour sites.  The parameters used
are $J_o = 1.0$ \/ and $K_o = 1.301$.

\item[Fig.III.1b] The phase diagram of Fig.III.1a in the magnetisation
($m$) and temperature ($T$) plane.  The full lines are second-order phase
boundaries and the dashed lines indicate first-order phase boundaries.
The hatched region is the two-phase coexistence region (SS and CDW)
corresponding to the first-order phase boundary between CDW and SS phases
in Fig.III.1a.

\item[Fig.III.2] The mean-field phase diagram when a next
nearest neighbour interaction is included in model II.5 (in the
large-$U$ limit). The phase denoted by I is the so called
intermediate phase (after K.-S. Liu and M.E. Fisher, J.
Low Temp. Phys. {\bf 10}, 655-683 (1973)).  

\item[Fig.III.3] The RPA excitation spectra of the pseudospin model shown
along different symmetry directions of the Brillouin zone of a cubic
lattice (see inset), with $J_o = 1.0$ and $K_o=1.301$.  The lines marked
CDW(+) and CDW(-) correspond to the upper and lower branches of the
spectrum in the CDW phase (for temperature $T=0.25$, pseudomagnetic field $B$
= 0.25, magnetisation $m = 0.4$ and staggered magnetisation $\eta = 0.96$).
The dashed line shows the spectrum for the SS phase (at $T = 0.25, B =
0.70, m = 0.61$ and \/ $\theta_A = \theta_B = 50.55^{o}$, where $\theta_A
\; \& \;  \theta_B$ are defined in the text).  The dash-dotted line shows
the spectrum in the NOM phase (at $T = 0.80, B = 0.40$ and $m = 0.27$).  The
spectrum in the SS phase is gapless whereas the spectrum in the CDW phase
has the anisotropy gap discussed in the text.

\item[Fig.III.4] The optical conductivity (in arbitrary units) in
the large-$U$ limit using RPA plotted as a function of
frequency. The parameter values taken are $zt=1$ eV, $\/U=2$ eV,
and $zV=0.05$ eV. The same phenomenological broadening ($\Gamma=0.09$)
has been used for both the channels. Note that the two-particle
absorption peak is a factor of about 9 down compared to the
one-particle peak.

\item[Fig.IV.1(a)] The mean-field phase diagrams of the negative-U,
extended Hubbard model in filling-temperature plane, for two different
values of the interaction parameters.  The solid lines and the hatched
region have the same meaning as in Fig.III.1.  The values of Coulomb
interactions and zt used to obtain this phase diagram are shown in the
figure. 

\item[Fig.IV.1b] The mean-field phase diagram of the negative-U,
extended Hubbard model in the chemical potential-temperature plane.  The
solid and the dashed lines denote continuous and first-order transition,
respectively as in Fig.III.1a.  Other symbols also have the same meanings
as in Fig.III.1.

\item[Fig.IV.2] (a) The single-particle excitation spectrum in the
mean-field CDW phase where $\Delta_c$ is the CDW gap parameter.  In (b)
are shown, schematically, the two-particle (hole) bound states that form
inside the gap of the two-particle spectrum.  The energy of these bound
states is the transport gap ($E_A\/$) in our theory.

\item[Fig.IV.3] The diagrammatic representation of the RPA  series for the
pair susceptibility in the mean-field CDW state.   Double lines denote
mean-field single-particle Green function in the CDW state and the wavy
line denotes the interaction {\large $\nu_{ij;i^{\prime}j^{\prime}
({\bf k},{\bf k}^{\prime})}$}. 

\item[Fig.IV.4] A schematic representation of the Cooper instability
equation as described in the text.

\item[Fig.IV.5a] The bare (empty circles) and the screened (empty
triangles) onsite Coulomb interactions as functions of (half) the band
width.  The screening mechanism is explained in the text (see also Fig.IV.7). 

\item[Fig.IV.5b] The extended Coulomb term $V$\/ ($\times 2z$)
as a function of (half) the band width $W (=zt)$.  The condition that we have
used to obtain this graph is that the bound state energy should be equal
to the transport gap (see the text).  Note that in the bandwidth
$\rightarrow$ 0 limit the value of 2$zV$ approaches 0.24eV, the result
obtainted in the large-$U$ limit.

\item[Fig.IV.6] The Coulomb interactions $U$ and $zV$ as a functions of the
bound-state energy.  This graph is obtained with the condition that the
CDW gap parameter is half the optical gap of BaBiO$_3$ (i.e.,
the CDW gap is 1 eV).  Note that $V$ goes to zero as the
bound-state energy approaches zero. 

\item[Fig.IV.7(a)] A diagrammatic representation of the screening (in RPA)
of the particle-particle interaction.  Note that the particle-particle
interaction is screened by the particle-hole excitations.

\item[Fig.IV.7b] Same as (a) for the particle-hole interaction.  The
particle-hole interaction is screened by the particle-particle processes.
Note that our calculation is not self consistent (see text).

\item[Fig.IV.8] (a) The superconducting transition temperature
with and without the screening described in the text. (b) The
CDW transition temeperature with and without the screening. 

\item[Fig.IV.9] The superconductor-nonordered and the CDW -nonordered
phase boundary evaluated with the same parameters as in IV.1b.  To show
that screening of $V$ does not much affect the superconductor-nonordered
phase boundary at the range of parameter values we are interested in, we
draw this phase boundary for $V=0$ also (dashed line).  Note that this
boundary is quite close to the boundary with $2zV=0.13eV$ (the solid and the
dashed line both represent continuous phase transitions.)
\end{description}

\end{document}